\documentclass[12pt]{article}

 \usepackage{graphicx}
 \usepackage{cite}

\begin{document}


\begin{center}
{\LARGE\bf Dispersion relation and spectral range of\\[2mm]
Karsten-Wilczek and Borici-Creutz fermions}
\end{center}

\vspace{20pt}

\begin{center}
{\large\bf Stephan D\"urr$\,^{a,b}$}
\,\,\,and\,\,\,
{\large\bf Johannes H.\ Weber$\,^{c}$}
\\[10pt]
${}^a${\sl Department of Physics, University of Wuppertal, 42119 Wuppertal, Germany}\\
${}^b${\sl J\"ulich Supercomputing Centre, Forschungszentrum J\"ulich, 52425 J\"ulich, Germany}\\
${}^c${\sl Department of Computational Mathematics, Science and Engineering and Department of\\
           Physics and Astronomy, Michigan State University, East Lansing, Michigan 48824, USA}
\end{center}

\vspace{10pt}

\begin{abstract}
We investigate some properties of Karsten-Wilczek and Borici-Creutz fermions, which are the best known varieties in the class of minimally doubled lattice fermion actions.
Our focus is on the dispersion relation and the distribution of eigenvalues in the free-field theory.
We consider the situation in two and four space-time dimensions, and we discuss how properties vary as a function of the Wilson-like lifting parameter $r$.
\end{abstract}

\vspace{20pt}

\newcommand{\pad}{\partial}
\newcommand{\hqu}{\hbar}
\newcommand{\til}{\tilde}
\newcommand{\pri}{^\prime}
\renewcommand{\dag}{^\dagger}
\newcommand{\<}{\langle}
\renewcommand{\>}{\rangle}
\newcommand{\gaf}{\gamma_5}
\newcommand{\nab}{\nabla}
\newcommand{\lap}{\triangle}
\newcommand{\dal}{{\sqcap\!\!\!\!\sqcup}}
\newcommand{\trc}{\mathrm{tr}}
\newcommand{\Trc}{\mathrm{Tr}}
\newcommand{\Mpi}{M_\pi}
\newcommand{\Fpi}{F_\pi}
\newcommand{\Mka}{M_K}
\newcommand{\Fka}{F_K}
\newcommand{\Met}{M_\et}
\newcommand{\Fet}{F_\et}
\newcommand{\Mss}{M_{\bar{s}s}}
\newcommand{\Fss}{F_{\bar{s}s}}
\newcommand{\Mcc}{M_{\bar{c}c}}
\newcommand{\Fcc}{F_{\bar{c}c}}

\newcommand{\al}{\alpha}
\newcommand{\be}{\beta}
\newcommand{\ga}{\gamma}
\newcommand{\de}{\delta}
\newcommand{\ep}{\epsilon}
\newcommand{\ve}{\varepsilon}
\newcommand{\ze}{\zeta}
\newcommand{\et}{\eta}
\renewcommand{\th}{\theta}
\newcommand{\vt}{\vartheta}
\newcommand{\io}{\iota}
\newcommand{\ka}{\kappa}
\newcommand{\la}{\lambda}
\newcommand{\rh}{\rho}
\newcommand{\vr}{\varrho}
\newcommand{\si}{\sigma}
\newcommand{\ta}{\tau}
\newcommand{\ph}{\phi}
\newcommand{\vp}{\varphi}
\newcommand{\ch}{\chi}
\newcommand{\ps}{\psi}
\newcommand{\om}{\omega}

\newcommand{\qhat}{\hat{q}}
\newcommand{\khat}{\hat{k}}

\newcommand{\bdm}{\begin{displaymath}}
\newcommand{\edm}{\end{displaymath}}
\newcommand{\bea}{\begin{eqnarray}}
\newcommand{\eea}{\end{eqnarray}}
\newcommand{\beq}{\begin{equation}}
\newcommand{\eeq}{\end{equation}}

\newcommand{\mr}{\mathrm}
\newcommand{\mb}{\mathbf}
\newcommand{\ri}{\mr{i}}
\newcommand{\Nf}{N_{\!f}}
\newcommand{\Nc}{N_{ c }}
\newcommand{\Nt}{N_{ t }}
\newcommand{\Nv}{N_{ v }}
\newcommand{\Nthr}{N_\mr{thr}}
\newcommand{\Dst}{D^\mr{st}}
\newcommand{\DS}{D_\mr{S}}
\newcommand{\DW}{D_\mr{W}}
\newcommand{\DKW}{D_\mr{KW}}
\newcommand{\DBC}{D_\mr{BC}}
\newcommand{\MeV}{\,\mr{MeV}}
\newcommand{\GeV}{\,\mr{GeV}}
\newcommand{\fm}{\,\mr{fm}}
\newcommand{\MSbar}{\overline{\mr{MS}}}



\section{Introduction}


The choice of any lattice fermion action is a bit of a compromise.
Ideally, one would want to realize ultra-locality, chiral symmetry, and absence of doublers (in addition to a correct continuum limit, of course).
But these are precisely the ingredients which, according to the Nielsen-Ninomiya theorem, cannot possibly coexist \cite{Nielsen:1981xu,Nielsen:1980rz}.
Furthermore, in real life computational expedience is an important criterion.
It follows that the choice of a lattice action which is well-suited to the specific needs of a planned numerical investigation
is an important decision which impacts the subsequent analysis of the lattice data in a profound manner.

Staggered fermions and Wilson fermions represent two popular choices in this context.
Staggered fermions put a focus on ultra-locality and chiral symmetry, at the expense of having $4$ species in the continuum \cite{Susskind:1976jm}.
Wilson fermions, on the other hand, prioritize ultra-locality and absence of doublers, at the expense of breaking chiral symmetry \cite{Wilson:1974sk}.

Staggered fermions seem ideally suited to study theories with four degenerate fermions (or a multiple thereof) in the continuum.
The details of taste-breaking \cite{Sharatchandra:1981si,KlubergStern:1983dg,Blum:1996uf,Orginos:1998ue,Lee:1999zxa}
and the non-commutativity of the continuum limit ($a\to0$) and the chiral limit ($m\to0$) \cite{Durr:2004ta} impose practical difficulties,
but to the best of our knowledge there is no concern about the theoretical soundness of this formulation of QCD with $\Nf\in4\,\mathbf{N}$ dynamical flavors.

Things are different, if one desires to study QCD with fewer than $4$ degenerate fermions, such as real-life QCD where $m_d,m_u,m_s,m_c$ are pairwise different.
Given the continuum argument that $\Nf$ degenerate fermions would raise the functional determinant of a single species to the $\Nf$-th power,
Marinari, Parisi and Rebbi suggested by means of ``reverse engineering'' that one would take the square-root of the staggered determinant to
simulate $2$ degenerate fermions or the quarter-root for a non-degenerate fermion species \cite{Marinari:1981qf}.

In practice, it seems the approach of ``rooting'' the staggered determinant yields convincing numerical results for real-life QCD,
with small statistical error bars even at physical values of the quark masses $m_d,m_u,m_s,m_c$ \cite{Bazavov:2009bb,Bazavov:2017lyh,Bazavov:2018omf},
and with some field-theoretic underpinning through rooted staggered chiral perturbation theory \cite{Sharpe:2004is,Bernard:2004ab,Bernard:2006zw}.
Nevertheless, this approach has been criticized \cite{Creutz:2007yg,Creutz:2008kb} on the grounds of the argument that --~in the presence of the cut-off~-- no local theory can be constructed
(or has been constructed) that would implement exactly (i.e.\ down to machine precision) the square-root of the staggered determinant as the determinant of a valid $2$ species formulation.
There have been several reviews of the issue at lattice conferences \cite{Adams:2004wp,Durr:2005ax,Sharpe:2006re,Kronfeld:2007ek,Golterman:2008gt}
which essentially collected pieces of evidence in favor of the approach.
But it holds true that no strict mathematical proof has been found, and no side has been able to convince the opponent.

Given this situation it is natural to ask whether a local formulation with chiral symmetry and just $2$ species (the minimum required by the Nielsen-Ninomiya theorem)
would shed some light on the issue.
Since staggered fermions emerge from the naive formulation by an ingenious procedure of ``thinning out'' the degrees of freedom by a factor $2^{d/2}$ in $d$ space-time dimensions,
one might dream of a similarly ingenious second step that would reduce the degeneracy from $4$ to $2$ in $d=4$ dimensions.
However, the eigenvalue spectrum of staggered fermions on interacting backgrounds shows a $4$-fold near-degeneracy (i.e.\ no exact degeneracy) \cite{Follana:2004sz,Durr:2004as},
and this means that such a second reduction step cannot possibly take place.

However, there is a better approach.
It is based on adding an extra term (of mass-dimension five) to the naive action which lifts $14$ of the $16$ species in $d=4$ dimensions,
albeit with the important difference to the Wilson term that it does not break chiral symmetry.
Such ``minimally doubled fermions'' have been proposed by Karsten \cite{Karsten:1981gd} and Wilczek \cite{Wilczek:1987kw},
and later by Creutz \cite{Creutz:2007af} and Borici \cite{Borici:2007kz}.
More recently, yet another variety with ``twisted ordering'' has been proposed by Creutz and Misumi \cite{Creutz:2010cz}.
Also the proposal of augmenting the Karsten-Wilczek action by a ``flavored chemical potential term'' has been made \cite{Misumi:2012uu,Misumi:2012ky}.
From the viewpoint of computational efficiency, all these formulations augur well, since they are ultra-local with on-axis links only.
In the literature issues of mixing with lower-dimensional operators have been addressed, and how one can defeat them with appropriate tuning strategies
\cite{Bedaque:2008xs,Cichy:2008gk,Capitani:2009yn,Capitani:2010nn,Creutz:2010qm,Kimura:2011ik,Weber:2013tfa,Weber:2016jug,Weber:2017eds}.
Also the consistency of these formulations with the index theorem has been verified \cite{Creutz:2010bm,Tiburzi:2010bm}.
Furthermore, some minimally doubled actions have been shown to possess an extra ``mirror fermion'' symmetry \cite{Pernici:1994yj},
and it has been demonstrated that the Karsten-Wilczek determinant is invariant under all of the discrete symmetries \cite{Weber:2016dgo}.

Still, some basic features of minimally doubled fermion actions have hardly been explored, for instance
the respective quark-level free-field dispersion relations (including the cut-off effects on a heavy quark mass) and spectral bounds.
In this article we try to fill some of these gaps for Karsten-Wilczek (KW) and Borici-Creutz (BC) fermions.
To understand how these formulations differ from the naive one we think it is useful to introduce a lifting parameter $r$, similar to what is done for Wilson fermions.
Hence for $r=0$ we start with the naive action, and we expect to see a cascade of species reductions as $r$ increases, eventually realizing $2$ species at $r=1$.
Since chiral symmetry holds throughout, the Nielsen-Ninomiya theorem demands that the reduction proceeds in steps of (integer multiples of) 2.

The remainder of this article is organized as follows.
In Sec.~\ref{sec:2} we give a brief review of the situation with naive and Wilson fermions, mostly to specify our notation.
The results for KW fermions are presented in Sec.~\ref{sec:3}, and those for BC fermions in Sec.~\ref{sec:4}.
We summarize our findings in Sec.~\ref{sec:5}, and more lengthy calculations are arranged in appendices A-E.


\section{Naive and Wilson fermions\label{sec:2}}


Throughout this article $\pad_\mu$ and $\pad_\mu^*$ denote the discrete forward an backward derivative, respectively,
and $\nab_\mu=(\pad_\mu+\pad_\mu^*)/2$ is the symmetric derivative.
These operators are gauged in the obvious manner; for instance the symmetric covariant derivative is
\beq
\nab_\mu\ps(x)=\frac{1}{2}\,\Big[U_\mu(x)\ps(x+\hat\mu)-U_\mu\dag(x-\hat\mu)\ps(x-\hat\mu)\Big]
\eeq
where $U_\mu(x)$ is the parallel transporter from $x+\hat\mu$ to $x$, and $\hat\mu$ denotes $a$ times the unit-vector in direction $\mu$.
Similarly, $\lap_\mu=\pad_\mu^*\pad_\mu=\pad_\mu\pad_\mu^*$ denotes the second discrete derivative, that is
\beq
\lap_\mu\ps(x)=U_\mu(x)\ps(x+\hat\mu)-2\ps(x)+U_\mu\dag(x-\hat\mu)\ps(x-\hat\mu)
\eeq
in the presence of a gauge field $U_\mu(x)$.

With this notation the naive Dirac operator is defined as
\beq
D_\mr{nai}(x,y)=\sum_\mu \ga_\mu \nab_\mu(x,y)+m\de_{x,y}
\label{def_nai}
\eeq
where the anti-hermitean behavior $\nab_\mu\dag=-\nab_\mu$ and the anti-commutation property $\{\ga_\mu,\gaf\}=0$
of the hermitean $\ga$-matrices imply that the naive Dirac operator is $\gaf$-hermitean,
i.e.\ $\gaf D_\mr{nai} \gaf=D_\mr{nai}\dag$.
In the free-field limit this operator assumes a diagonal form in momentum space,
\bea
D_\mr{nai}(p)&=&\ri \sum_\mu \ga_\mu \frac{1}{a}\sin(ap_\mu)+m
\nonumber
\\
&=&\ri\sum_\mu \ga_\mu \bar{p}_\mu+m
\quad \mbox{with} \quad \bar{p}_\mu=\frac{1}{a}\sin(ap_\mu)
\label{momrep_nai}
\eea
which again highlights the anti-hermitean nature of the derivative (momentum) term.

The Wilson Dirac operator follows by adding a hermitean and positive semi-definite term of dimension $5$ to the naive Dirac operator
\beq
D_\mr{W}(x,y)=\sum_\mu \ga_\mu \nab_\mu(x,y)
-\frac{ar}{2}\sum_\mu \lap_\mu(x,y)
+m\de_{x,y}
\;.
\label{def_wils}
\eeq
The hermitean behavior $\lap_\mu\dag=\lap_\mu$ together with the properties used in the naive case
imply that the Wilson (W) operator is $\gaf$-hermitean, i.e.\ $\gaf D_\mr{W} \gaf=D_\mr{W}\dag$.
An unpleasant feature is that the term $\sum_\mu \lap_\mu(x,y)$ mixes (on interacting gauge backgrounds) with the identity.
As a result, the bare mass $m$ in (\ref{def_wils}) gets renormalized, and chiral symmetry is broken \cite{BOOK_MM}.
In the free-field limit the Wilson operator assumes a diagonal form in momentum space,
\bea
D_\mr{W}(p)&=&\ri \sum_\mu \ga_\mu \frac{1}{a}\sin(ap_\mu)+ar\sum_\mu \{1-\cos(ap_\mu)\}+m
\nonumber
\\
&=&\ri \sum_\mu \ga_\mu \bar{p}_\mu
+\frac{ar}{2} \sum_\mu \hat{p}_\mu^2+m
\quad \mbox{with} \quad \hat{p}_\mu=\frac{2}{a}\sin(\frac{ap_\mu}{2})
\label{momrep_wils}
\eea
which again highlights the anti-hermitean and hermitean positive semi-definite nature of the two terms, respectively.
Specifically for $r=1$ the $15$ unwanted species do not propagate in any of the on-axis directions \cite{BOOK_MM}.

\begin{figure}[p]
\centering
\includegraphics[width=0.78\textwidth]{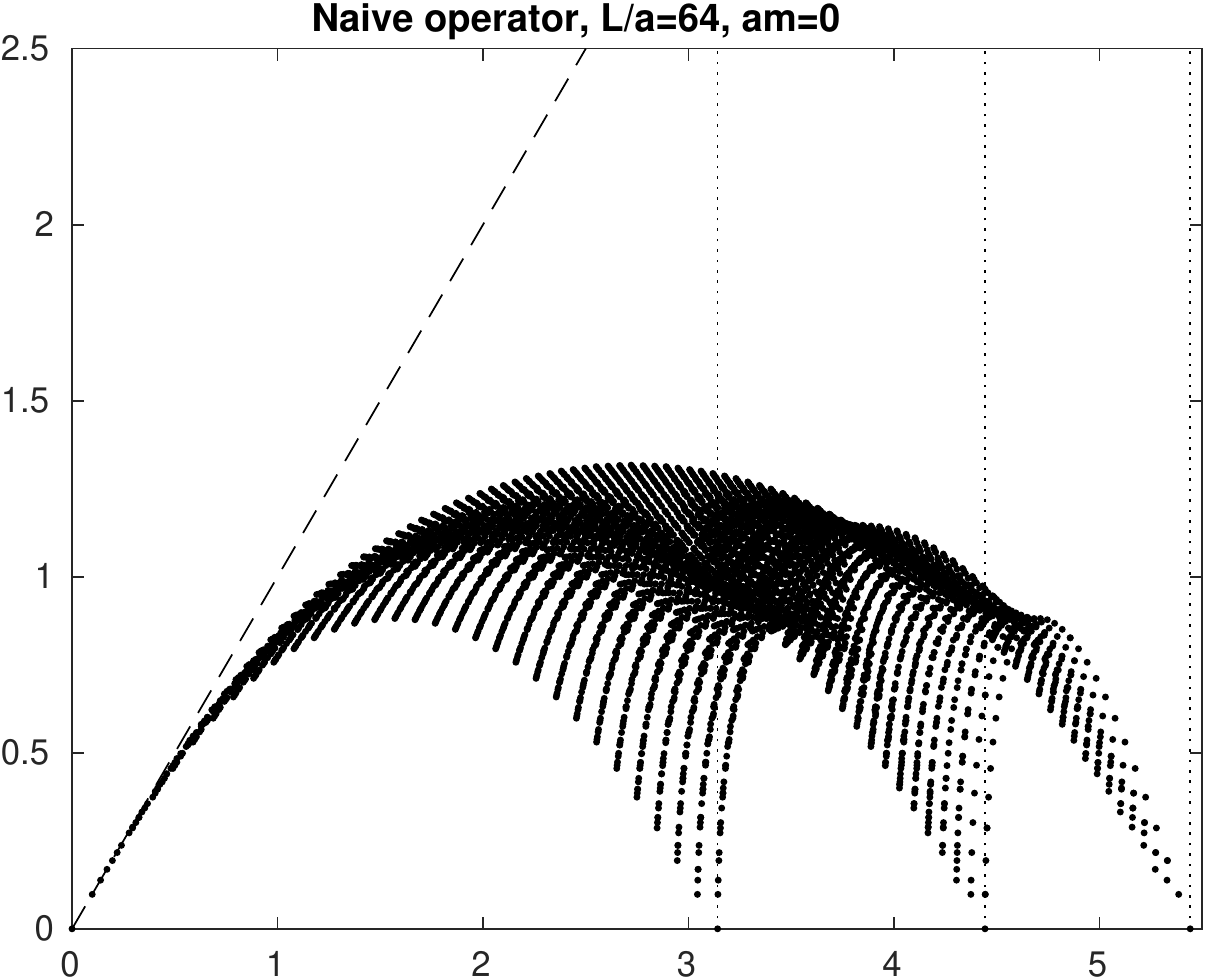}\\[1mm]
\includegraphics[width=0.78\textwidth]{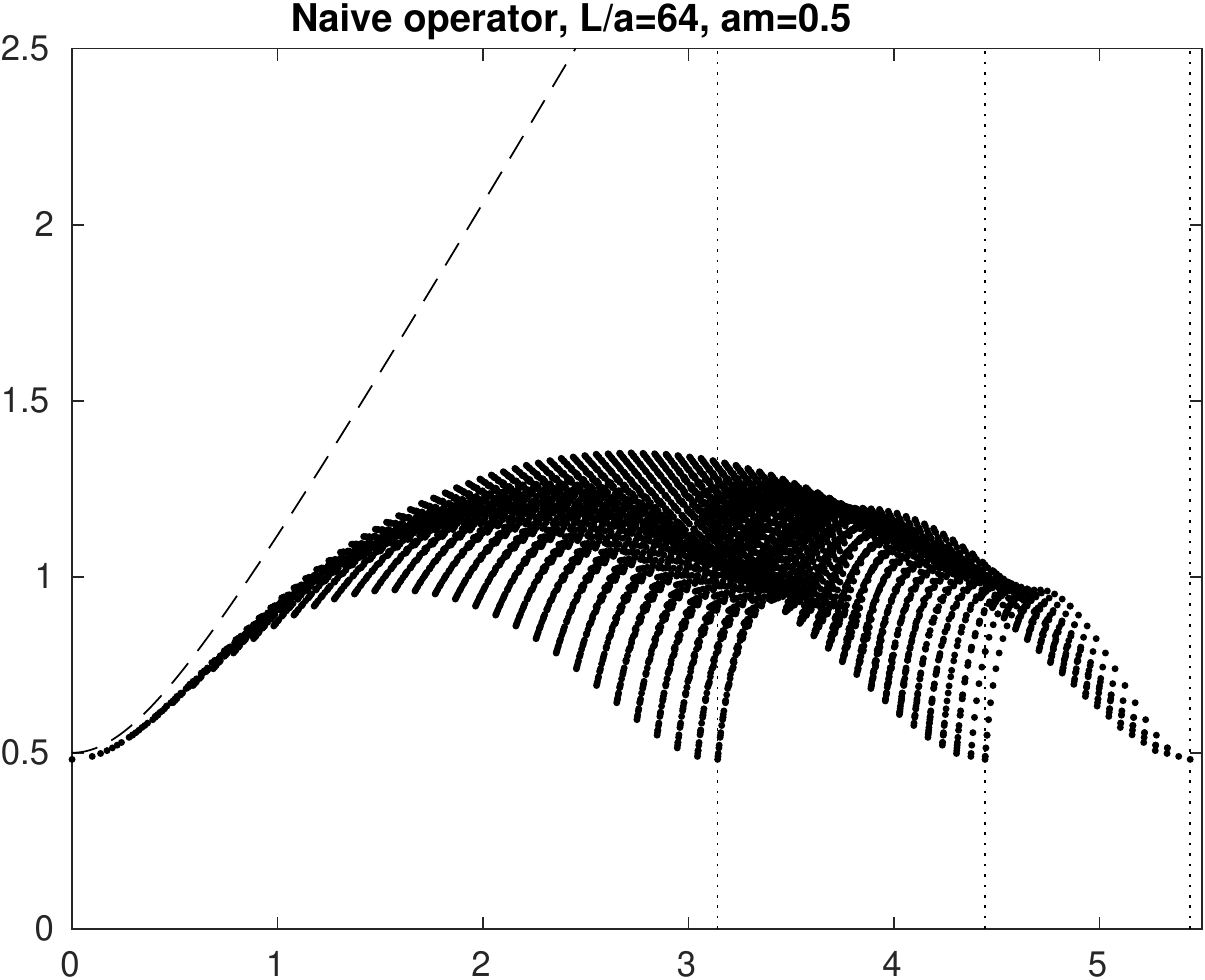}%
\caption{\label{fig:disp_naiv}
Free-field dispersion relation $aE$ versus $a|\vec{p}|$ of the naive Dirac operator at $am=0$ and $am=0.5$.
The dashed curves give the continuum dispersion relations, and the vertical lines show
the end of the Brillouin zone in the $(1,0,0)$, $(1,1,0)$, and $(1,1,1)$ directions, respectively.}
\end{figure}

\begin{figure}[p]
\centering
\includegraphics[width=0.78\textwidth]{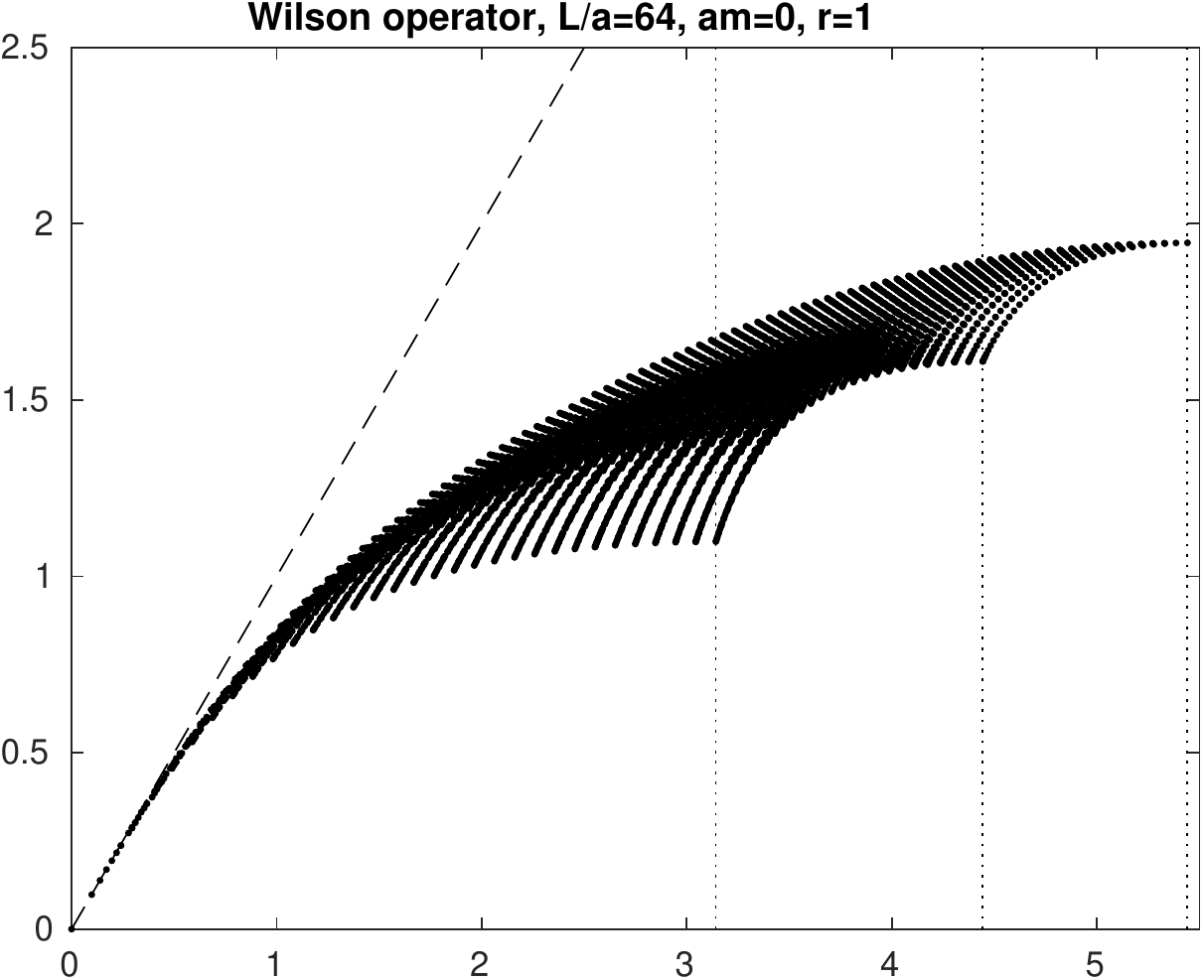}\\[1mm]
\includegraphics[width=0.78\textwidth]{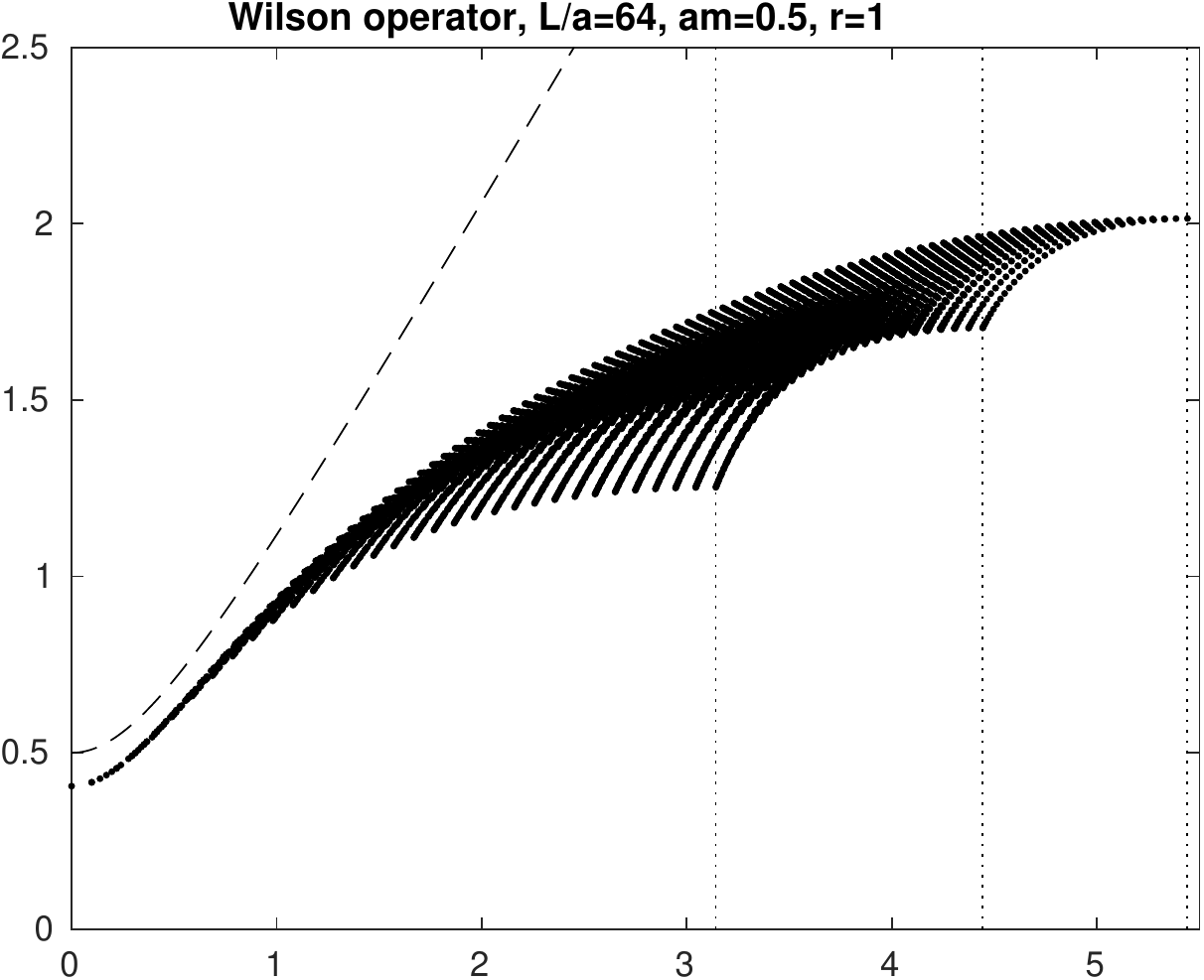}%
\caption{\label{fig:disp_wils}
Same as Fig.~\ref{fig:disp_naiv}, but for the Wilson operator at $r=1$.}
\end{figure}

With these expressions in hand, we are in a position to study the quark-level free-field dispersion relations.
Both for naive and Wilson fermions, the energy $aE$ can be given as an analytic function of the spatial momentum $a\vec{p}$,
see App.~\ref{app:A} for a brief account of this standard calculation.
The results are shown in Fig.~\ref{fig:disp_naiv} for the naive formulation and in Fig.~\ref{fig:disp_wils} for the Wilson action at $r=1$.
In either figure the situation at $am=0$ and at $am=0.5$ is compared to the respective continuum dispersion relation.
Apart from the unwanted zeros (or minima) at $a|\vec{p}|=\pi,\sqrt{2}\pi,\sqrt{3}\pi$ the naive action features well for small enough momenta.
In particular at $a|\vec{p}|=0$ it features better than the Wilson action, since the gap to the continuum curve is smaller.
This can be understood on analytical grounds, too.
The energy at zero momentum is nothing but the heavy quark mass.
As detailed in App.~\ref{app:B}, it is known to be afflicted with cut-off effects $O((am)^2)$ in the naive case,
but $O(am)$ in the Wilson case.

\begin{figure}[tb]
\centering
\includegraphics[width=0.78\textwidth]{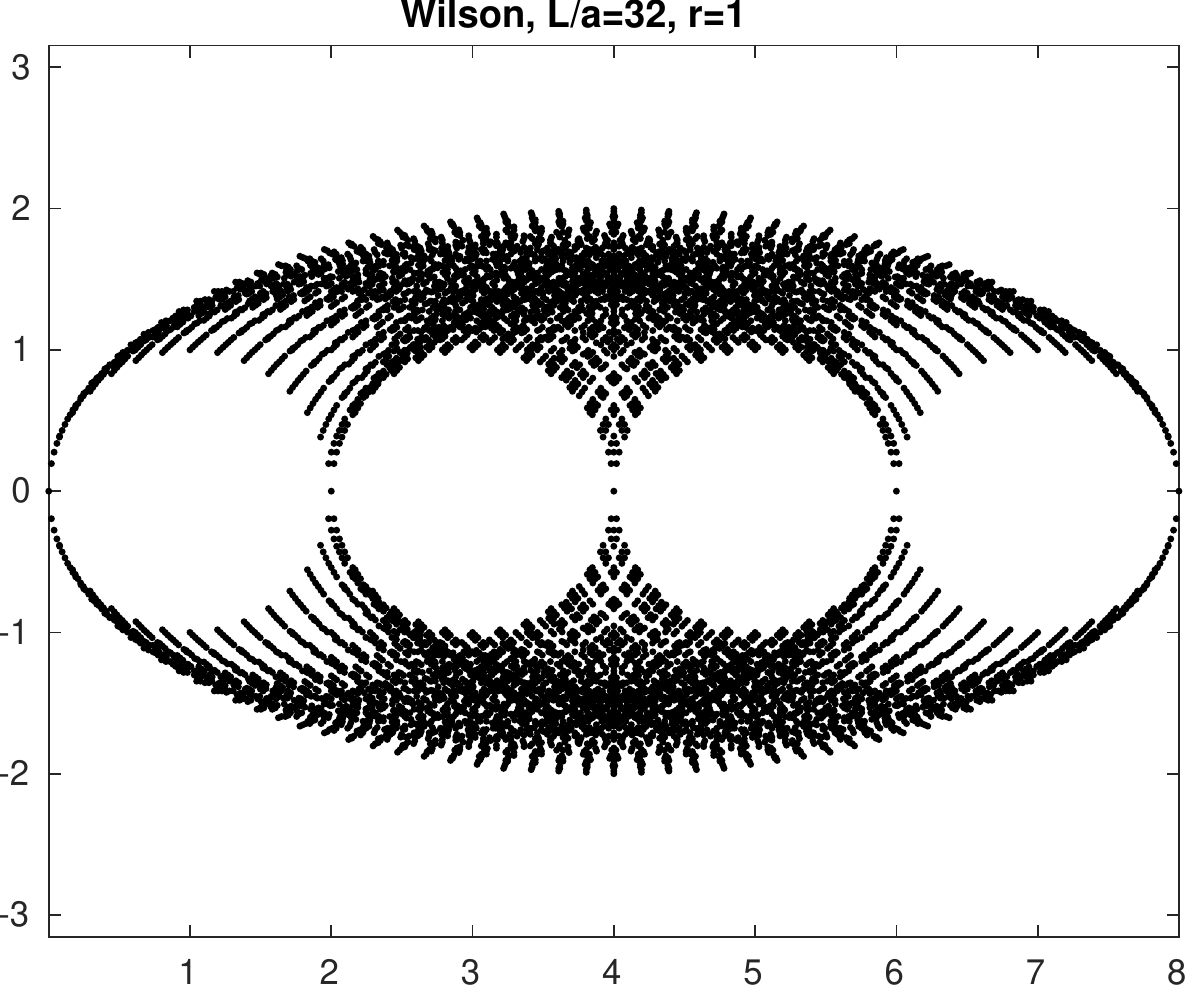}
\caption{\label{fig:eig_wils}
Free-field Wilson eigenvalue spectrum at $r=1$ in the complex plane.}
\end{figure}

From expression (\ref{momrep_nai}) or (\ref{momrep_wils}) one finds the eigenvalues of the free-field operators.
Since each $\ga_\mu$ ($\mu=1,..,4$) has eigenvalues $\pm1$, it follows that the upper end of the massless naive eigenvalue
spectrum is realized for $ap_\mu$ along the hyperdiagonal $(1,1,1,1)$, or flipped versions thereof, so
\beq
|\mr{Im}(\la_\mr{nai})|\leq2
\eeq
and similarly it follows that the Wilson eigenvalue spectrum is contained in the rectangle
\beq
0\leq\mr{Re}(\la_\mr{W})\leq8r
\;,\quad
|\mr{Im}(\la_\mr{W})|\leq2
\;.
\eeq
The complex eigenvalue spectrum of the Wilson operator at $am=0$ is shown in Fig.~\ref{fig:eig_wils}.
The symmetry about the real axis reflects the pairing property imposed by the $\gaf$-hermiticity.
As is well known, the five branches in the Wilson eigenvalue spectrum correspond to species with multiplicities $1,4,6,4,1$, and chiralities $+,-,+,-,+$, respectively~\cite{BOOK_MM}.
In total $8$ species thus have correct chirality, and $8$ have opposing chirality.
The free-field eigenvalue spectrum of the naive (or staggered) operator follows by horizontally projecting the $\la_\mr{W}$ onto the imaginary axis
(and reducing degeneracies by a factor $4$ in the staggered case).
Under this operation the separation between right-chirality and opposite-chirality species gets lost, and this feature will carry on to minimally doubled actions.


\section{Karsten-Wilczek fermions\label{sec:3}}


The Karsten-Wilczek proposal is to restrict the Wilson term in (\ref{def_wils}) to the spatial components
\beq
D_\mr{KW}(x,y)=\sum_\mu \ga_\mu \nab_\mu(x,y)
-\ri\frac{ar}{2}\ga_4 \sum_{i=1}^{3} \lap_i(x,y)
+m\de_{x,y}
\label{def_KW}
\eeq
with an extra factor $\ri\ga_4$ to make it anti-hermitean and anti-commuting with $\ga_5$ \cite{Karsten:1981gd,Wilczek:1987kw}.
As a result the Karsten-Wilczek (KW) operator is $\gaf$-hermitean, i.e.\ $\gaf D_\mr{KW} \gaf=D_\mr{KW}\dag$.
An issue discussed in the literature is that $\ga_4 \sum_i \lap_i(x,y)$ mixes (on interacting backgrounds) with $\ga_4$
\cite{Bedaque:2008xs,Cichy:2008gk,Capitani:2009yn,Capitani:2010nn,Creutz:2010qm,Kimura:2011ik,Weber:2013tfa,Weber:2016jug,Weber:2017eds}.
In the free-field limit the KW operator assumes a diagonal form in momentum space,
\bea
D_\mr{KW}(p)&=&\ri \sum_\mu \ga_\mu \frac{1}{a}\sin(ap_\mu)
+\ri ar\ga_4 \sum_{i=1}^{3} \{1-\cos(ap_i)\}+m
\nonumber
\\
&=&\ri \sum_\mu \ga_\mu \bar{p}_\mu
+\ri \frac{ar}{2}\ga_4 \sum_{i=1}^{3} \hat{p}_i^2
+m
\label{momrep_KW}
\eea
which again highlights the anti-hermitean nature of either term.
This formulation was shown to have $2$ species for $r=1$ in the original works \cite{Karsten:1981gd,Wilczek:1987kw},
but how this number decreases from $16$, at $r=0$, to the minimally doubled value has, to the best of our knowledge, not been investigated.
We find that the number of species is reduced in three steps (in $d=4$ dimensions).
At $r=1/6,1/4,1/2$ the number of species is reduced by $2,6,6$, respectively, so the species chain is $16\to14\to8\to2$.
See App.~\ref{app:C} for details, e.g.\ the situation with $d\neq4$.

\begin{figure}[p]
\centering
\includegraphics[width=0.78\textwidth]{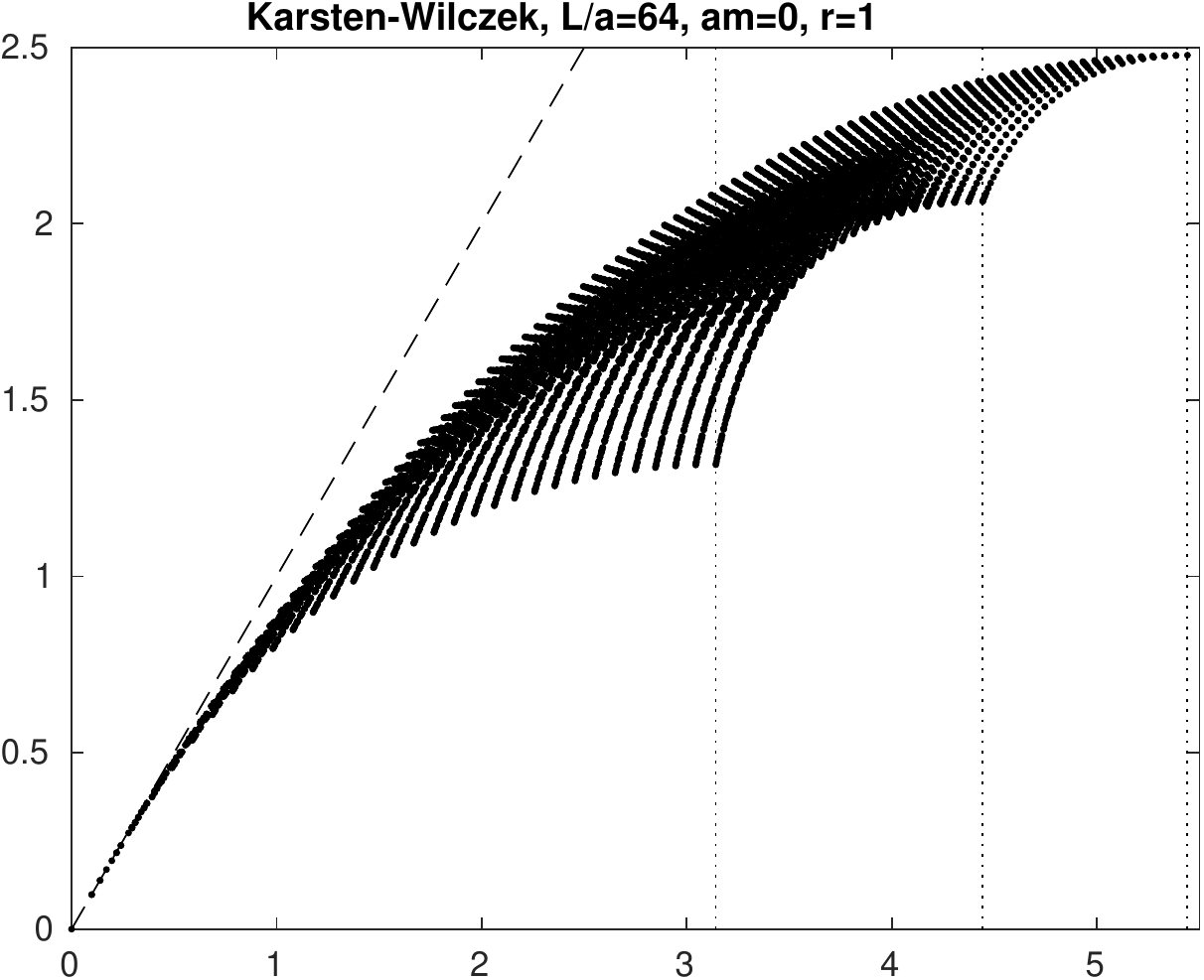}\\[1mm]
\includegraphics[width=0.78\textwidth]{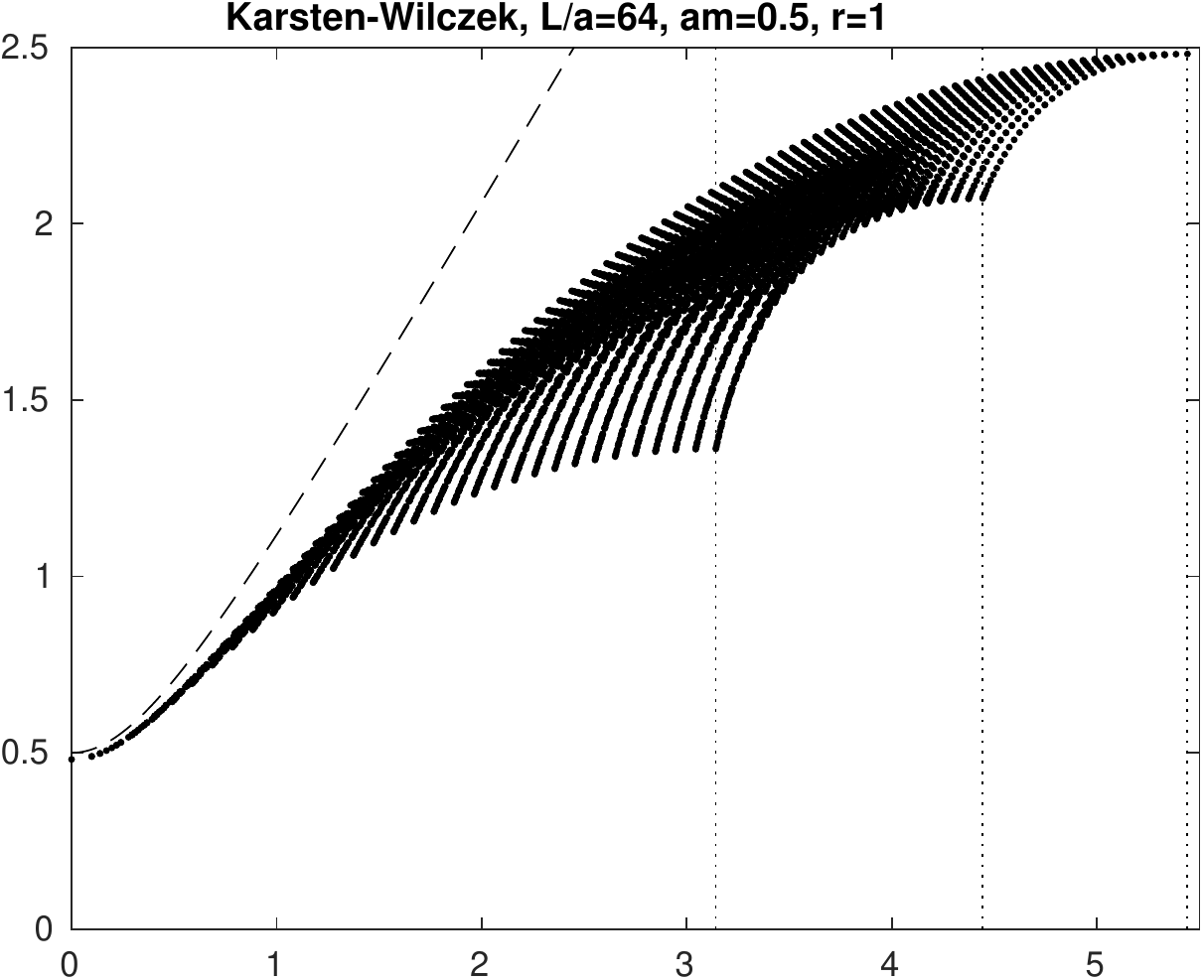}%
\caption{\label{fig:disp_kawi}
Same as Fig.~\ref{fig:disp_naiv}, but for the Karsten-Wilczek (KW) operator at $r=1$.}
\end{figure}

Starting from eqn.~(\ref{momrep_KW}) one can work out the free-field dispersion relation of KW fermions, see App.~\ref{app:A} for details.
For a given momentum configuration $a\vec{p}$ the Euclidean energy $aE$ is, in general, complex valued,
and its real part is plotted in Fig.~\ref{fig:disp_kawi} for $r=1$.
Again, two values of the quark mass are used, $am=0$ and $am=0.5$.
In either case the KW dispersion relation follows the continuum curve faithfully, out to momentum values $a|\vec{p}|\simeq1$.
In particular at $a|\vec{p}|=0$ it features much better than the Wilson action, reminiscent of the naive action.
This is not a coincidence, since the rest energy has the same functional dependence on $am$ as the naive action, see App.~\ref{app:B} for details.
In other words, cut-off effects on this quantity start at $O((am)^2)$ only, unlike the $O(am)$ signature of Wilson fermions.

\begin{figure}[tb]
\centering
\includegraphics[width=0.78\textwidth]{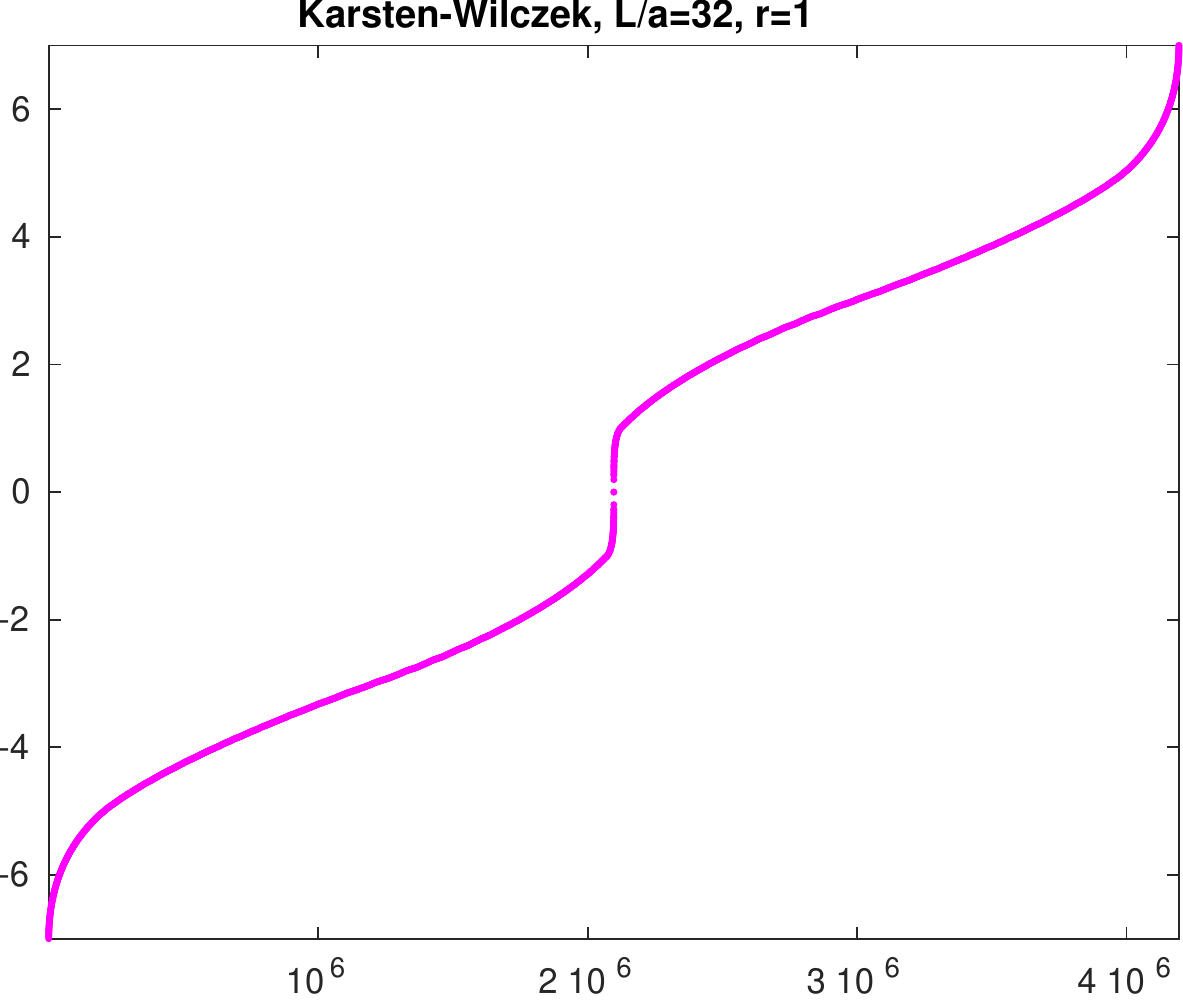}
\caption{\label{fig:eig_KW}
Free-field eigenvalue spectrum of the KW operator at $r=1$.
The imaginary part is plotted against the index.}
\end{figure}

From eqn.~(\ref{momrep_KW}) one finds the eigenvalues of the free-field KW operator.
The result for $r=1$ and $am=0$ is shown in Fig.~\ref{fig:eig_KW}.
On a $32^4$ lattice one finds $4\cdot32^4$ purely imaginary and $\gaf$-paired eigenvalues, as expected.
Plotting the eigenvalues against the index means that the inverse slope encodes for the density of the $\la_\mr{KW}/\ri$ on the imaginary axis.

\begin{figure}[p]
\centering
\includegraphics[width=0.85\textwidth]{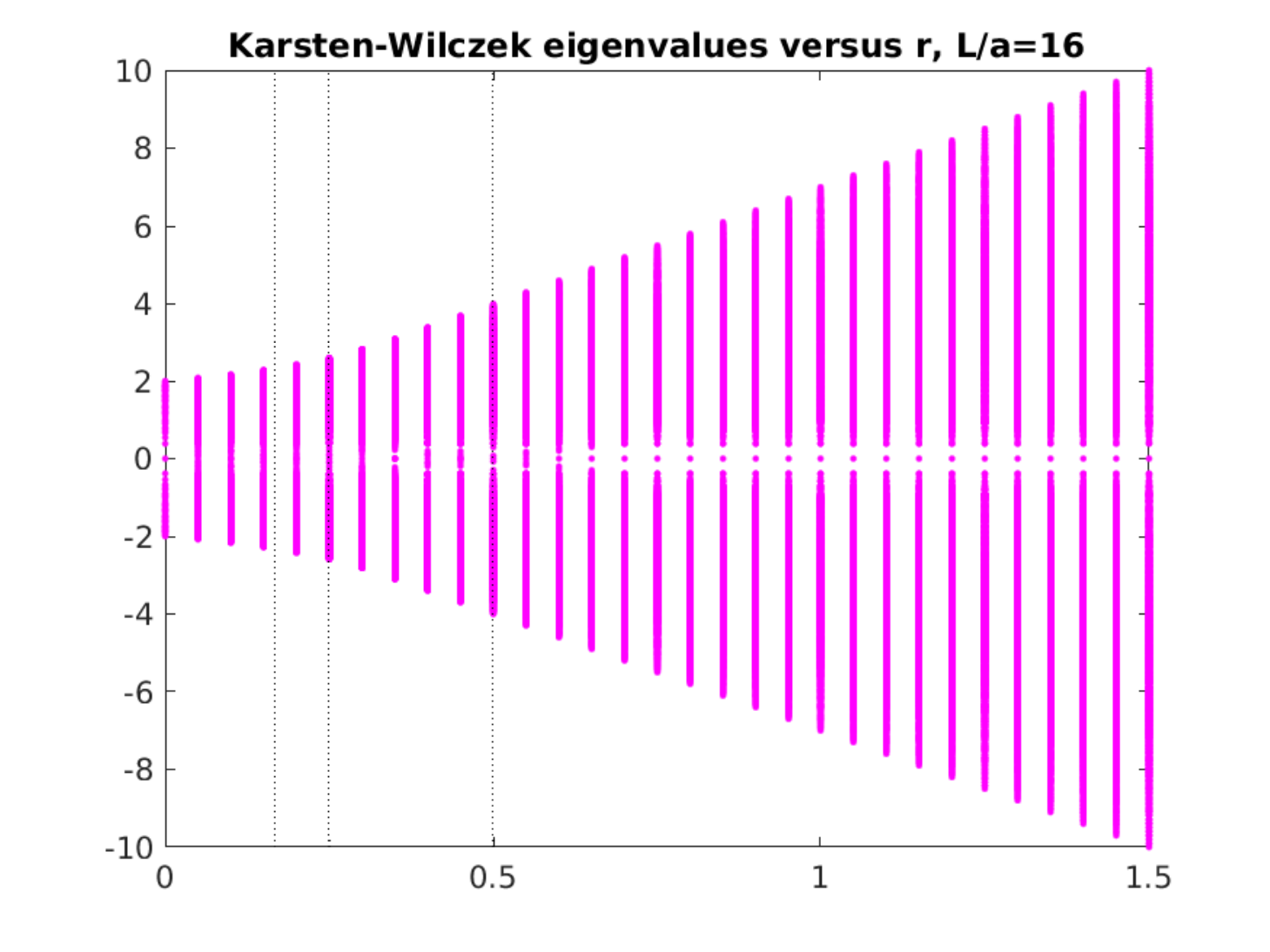}\\
\includegraphics[width=0.85\textwidth]{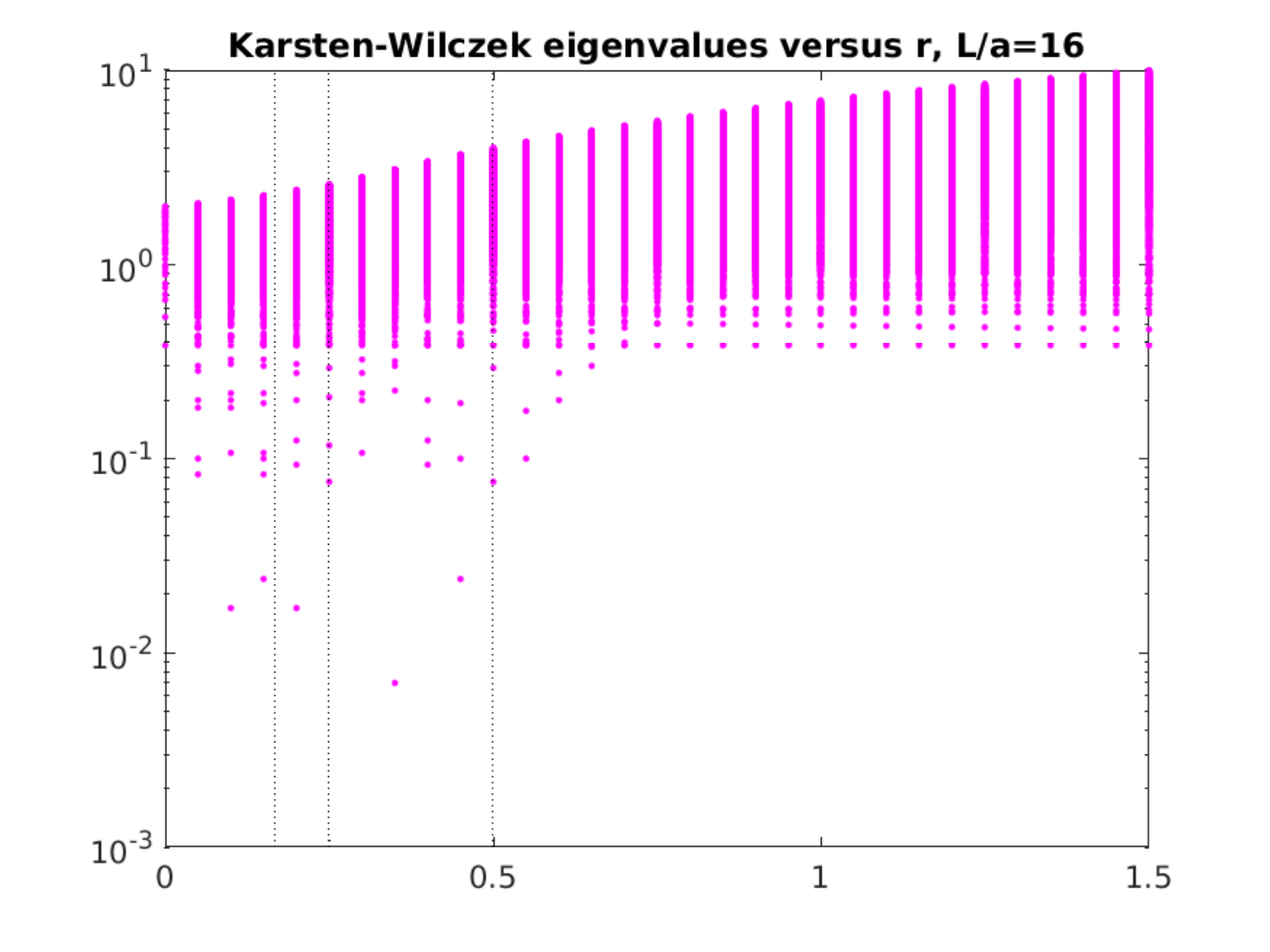}
\caption{\label{fig:linlog_KW}
Imaginary part of the free eigenvalues of the KW operator in linear and logarithmic
representation (for the upper half-spectrum) versus the lifting parameter $r$.
The vertical lines at $r=1/6,1/4,1/2$ mark the transitions to $14$, $8$, and $2$ species, respectively.}
\end{figure}

It is instructive to repeat this for a series of $r$ values; the result is shown in Fig.~\ref{fig:linlog_KW},
with vertical lines marking the abscissa values $r=1/6,1/4,1/2$ where the number of species changes.
The spectral range is seen to increase with growing $r$.
In addition, the low-energy end of the eigenvalue spectrum seems unstable for small $r$, but stable in a broad range around $r=1$.

\begin{figure}[tb]
\centering
\includegraphics[width=0.78\textwidth]{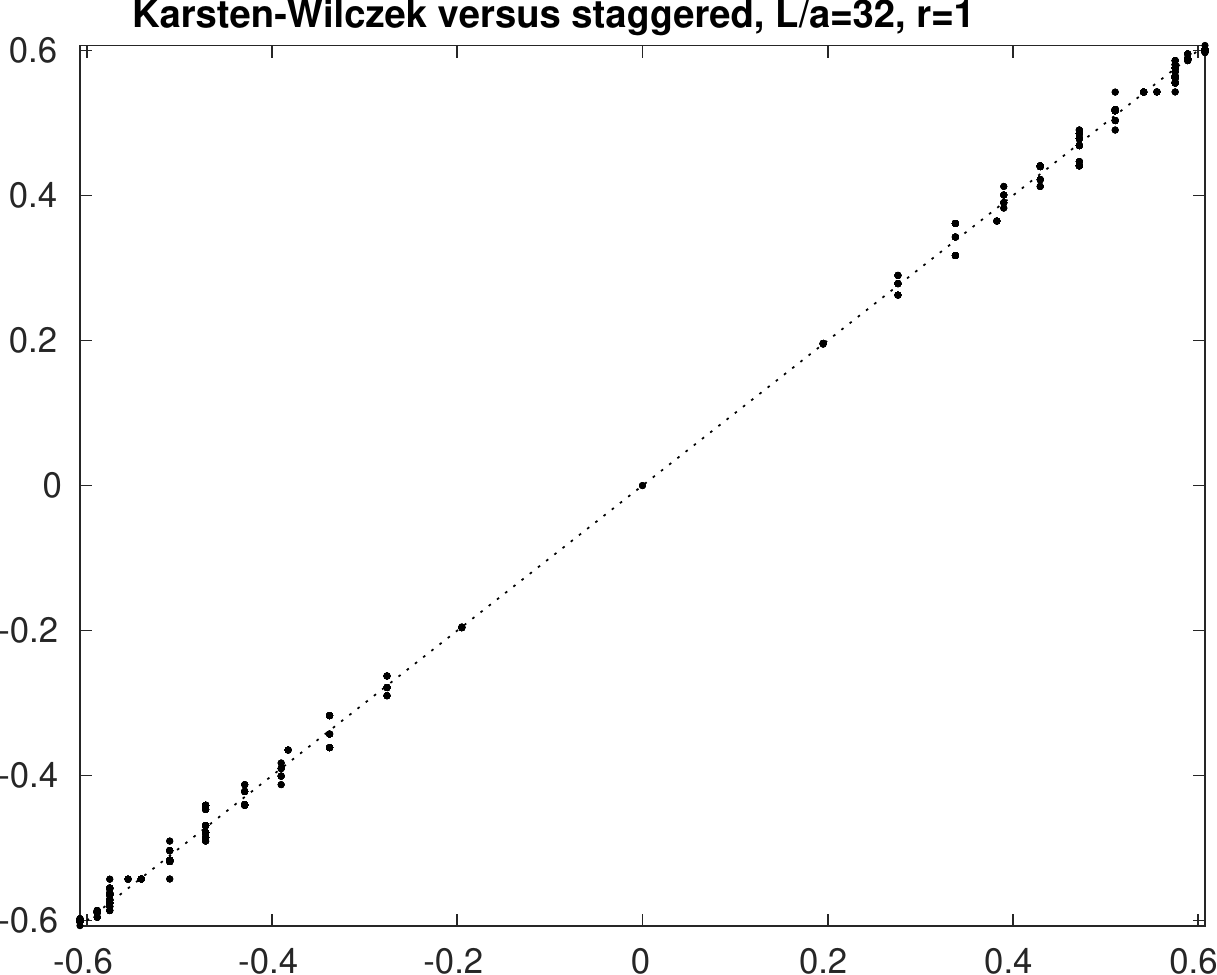}%
\caption{\label{fig:KW_versus_stag}
Sorted free-field eigenvalues (with a $2$-fold degeneracy removed) of the KW operator at $r=1$
plotted versus sorted staggered eigenvalues (with a $4$-fold degeneracy removed).
In both cases the imaginary part $\mr{Im}(\la)=\la/\ri$ at $am=0$ is used, and plenty of degeneracies remain.
The dotted line shows the identity for comparison.}
\end{figure}

Given Fig.~\ref{fig:linlog_KW}, one may wonder about the existence of an analytic function which describes the upper end as a function of $r$.
In App.~\ref{app:E} we derive the free-field spectral bound
\beq
|\mr{Im}(\la_\mr{KW})|\leq
\left\{
\begin{array}{ll}
\sqrt{(4+6r)/(1-3r^2)} & (r\leq1/3)\\
1+6r & (r\geq1/3)
\end{array}
\right.
\label{specbound_KW}
\eeq
in $d=4$ dimensions.
Hence at $r=1$ the imaginary parts $\la_\mr{KW}/\ri$ cover the range $[-7,7]$, to be compared to $[-2,2]$ for naive and staggered fermions.
On the other hand, the smallest non-zero KW eigenvalue is found in essentially the same place~%
\footnote{In Fig.~\ref{fig:KW_versus_stag} one finds the small (in absolute magnitude) KW eigenvalues
by projecting the black dots onto the $y$-axis, and the staggered counterparts by projecting them onto the $x$-axis.
Hence, $\min(|\la_\mr{KW}|)\simeq0.2$, and $\min(|\la_\mr{stag}|)\simeq0.2$ in a $32^4$ box, if we disregard the non-topological zero-modes.
In large boxes the spectral gap decreases as $1/L$, so we anticipate $\min(|\la|)\simeq0.1$ in a $64^4$ box for both KW and staggered fermions.}
as the smallest staggered eigenvalue, see Fig.~\ref{fig:KW_versus_stag}.
This amounts to an enhancement of the condition number of $D\dag D$, compared to the staggered formulation at the same $am$, by a factor up to $3.5^2=12.25$ (in the chiral limit).


\section{Borici-Creutz fermions\label{sec:4}}


The basis for Borici-Creutz fermions in $d$ space-time dimensions is the idempotent operator
\beq
\Gamma=\frac{1}{\sqrt{d}}\sum_\mu\ga_\mu
\quad \mbox{with} \quad
\Gamma^2
=\frac{1}{2d}\{\sum_\al\ga_\al,\sum_\be\ga_\be\}
=\frac{2d}{2d}=1
\label{def_big}
\eeq
and $\{\Gamma,\ga_\mu\}=\frac{2}{\sqrt{d}}$ and $\{\Gamma,\ga_5\}=0$.
This suggests to define the set of dual gamma-matrices
\beq
\ga_\mu'=\Gamma \ga_\mu \Gamma
=\Big( \frac{2}{\sqrt{d}} - \ga_\mu \Gamma \Big)\Gamma
=\frac{2}{\sqrt{d}}\Gamma-\ga_\mu
\label{def_dualgamma}
\eeq
which are hermitean and satisfy the Dirac-Clifford algebra, since (\ref{def_dualgamma})
implies $\{\ga_\mu',\ga_\nu'\}=2\de_{\mu\nu}$ and $\{\Gamma,\ga_\mu'\}=\frac{2}{\sqrt{d}}$.
Furthermore, one finds
\bea
\{\ga_\mu,\ga_\nu'\}&=&\ga_\mu(\frac{2}{\sqrt{d}}\Gamma-\ga_\nu)+(\frac{2}{\sqrt{d}}\Gamma-\ga_\nu)\ga_\mu
=\frac{2}{\sqrt{d}}\{\ga_\mu,\Gamma\}-\{\ga_\mu,\ga_\nu\}=\frac{4}{d}-2\de_{\mu\nu}
\label{gagaprime}
\\
\{\ga_\mu',\ga_\nu\}&=&(\frac{2}{\sqrt{d}}\Gamma-\ga_\mu)\ga_\nu+\ga_\nu(\frac{2}{\sqrt{d}}\Gamma-\ga_\mu)
=\frac{2}{\sqrt{d}}\{\Gamma,\ga_\nu\}-\{\ga_\mu,\ga_\nu\}=\frac{4}{d}-2\de_{\mu\nu}
\label{gaprimega}
\;.
\eea

The Borici-Creutz (BC) proposal is to dress the Wilson term in (\ref{def_wils}) with $\ri$ times (\ref{def_dualgamma}), i.e.\
\beq
D_\mr{BC}(x,y)=\sum_\mu \ga_\mu \nab_\mu(x,y)
-\ri\frac{ar}{2}\sum_\mu \ga_\mu' \lap_\mu(x,y)
+m\de_{x,y}
\label{def_BC}
\eeq
where our second term differs in sign from the original proposal \cite{Borici:2007kz}.
Note that the second term is anti-hermitean and anti-commutes with $\ga_5$, since
\beq
\ga_\mu'\ga_5=\Gamma\ga_\mu\Gamma\ga_5=-\Gamma\ga_\mu\ga_5\Gamma=\Gamma\ga_5\ga_\mu\Gamma=-\ga_5\Gamma\ga_\mu\Gamma=-\ga_5\ga_\mu'
\eeq
and this renders the BC operator $\gaf$-hermitean, i.e.\ $\gaf D_\mr{BC} \gaf=D_\mr{BC}\dag$.
An issue discussed in the literature is whether $\sum_\mu \ga_\mu'\lap_\mu$ mixes (on interacting backgrounds) with $\Gamma$
\cite{Bedaque:2008xs,Cichy:2008gk,Capitani:2009yn,Capitani:2010nn,Creutz:2010qm,Kimura:2011ik,Weber:2013tfa,Weber:2016jug,Weber:2017eds}.
In the free-field limit the BC operator assumes a diagonal form in momentum space,
\bea
D_\mr{BC}(p)&=&\ri \sum_\mu \ga_\mu \bar{p}_\mu
+\ri ar \sum_\mu \ga_\mu' \{1-\cos(ap_\mu)\}
+m
\nonumber
\\
&=&\ri \sum_\mu \ga_\mu \bar{p}_\mu
+\ri \frac{ar}{2} \sum_\mu \ga_\mu' \hat{p}_\mu^2
+m
\label{momrep_BC}
\eea
in which the bracket $\{1-\cos(ap_\mu)\}$ may be split and the sum over $\ga_\mu'$ performed by means of
\beq
\sum_\mu \ga_\mu'=2\sqrt{d}\Gamma-\sum_\mu\ga_\mu
=2\sqrt{d}\Gamma-\sqrt{d}\Gamma=\sqrt{d}\Gamma
\;.
\eeq
Furthermore, the free-field form (\ref{momrep_BC}) highlights the invariance under any permutation of the $d$ axes.
In App.~\ref{app:C} we discuss, for $d=4$, how the number of species is reduced by $6$ at $r=1/\sqrt{3}$,
and by $8$ at $r=1/\sqrt{2}$; so the species chain is $16\to10\to2$.
Of course, the number of species is unchanged by a sign flip of $r$.

In two (Euclidean) space-time dimensions it is customary to use $\ga_1=\si_1,\ga_2=\si_2$.
Similarly, the chirality operator is defined as $\ga_5=-\ri\ga_1\ga_2=-\ri\si_1\si_2=\si_3$.
Upon using the simplifications
\bea
\Gamma&=&
\frac{1}{\sqrt{2}}
\Big(\si_1+\si_2\Big)
=
\frac{1}{\sqrt{2}}\,
\bigg(\!
\begin{array}{cc}
0&1-\ri \\ 1+\ri&0
\end{array}
\!\bigg)
=
\bigg(\!
\begin{array}{cc}
0&e^{-\ri\pi/4} \\ e^{+\ri\pi/4}&0
\end{array}
\!\bigg)
\\
\si_1'&=&\Gamma\si_1\Gamma=\frac{1}{2}(\si_1+\si_2)\si_1(\si_1+\si_2)
=\frac{1}{2}(\si_1+\si_2+\si_2+\si_2\si_1\si_2)
=\si_2
\\
\si_2'&=&\Gamma\si_2\Gamma=\frac{1}{2}(\si_1+\si_2)\si_2(\si_1+\si_2)
=\frac{1}{2}(\si_1\si_2\si_1+\si_1+\si_1+\si_2)
=\si_1
\eea
the operators (\ref{def_KW}) and (\ref{def_BC}) are seen to take the simple form
\bea
D_\mr{KW}(x,y)&=&\sum_\mu \si_\mu \nab_\mu(x,y)
-\ri\frac{ar}{2}\si_2 \lap_1(x,y)
+m\de_{x,y}
\label{twodim_KW}
\\
D_\mr{BC}(x,y)&=&\sum_\mu \si_\mu \nab_\mu(x,y)
-\ri\frac{ar}{2} \si_2 \lap_1(x,y)
-\ri\frac{ar}{2} \si_1 \lap_2(x,y)
+m\de_{x,y}
\label{twodim_BC}
\eea
which shows that the BC operator is not a symmetrized form of the KW operator; it has an extra term.
This explains why we deviate, in the sign of the mass-dimension $5$ term in eqns.~(\ref{def_BC}, \ref{momrep_BC}), from the literature.
With our convention the joint terms in eqns.~(\ref{twodim_KW}, \ref{twodim_BC}) have like sign.

\begin{figure}[p]
\centering
\includegraphics[width=0.78\textwidth]{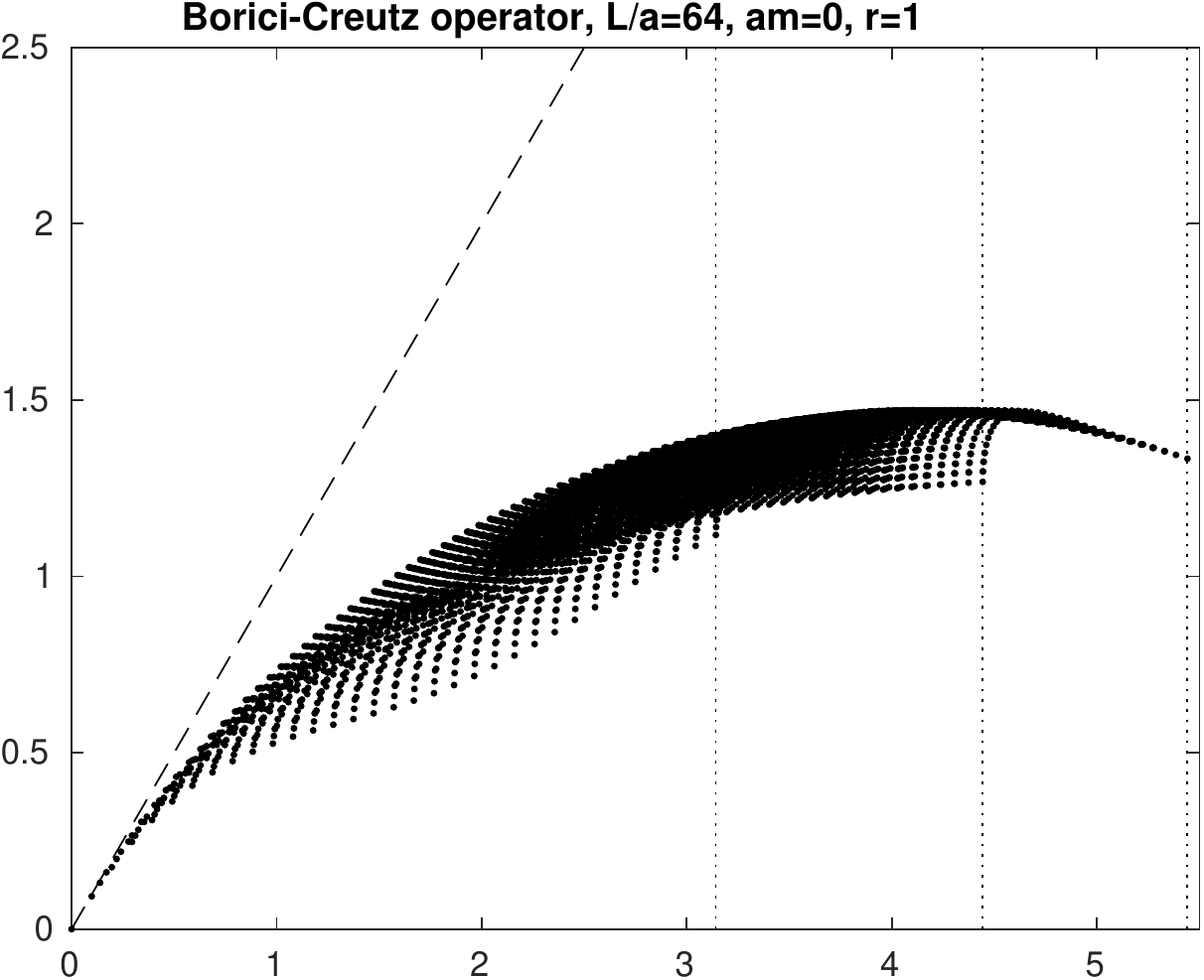}\\[1mm]
\includegraphics[width=0.78\textwidth]{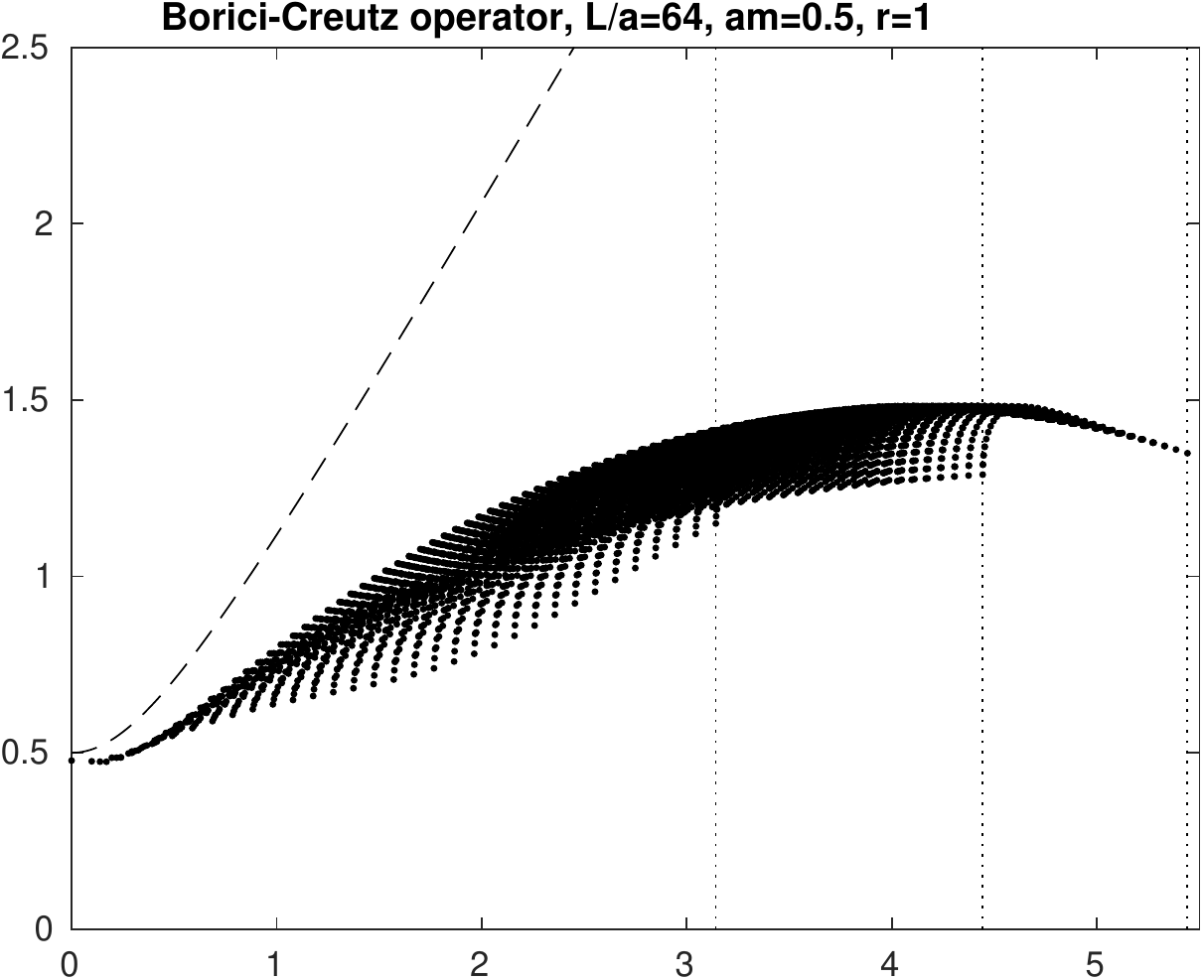}%
\caption{\label{fig:disp_bocr}
Same as Fig.~\ref{fig:disp_naiv}, but for the Borici-Creutz (BC) operator at $r=1$.}
\end{figure}

Starting from eqn.~(\ref{momrep_BC}) one can work out the free-field dispersion relation of BC fermions, see App.~\ref{app:A} for details.
For a given momentum configuration $a\vec{p}$ the Euclidean energy $aE$ is, in general, complex valued,
and its real part is plotted in Fig.~\ref{fig:disp_bocr} for $r=1$.
Again, two values of the quark mass are used, $am=0$ and $am=0.5$.
In either case the BC dispersion relation follows the continuum curve reasonably well,
out to momentum values $a|\vec{p}|\simeq\frac{1}{2}$, a range slightly narrower than what was found in the KW case.
Specifically at $a|\vec{p}|=0$ it features much better than the Wilson action, though a little worse than the KW action.
In App.~\ref{app:B} the rest energy of a heavy BC fermion is found to have a contribution $\propto\frac{19}{96}(am)^2$,
nearly as good as $\propto\frac{1}{6}(am)^2$ of the KW action (both values are for $r=1$ in $d=4$ dimensions).
Unlike the KW energy, the BC energy has an imaginary contribution $\propto\frac{1}{4}am$, which is not a desirable property.
What is particularly disconcerting in the free-field dispersion relation of a BC fermion at heavy quark mass is that the global minimum is not necessarily at $a|\vec{p}|=0$.
For $am=0.5$ the effect happens to be numerically small, but a spontaneous breaking of translation invariance
(even if confined to Weiss-type sub-domains of the Brillouin zone) would cause headache.

\begin{figure}[tb]
\centering
\includegraphics[width=0.78\textwidth]{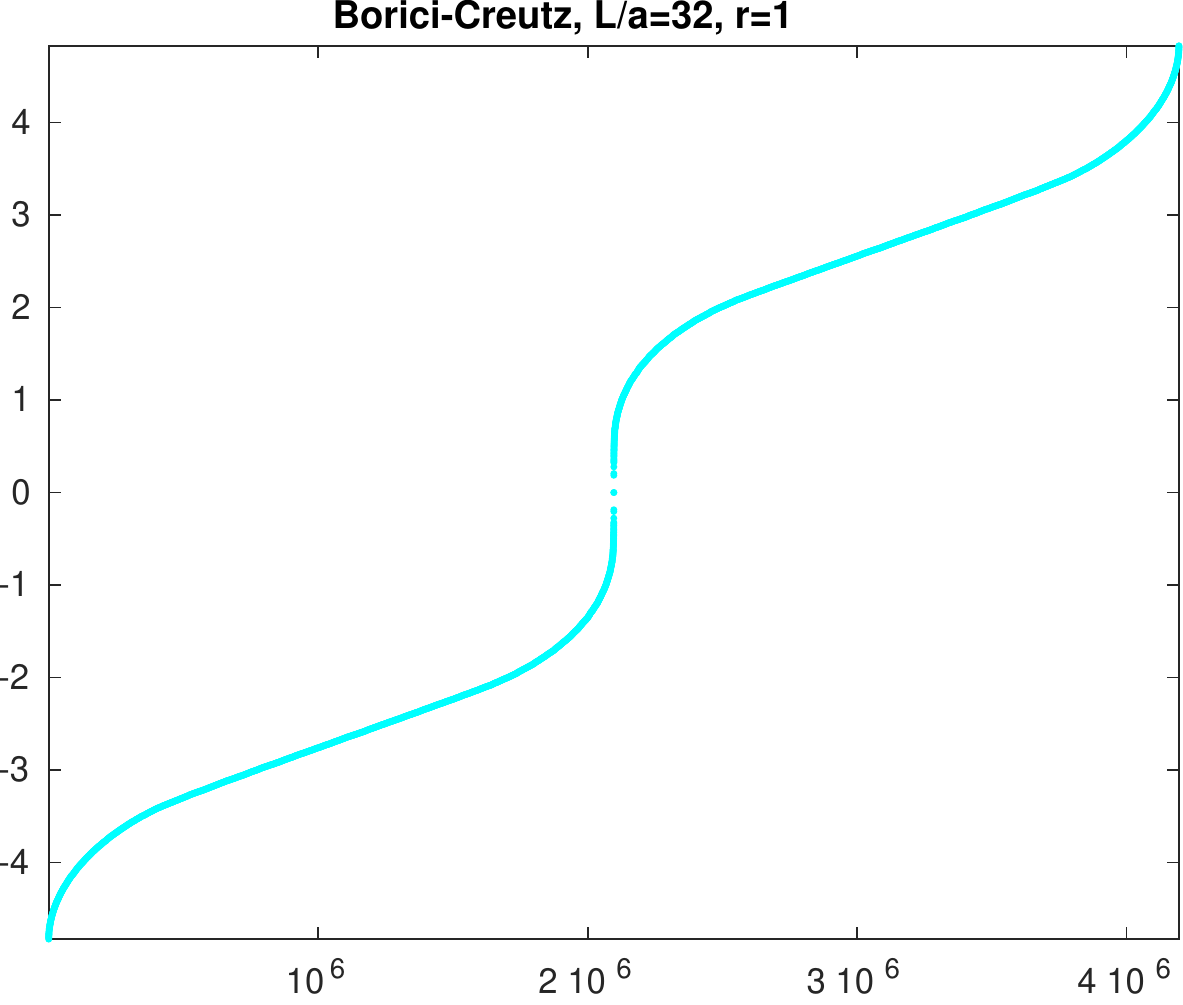}%
\caption{\label{fig:eig_BC}
Free-field eigenvalue spectrum of the BC operator at $r=1$.
The imaginary part is plotted against the index.}
\end{figure}

From eqn.~(\ref{momrep_BC}) one finds the eigenvalues of the free-field BC operator.
The result for $r=1$ and $am=0$ is shown in Fig.~\ref{fig:eig_BC}.
Similar to the KW case, eigenvalues come in $\gaf$-pairs and are purely imaginary.
The only difference to Fig.~\ref{fig:eig_KW} is that the BC range is slightly narrower.

\begin{figure}[p]
\centering
\includegraphics[width=0.85\textwidth]{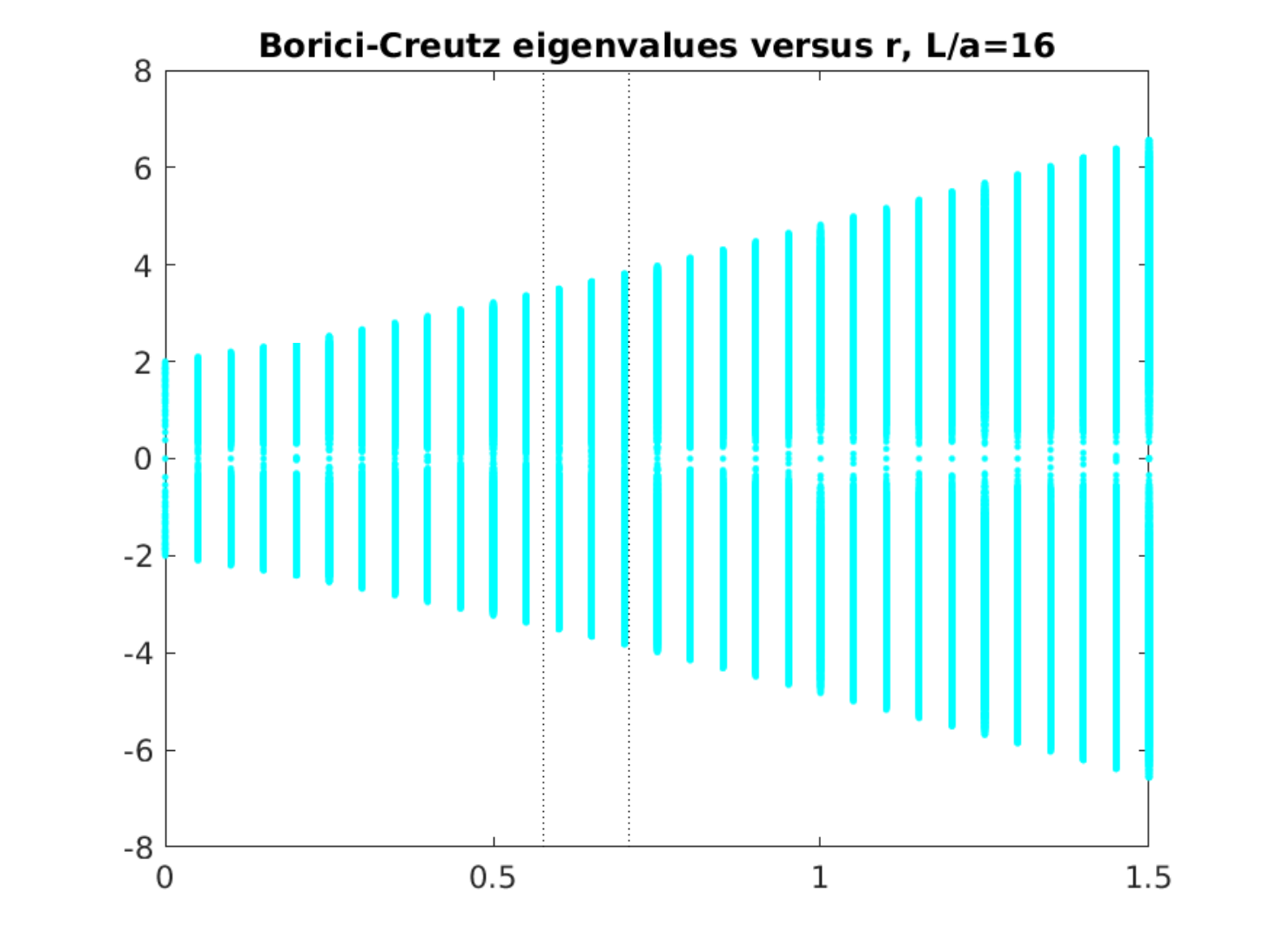}\\
\includegraphics[width=0.85\textwidth]{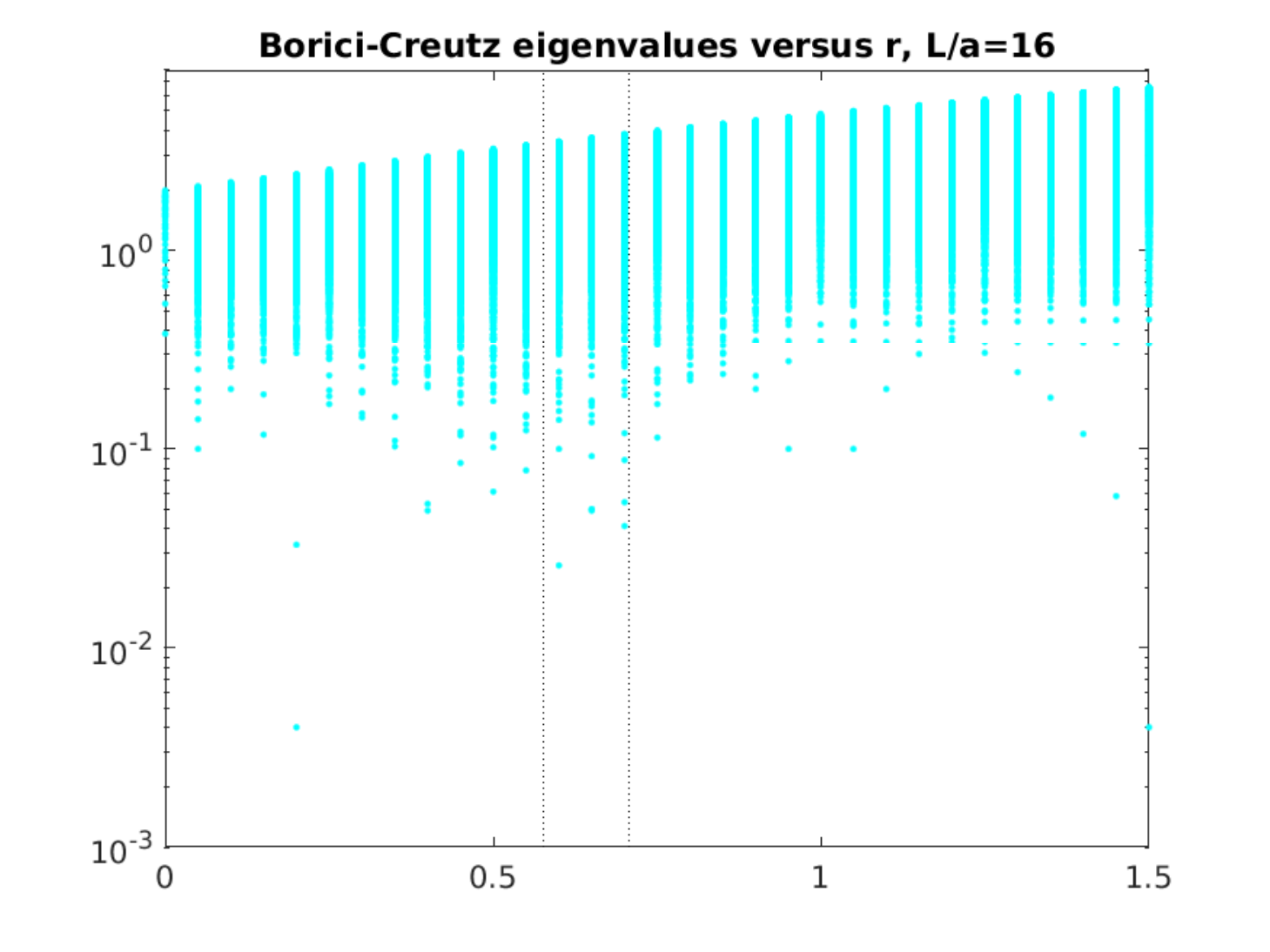}%
\caption{\label{fig:linlog_BC}
Imaginary part of the free eigenvalues of the BC operator in linear and logarithmic
representation (for the upper half-spectrum) versus the lifting parameter $r$.
The vertical lines at $r=1/\sqrt{3},1/\sqrt{2}$ mark the transitions to $10$, and $2$ species, respectively.}
\end{figure}

It is instructive to repeat this for a series of $r$ values; the result is shown in Fig.~\ref{fig:linlog_BC},
with vertical lines marking the abscissa values $r=1/\sqrt{3},1/\sqrt{2}$ where the number of species changes.
The spectral range is seen to increase with growing $r$.
In addition, the low-energy end of the eigenvalue spectrum seems far less stable than in the KW case,
the canonical choice $r=1$ seems to represent a small island of stability.

\begin{figure}[tb]
\centering
\includegraphics[width=0.78\textwidth]{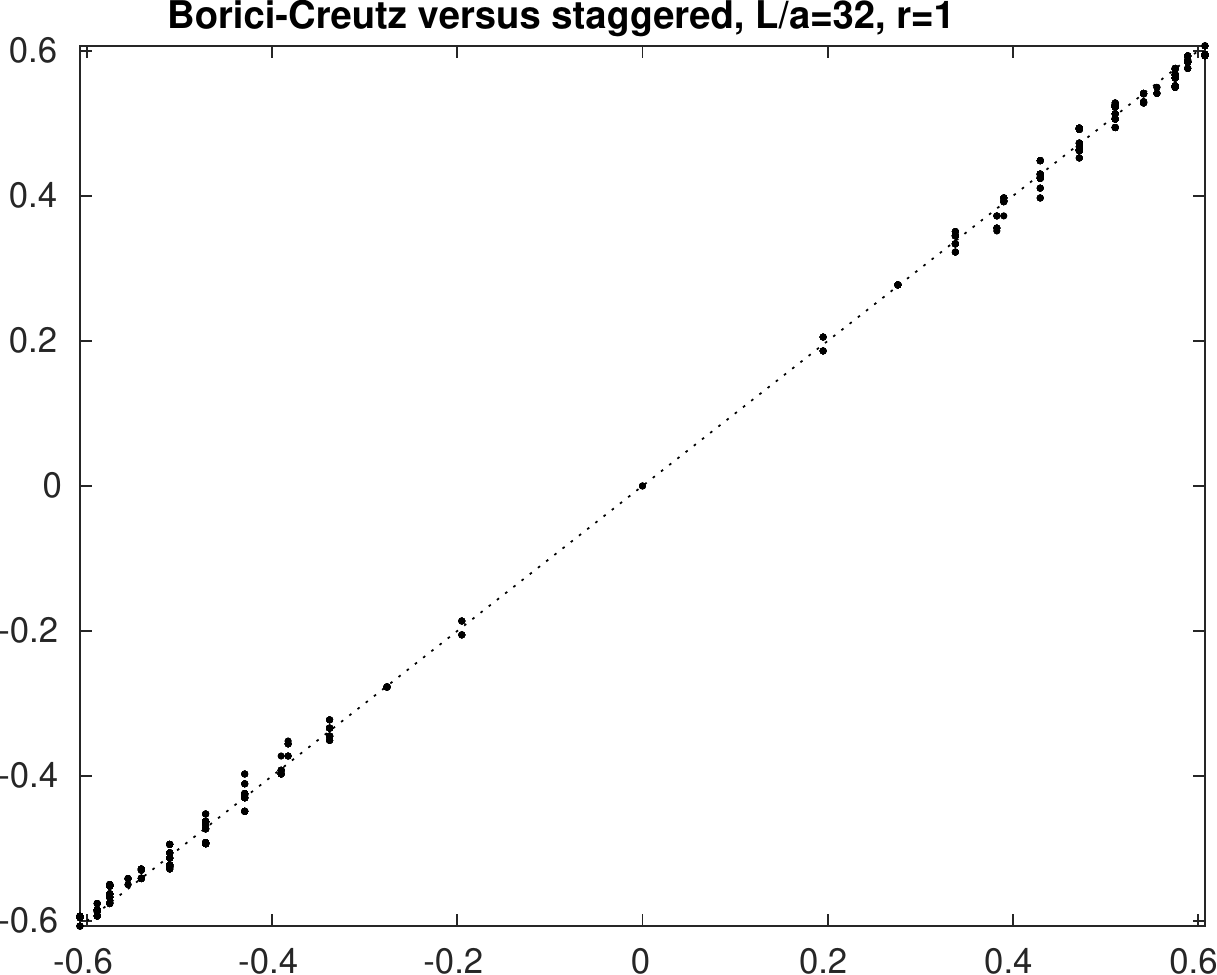}%
\caption{\label{fig:BC_versus_stag}
Sorted free-field eigenvalues (with a $2$-fold degeneracy removed) of the BC operator at $r=1$
plotted versus sorted staggered eigenvalues (with a $4$-fold degeneracy removed).
In both cases the imaginary part $\mr{Im}(\la)=\la/\ri$ at $am=0$ is used, and plenty of degeneracies remain.
The dotted line shows the identity for comparison.}
\end{figure}

Given Fig.~\ref{fig:linlog_BC}, one may wonder about the existence of an analytic function which describes the upper end as a function of $r$.
In App.~\ref{app:E} we derive the free-field spectral bound
\beq
|\mr{Im}(\la_\mr{BC})| \leq 2(r+\sqrt{1+r^2})
\label{specbound_BC}
\eeq
in $d=4$ dimensions.
Hence at $r=1$ the imaginary parts $\la_\mr{BC}/\ri$ cover the symmetric range $[-4.8284,2+\sqrt{8}]$, to be compared to $[-2,2]$ for naive and staggered fermions.
For $r=1$ the smallest non-zero BC eigenvalue is found in essentially the same place~%
\footnote{In Fig.~\ref{fig:BC_versus_stag} one finds the small (in absolute magnitude) BC eigenvalues
by projecting the black dots onto the $y$-axis, and the staggered counterparts by projecting them onto the $x$-axis.
Hence, $\min(|\la_\mr{BC}|)\simeq0.2$, and $\min(|\la_\mr{stag}|)\simeq0.2$ in a $32^4$ box, if we disregard the non-topological zero-modes.
In large boxes the spectral gap decreases as $1/L$, so we anticipate $\min(|\la|)\simeq0.1$ in a $64^4$ box for both BC and staggered fermions.}
as the smallest staggered eigenvalue, see Fig.~\ref{fig:BC_versus_stag}.
This amounts to an enhancement of the condition number of $D\dag D$, compared to the staggered formulation at the same $am$, by a factor up to $(1+\sqrt{2})^2=5.8284$ (in the chiral limit).
Hence, the slowdown (relative to the staggered formulation) implied by the larger condition number is not as pronounced as in the KW case, but still significant.


\section{Summary\label{sec:5}}


In this paper we tried to fill some of the most obvious gaps in the knowledge about the two most popular minimally doubled fermion actions,
namely the formulations due to Karsten-Wilczek (KW) and Borici-Creutz (BC), respectively.
The gaps concern the eigenvalue spectra and the dispersion relations (including the leading cut-off effects on the heavy fermion mass) in the free-field limit.
We studied these issues as a function of the lifting parameter $r$, in order to see how the number of species gets reduced from $16$ (at $r=0$) to $2$ (at $r=1$).
Our investigation was limited to KW and BC fermions, but there are two more approaches, ``twisted ordering'' and ``flavored chemical potential term''
where a Wilson-like parameter $r$ can be introduced to study how the number of species gets reduced \cite{Creutz:2010cz,Misumi:2012uu,Misumi:2012ky}.

Regarding the eigenvalue spectra we find an extension, relative to the staggered/naive one,
by a factor $3.5$ for KW fermions, or $1+\sqrt{2}\simeq2.4142$ for BC fermions (both at $r=1$ and vanishing quark mass).
This leads to an enhancement of the condition number of $D\dag D$ (as relevant for generating dynamical ensembles)
by a factor up to $12.25$, or $5.8284$, respectively, compared to the staggered/naive case.
This, together with the matrix size being a factor $4$ larger than for staggered fermions,
limits our optimism regarding the computational efficiency of these two formulations.
At finite quark mass the spectral bounds (\ref{specbound_KW}, \ref{specbound_BC}) generalize to $\max(|\la|^2)=\mr{Im}(\la)^2+(am)^2$,
with $\mr{Im}(\la)$ given by (\ref{specbound_KW}, \ref{specbound_BC}), respectively.

In addition, we studied the dispersion relations.
On the one hand, we find that the KW operator features very well in this respect.
It follows the continuum dispersion relation more closely than the Wilson operator.
In particular at $a\vec{p}=\vec{0}$ the cut-off effects on the heavy quark mass start at $O((am)^2)$,
just like the naive/staggered action, not at $O(am)$ like the Wilson operator.
On the other hand, the dispersion relation of the BC operator in $d=4$ dimensions shows some more problematic features,
including a funny behavior at small $a|\vec{p}|$ and an imaginary part of the heavy quark rest mass which starts at $O(am)$.

Obviously, there remain many unexplored issues with these fermion formulations.
We think it would be interesting to study the behavior of small eigenvalues on interacting backgrounds
(especially some with non-zero topological charge), and how they implement the constraints imposed by the Nielsen-Ninomiya theorem.
Our Figs.~\ref{fig:KW_versus_stag} and \ref{fig:BC_versus_stag} are inspired by Figs.~$6$ and $7$ of Ref.~\cite{Durr:2004as},
and we hope to see the ``fingerprint property'' of low-energy fermion eigenvalues~%
\footnote{By this we mean that the pattern of low-energy Dirac operator eigenvalues is characteristic of the gauge background
and nearly independent of the fermion formulation, at least at small enough lattice spacings.}
confirmed with the KW and BC formulations, too.
Also some more light on the mixing pattern with lower-dimensional operators (beyond what was found in
\cite{Bedaque:2008xs,Cichy:2008gk,Capitani:2009yn,Capitani:2010nn,Creutz:2010qm,Kimura:2011ik,Weber:2013tfa,Weber:2016jug,Weber:2017eds}) might prove useful.
Overall, we feel a collaboration aiming for exploratory large-scale production runs with minimally doubled fermions
would be well advised to give first priority to the KW formulation.

\bigskip

\noindent{\bf Acknowledgements}:
This work was supported by the German DFG through the collaborative research grant SFB-TRR-55.
This work was supported by the U.S.\ Department of Energy, Office of Science, Office of Nuclear Physics and Office of Advanced Scientific Computing Research
within the framework of Scientific Discovery through Advance Computing (SciDAC) award Computing the Properties of Matter with Leadership Computing Resources.


\clearpage

\appendix


\section{Dispersion relations\label{app:A}}


\subsection{Naive fermions}

The naive operator and its Green's function take the form
\bea
D_\mr{nai}&=&
\sum_\mu\ga_\mu\nab_\mu+m\;=\;\ri\sum_\mu\ga_\mu\bar{p}_\mu+m
\\
G_\mr{nai}&=&\frac
{-\ri\sum_\mu\ga_\mu\bar{p}_\mu+m}
{(\ri\sum_\rh\ga_\rh\bar{p}_\rh+m)
(-\ri\sum_\si\ga_\si\bar{p}_\si+m)}
=\frac
{-\ri\sum_\mu\ga_\mu\bar{p}_\mu+m}
{\bar{p}^2+m^2}
\;.
\eea
The dispersion relation follows from searching for zeros of the denominator with $p_4\to\ri E$, so
\beq
0=\sum_i\sin^2(ap_i)-\sinh^2(aE)+(am)^2
\eeq
means that the physical solution is given by the positive root
\beq
aE=\sqrt{\mr{asinh}\Big(\sum_i\sin^2(ap_i)+(am)^2\Big)}
\;.
\eeq

\subsection{Wilson fermions}

The Wilson operator and its Green's function take the form
\bea
D_\mr{W}&=&
\sum_\mu\ga_\mu\nab_\mu-\frac{ar}{2}\lap+m\;=\;\ri\sum_\mu\ga_\mu\bar{p}_\mu+\frac{ar}{2}\hat{p}^2+m
\\
G_\mr{W}&=&\frac
{-\ri\sum_\mu\ga_\mu\bar{p}_\mu+\frac{ar}{2}\hat{p}^2+m}
{(\ri\sum_\rh\ga_\rh\bar{p}_\rh+\frac{ar}{2}\hat{p}^2+m)
(-\ri\sum_\si\ga_\si\bar{p}_\si+\frac{ar}{2}\hat{p}^2+m)}
=\frac
{-\ri\sum_\mu\ga_\mu\bar{p}_\mu+\frac{ar}{2}\hat{p}^2+m}
{\bar{p}^2+(\frac{ar}{2}\hat{p}^2+m)^2}
\eea
with $\bar{p}_\mu=\frac{1}{a}\sin(ap_\mu)$ and $\hat{p}_\mu=\frac{2}{a}\sin(\frac{ap_\mu}{2})$.
It follows that
\beq
\hat{p}^2=\sum_\mu\hat{p}_\mu^2=\frac{4}{a^2}\sum_\mu\sin^2(\frac{ap_\mu}{2})=\frac{2d}{a^2}-\frac{2}{a^2}\sum_\mu\cos(ap_\mu)
\eeq
or $\frac{ar}{2}\hat{p}^2=\frac{dr}{a}-\frac{r}{a}\sum_\mu\cos(ap_\mu)$,
and searching for a zero of the denominator with $p_4\to\ri E$ yields
\bea
\sinh^2(aE)&=&\sum_i\sin^2(ap_i)+\Big(dr-r\cosh(aE)-r\sum_i\cos(ap_i)+am\Big)^2
\\
&=&\sum_i\sin^2(ap_i)+r^2\cosh^2(aE)-2r\cosh(aE)\Big[dr+am-r\sum_i\cos(ap_i)\Big]+\Big[...\Big]^2
\nonumber
\;.
\eea
For $r=1$ the identity $\cosh^2-\sinh^2=1$ turns this into a linear equation in $\cosh(aE)$
\beq
2\cosh(aE)\Big[d+am-\sum_i\cos(ap_i)\Big]=1+\sum_i\sin^2(ap_i)+\Big[d+am-\sum_i\cos(ap_i)\Big]^2
\eeq
which one solves for $aE>0$ by means of $\mr{acosh}(x)=\ln(x+\sqrt{x^2-1})$ for $x>1$.
For $r\neq1$ one stays with a quadratic equation in $\cosh(aE)$
\beq
0=1+\sum_i\sin^2(ap_i)+(r^2-1)\cosh^2(aE)-2r\cosh(aE)\Big[dr+am-r\sum_i\cos(ap_i)\Big]+\Big[...\Big]^2
\eeq
which one addresses by first solving for a real positive $\cosh(aE)$ and then inverting the cosh.


\subsection{Karsten-Wilczek fermions}

The KW operator and its Green's function take the form
\bea
D_\mr{KW}&=&
\sum_\mu\ga_\mu\nab_\mu-\ri\frac{ar}{2}\ga_d\sum_{i=1}^{d-1}\lap_i+m\;=\;\ri\sum_\mu\ga_\mu\bar{p}_\mu+\ri\frac{ar}{2}\ga_d\sum_{i=1}^{d-1}\hat{p}_i^2+m
\\
G_\mr{KW}&=&\frac
{-\ri\sum_\mu\ga_\mu\bar{p}_\mu-\ri\frac{ar}{2}\ga_d\sum_{i=1}^{d-1}\hat{p}_i^2+m}
{(\ri\sum_\rh\ga_\rh\bar{p}_\rh+\ri\frac{ar}{2}\ga_d\sum_{i=1}^{d-1}\hat{p}_i^2+m)
(-\ri\sum_\si\ga_\si\bar{p}_\si-\ri\frac{ar}{2}\ga_d\sum_{j=1}^{d-1}\hat{p}_j^2+m)}
\nonumber
\\
&=&\frac
{-\ri\sum_\mu\ga_\mu\bar{p}_\mu-\ri\frac{ar}{2}\ga_d\sum_{i=1}^{d-1}\hat{p}_i^2+m}
{(\sum_{i=1}^{d-1}\ga_i\bar{p}_i+\ga_d\bar{p}_d+\frac{ar}{2}\ga_d\sum_{i=1}^{d-1}\hat{p}_i^2)^2+m^2}
\nonumber
\\
&=&\frac
{-\ri\sum_\mu\ga_\mu\bar{p}_\mu-\ri\frac{ar}{2}\ga_d\sum_{i=1}^{d-1}\hat{p}_i^2+m}
{\sum_{i=1}^{d-1}\bar{p}_i^2+(\bar{p}_d+\frac{ar}{2}\sum_{i=1}^{d-1}\hat{p}_i^2)^2+m^2}
\eea
where in the last step specific properties of the Dirac-Clifford algebra were used.
Searching for a zero of the denominator with  $\frac{ar}{2}\sum_i\hat{p}_i^2=\frac{r}{a}\sum_i\{1-\cos(ap_i)\}$ and $p_4\to\ri E$ yields
\beq
0=\sum_{i=1}^{d-1}\sin^2(ap_i)+
\Big(
\ri\sinh(aE)+r\sum_{i=1}^{d-1}\{1-\cos(ap_i)\}
\Big)^2
+(am)^2
\eeq
which does not necessarily yield a real solution for $E$.
In such a situation one should go for a complex $E$, and treat its real part as the ``energy'' of the respective mode.
In other words
\beq
\sinh(aE)=\ri r\sum_{i=1}^{d-1}\{1-\cos(ap_i)\}\pm\sqrt{\sum_{i=1}^{d-1}\sin^2(ap_i)+(am)^2}
\eeq
yields a complex $\sinh(aE)$, and through the $\mr{asinh}$ function the definition of a complex $aE$ is obtained,
whose positive real part is plotted against $\sqrt{\sum_{i=1}^{d-1}p_i^2}$.


\subsection{Borici-Creutz fermions}

The BC operator and its Green's function take the form
\bea
D_\mr{BC}&=&
\sum_\mu\ga_\mu\nab_\mu-\ri\frac{ar}{2}\sum_\mu\ga_\mu'\lap_\mu+m\;=\;\ri\sum_\mu\ga_\mu\bar{p}_\mu+\ri\frac{ar}{2}\sum_\mu\ga_\mu'\hat{p}_\mu^2+m
\\
G_\mr{BC}&=&\frac
{-\ri\sum_\mu\ga_\mu\bar{p}_\mu-\ri\frac{ar}{2}\sum_\mu\ga_\mu'\hat{p}_\mu^2+m}
{(\ri\sum_\rh\ga_\rh\bar{p}_\rh+\ri\frac{ar}{2}\sum_\rh\ga_\rh'\hat{p}_\rh^2+m)
(-\ri\sum_\si\ga_\si\bar{p}_\si-\ri\frac{ar}{2}\sum_\si\ga_\si'\hat{p}_\si^2+m)}
\\
&=&\frac
{-\ri\sum_\mu\ga_\mu\bar{p}_\mu-\ri\frac{ar}{2}\sum_\mu\ga_\mu'\hat{p}_\mu^2+m}
{                \sum_{\rh,\si}  \ga_\rh  \ga_\si   \bar{p}_\rh  \bar{p}_\si  
+\frac{a  r  }{2}\sum_{\rh,\si}  \ga_\rh  \ga_\si'  \bar{p}_\rh  \hat{p}_\si^2
+\frac{a  r  }{2}\sum_{\rh,\si}  \ga_\rh' \ga_\si   \hat{p}_\rh^2\bar{p}_\si  
+\frac{a^2r^2}{4}\sum_{\rh,\si}  \ga_\rh' \ga_\si'  \hat{p}_\rh^2\hat{p}_\si^2
+m^2}
\nonumber
\eea
and our task is to further simplify the denominator.
The first term is symmetric in $\bar{p}_\rh\leftrightarrow\bar{p}_\si$; it may be rewritten as
$\frac{1}{2}\sum_{\rh,\si}\{\ga_\rh ,\ga_\si \}\bar{p}_\rh\bar{p}_\si=\sum_\la\bar{p}_\la^2$,
where the Dirac-Clifford property of the $\ga$-matrices has been used.
For exactly the same reason the fourth term may be rewritten as
$\frac{a^2r^2}{8}\sum_{\rh,\si}\{\ga_\rh',\ga_\si'\}\hat{p}_\rh^2\hat{p}_\si^2=\frac{a^2r^2}{4}\sum_\la\hat{p}_\la^4$,
where the Dirac-Clifford property of the $\ga'$-matrices has been used.
The two cross-terms are a bit trickier to deal with.
It proves useful to notice that the second term can be inflated to look like
$\frac{a  r  }{4}\sum_{\rh,\si}  \ga_\rh  \ga_\si'  \bar{p}_\rh  \hat{p}_\si^2+\frac{a  r  }{4}\sum_{\rh,\si}  \ga_\si  \ga_\rh'  \bar{p}_\si  \hat{p}_\rh^2$.
Similarly, the third term can be brought into the form
$\frac{a  r  }{4}\sum_{\rh,\si}  \ga_\rh' \ga_\si   \hat{p}_\rh^2\bar{p}_\si  +\frac{a  r  }{4}\sum_{\rh,\si}  \ga_\si' \ga_\rh   \hat{p}_\si^2\bar{p}_\rh$.
Accordingly, the second and third terms can be combined into
$\frac{a  r  }{4}\sum_{\rh,\si}\{\ga_\rh ,\ga_\si'\}\bar{p}_\rh  \hat{p}_\si^2+\frac{a  r  }{4}\sum_{\rh,\si}\{\ga_\rh',\ga_\si \}\hat{p}_\rh^2\bar{p}_\si$,
and the relations (\ref{gagaprime},~\ref{gaprimega}) suggest replacing the latter expression by
$\frac{ar}{d}\sum_{\rh,\si}\bar{p}_\rh\hat{p}_\si^2-\frac{ar}{2}\sum_\la\bar{p}_\la\hat{p}_\la^2+\frac{ar}{d}\sum_{\rh,\si}\hat{p}_\rh^2\bar{p}_\si-\frac{ar}{2}\sum_\la\hat{p}_\la^2\bar{p}_\la$.
Putting everything together we thus arrive at
\beq
G_\mr{BC}=
\frac
{-\ri\sum_\mu\ga_\mu\bar{p}_\mu-\ri\frac{ar}{2}\sum_\mu\ga_\mu'\hat{p}_\mu^2+m}
{\sum_\la\bar{p}_\la^2
-ar\sum_\la\bar{p}_\la  \hat{p}_\la^2
+\frac{a^2r^2}{4}\sum_\la\hat{p}_\la^4
+\frac{2ar}{d}\sum_{\rh,\si}\bar{p}_\rh\hat{p}_\si^2
+m^2}
\eeq
and our task is to search for a zero of the denominator, i.e.\ to solve
\bea
0&=&\sum_\la\bar{p}_\la^2
-       a  r      \sum_\la      \bar{p}_\la  \hat{p}_\la^2
+\frac{ a^2r^2}{4}\sum_\la      \hat{p}_\la^4
+\frac{2a  r  }{d}\sum_{\rh,\si}\bar{p}_\rh  \hat{p}_\si^2
+m^2
\nonumber
\\
&=&\sum_\la\Big[\bar{p}_\la-\frac{ar}{2}\hat{p}_\la^2\Big]^2
+\frac{2a  r  }{d}\sum_{\rh,\si}\bar{p}_\rh  \hat{p}_\si^2
+m^2
\eea
with the substitution $p_4\to\ri E$ for $aE$.
Using $\bar{p}_\rh=\frac{1}{a}\sin(ap_\rh)$ and $\hat{p}_\si^2=\frac{2}{a^2}\{1-\cos(ap_\si)\}$ yields
\beq
0=\sum_\la\Big[\sin(ap_\la)-r\{1-\cos(ap_\la)\}\Big]^2
+\frac{4r}{d}\sum_{\rh,\si}\sin(ap_\rh)\{1-\cos(ap_\si)\}
+(am)^2
\label{BC_zero_original}
\eeq
which the substitution then brings into the form (with $i,j$ running from $1$ to $d-1$)
\bea
0&=&\sum_i\Big[\sin(ap_i)-r\{1-\cos(ap_i)\}\Big]^2+\Big[\ri\sinh(aE)-r\{1-\cosh(aE)\}\Big]^2
\nonumber
\\
&+&\frac{4r}{d}\sum_{i,j}\sin(ap_i)\{1-\cos(ap_j)\}+\frac{4\ri r}{d}\sinh(aE)\sum_j\{1-\cos(ap_j)\}
\nonumber
\\
&+&\frac{4r}{d}\sum_i\sin(ap_i)\{1-\cosh(aE)\}+\frac{4\ri r}{d}\sinh(aE)\{1-\cosh(aE)\}
+(am)^2
\;.
\label{BC_zero_standard}
\eea
In $d=2$ space-time dimensions this expression simplifies to (each sum contains a single term)
\bea
0&=&\sum_i\Big[\sin(ap_i)-r\{1-\cos(ap_i)\}\Big]^2 -\sinh^2(aE)+r^2\{1-\cosh(aE)\}^2
\nonumber
\\
&+&2r\sum_{i,j}\sin(ap_i)\{1-\cos(ap_j)\}+2\ri r\sinh(aE)\sum_j\{1-\cos(ap_j)\}
\nonumber
\\
&+&2r\sum_i\sin(ap_i)\{1-\cosh(aE)\}
+(am)^2
\eea
while in $d=4$ space-time dimensions one finds
\bea
0&=&\sum_i\Big[\sin(ap_i)-r\{1-\cos(ap_i)\}\Big]^2 -\sinh^2(aE)+r^2\{1-\cosh(aE)\}^2
\nonumber
\\
&+&r\sum_{i,j}\sin(ap_i)\{1-\cos(ap_j)\}+\ri r\sinh(aE)\sum_j\{1-\cos(ap_j)\}
\nonumber
\\
&+&r\sum_i\sin(ap_i)\{1-\cosh(aE)\} -\ri r\sinh(aE)\{1-\cosh(aE)\}
+(am)^2
\;.
\eea
In the special case $r=1$ the $d=2$ version simplifies to
\bea
0&=&\sum_i\Big[\sin(ap_i)-\{1-\cos(ap_i)\}\Big]^2
+\Big[2+2\sum_i\sin(ap_i)\Big]\{1-\cosh(aE)\}
\nonumber
\\
&+&2\sum_{i,j}\sin(ap_i)\{1-\cos(ap_j)\}+2\ri\sinh(aE)\sum_j\{1-\cos(ap_j)\}
+(am)^2
\label{simplified_r1_d2}
\eea
while the $d=4$ version takes the form
\bea
0
&=&\sum_i\Big[\sin(ap_i)-\{1-\cos(ap_i)\}\Big]^2
+\Big[2+\sum_i\sin(ap_i)-\ri\sinh(aE)\Big]\{1-\cosh(aE)\}
\nonumber
\\
&+&\sum_{i,j}\sin(ap_i)\{1-\cos(ap_j)\}+\ri\sinh(aE)\sum_j\{1-\cos(ap_j)\}
+(am)^2
\;.
\label{simplified_r1_d4}
\eea
These equations look complicated, and this is why we shall work our way backwards, from the simplest case to the more complicated case.

A peculiar feature of the $d=2,r=1$ case is that the equation is linear in $\sinh(aE)$ and $\cosh(aE)$.
This suggests multiplying eqn.~(\ref{simplified_r1_d2}) with $\exp(aE)$ to obtain
\bea
0&=&\sum_i\Big[\sin(ap_i)-\{1-\cos(ap_i)\}\Big]^2e^{aE}
+\Big[1+\sum_i\sin(ap_i)\Big]\{2e^{aE}-e^{2aE}-1\}
\nonumber
\\
&+&2\sum_{i,j}\sin(ap_i)\{1-\cos(ap_j)\}e^{aE}
+\ri[e^{2aE}-1]\sum_j\{1-\cos(ap_j)\}
+(am)^2e^{aE}
\label{solution_r1_d2}
\eea
which is a quadratic equation in $e^{aE}$.
Evidently, this means that we should go for the two complex $e^{aE}$ as function of $ap_1$,
to obtain a complex $aE$ whose positive real part is plotted against $|p_1|$.
By contrast, the $d=4,r=1$ case has a mixed term in $\sinh(aE)\cosh(aE)$.
The hyperbolic semi-angle substitution $t=\tanh(aE/2)$, whereupon $\sinh(aE)=2t/(1-t^2)$, $\cosh(aE)=(1+t^2)/(1-t^2)$
and $1-\cosh(aE)=-2t^2/(1-t^2)$, turns eqn.~(\ref{simplified_r1_d4}) into
\bea
0&=&\sum_i\Big[\sin(ap_i)-\{1-\cos(ap_i)\}\Big]^2
-\Big[2+\sum_i\sin(ap_i) -\frac{2\ri t}{1-t^2} \Big]\frac{2t^2}{1-t^2}
\nonumber
\\
&+&\sum_{i,j}\sin(ap_i)\{1-\cos(ap_j)\}
+\frac{2\ri t}{1-t^2}\sum_j\{1-\cos(ap_j)\}
+(am)^2
\eea
and upon multiplying this equation with $(1-t^2)^2$ one finds the (possibly modified) condition
\bea
0
&=&\sum_i\Big[\sin(ap_i)-\{1-\cos(ap_i)\}\Big]^2(1-t^2)^2
-2\Big[2+\sum_i\sin(ap_i)\Big]t^2(1-t^2)
+4\ri t^3
\nonumber
\\
&+&\sum_{i,j}\sin(ap_i)\{1-\cos(ap_j)\}(1-t^2)^2
+2\ri\sum_j\{1-\cos(ap_j)\}t(1-t^2)
+(am)^2(1-t^2)^2\qquad
\label{solution_r1_d4}
\eea
which amounts to a fourth-order polynomial in $t$.

For generic $r$ we resort to the hyperbolic semi-angle substitution, regardless of the space-time dimension.
For $d=2$ we obtain the relation
\bea
0&=&\sum_i\Big[\sin(ap_i)-r\{1-\cos(ap_i)\}\Big]^2 +\frac{4[r^2-1]t^4}{(1-t^2)^2}
\nonumber
\\
&+&2r\sum_{i,j}\sin(ap_i)\{1-\cos(ap_j)\}
+2\ri r\frac{2t}{1-t^2}\sum_j\{1-\cos(ap_j)\}
\nonumber
\\
&-&2r\sum_i\sin(ap_i)\frac{2t^2}{1-t^2}
+(am)^2
\label{startingpoint_generic_d2}
\eea
and upon multiplying this equation with $(1-t^2)^2$ one finds the (possibly modified) condition
\bea
0&=&\sum_i\Big[\sin(ap_i)-r\{1-\cos(ap_i)\}\Big]^2(1-t^2)^2 +4[r^2-1]t^4
\nonumber
\\
&+&2r\sum_{i,j}\sin(ap_i)\{1-\cos(ap_j)\}(1-t^2)^2
+4\ri r\sum_j\{1-\cos(ap_j)\}t(1-t^2)
\nonumber
\\
&-&4r\sum_i\sin(ap_i)t^2(1-t^2)
+(am)^2(1-t^2)^2
\label{solution_generic_d2}
\eea
which amounts to a fourth-order polynomial in $t$.
Note that for $r^2=1$ the second term in eqn.~(\ref{startingpoint_generic_d2}) vanishes.
It is then sufficient to multiply the equation with $1-t^2$, and one ends up with a quadratic polynomial in $t$
(equivalent to the procedure used above).
In other words, after setting $r=1$ and dropping a factor $1-t^2$ eqn.~(\ref{solution_generic_d2}) is equivalent to eqn.~(\ref{solution_r1_d2}).
For $d=4$ the same semi-angle substitution yields
\bea
0&=&\sum_i\Big[\sin(ap_i)-r\{1-\cos(ap_i)\}\Big]^2
-\frac{4t^2}{(1-t^2)^2}
+\frac{4r^2t^4}{(1-t^2)^2}
\nonumber
\\
&+&r\sum_{i,j}\sin(ap_i)\{1-\cos(ap_j)\}
+\ri r\sum_j\{1-\cos(ap_j)\}\frac{2t}{1-t^2}
\nonumber
\\
&-&r\sum_i\sin(ap_i)\frac{2t^2}{1-t^2}
+\ri r\frac{2t}{1-t^2}\frac{2t^2}{1-t^2}
+(am)^2
\eea
and upon multiplying this equation with $(1-t^2)^2$ one finds the (possibly modified) condition
\bea
0&=&\sum_i\Big[\sin(ap_i)-r\{1-\cos(ap_i)\}\Big]^2(1-t^2)^2 
-4t^2
+4r^2t^4
\nonumber
\\
&+&r\sum_{i,j}\sin(ap_i)\{1-\cos(ap_j)\}(1-t^2)^2
+2\ri r\sum_j\{1-\cos(ap_j)\}t(1-t^2)
\nonumber
\\
&-&2r\sum_i\sin(ap_i)t^2(1-t^2)
+4\ri rt^3
+(am)^2(1-t^2)^2\qquad
\label{solution_generic_d4}
\eea
which amounts to a fourth-order polynomial in $t$.
Upon setting $r=1$ eqn.~(\ref{solution_generic_d4}) simplifies to eqn.~(\ref{solution_r1_d4}) without further ado.

Using the built-in capabilities of a computer algebra program or a numerical package such as matlab/octave,
it is straight-forward to find all (in general complex-valued) solutions to a fourth-order polynomial with given numerical coefficients.
In this spirit we evaluate, for a given $(p_1,p_2,p_3)$ configuration, the four solutions $t$ and apply $aE=2\,\mr{atanh}(t)$ to obtain the energies.
The one with the smallest positive real part is interpreted as the energy of the fermion in that momentum configuration,
and its imaginary part gives the damping of the pertinent mode.
This is the numerical basis of all dispersion relations shown in this article.
On the analytical side, one may proceed one step further upon expanding the physical solution in powers of $am$.
This yields results relevant to assess the suitability of these actions for heavy-quark physics, as discussed in the main part of the article and App.~\ref{app:B}.


\section{Suitability for heavy-quark physics\label{app:B}}


\subsection{Naive fermions}

At $a\vec{p}=\vec{0}$ the naive dispersion relation simplifies to
\beq
\sinh(aE)=am
\label{static_naive_full}
\eeq
and this means that the series expansion in powers of $am$ takes the form
\beq
aE=am
\bigg\{
1-\frac{1}{6}(am)^2+\frac{3}{40}(am)^4+O((am)^6)
\bigg\}
\;.
\label{static_naive_series}
\eeq
Hence, the rest-mass of a fermion in the naive discretization has cut-off effects $O((am)^2)$.


\subsection{Wilson fermions}

At $a\vec{p}=\vec{0}$ the Wilson dispersion relation for arbitrary $d$ and $r=1$ simplifies to
\beq
\cosh(aE)=\frac{1}{2(1+am)}+\frac{1+am}{2}
\eeq
which is solved if $\exp(aE)=1+am$, that is for $aE=\log(1+am)$.
The series expansion
\beq
aE=am
\bigg\{
1-\frac{1}{2}am+\frac{1}{3}(am)^2-\frac{1}{4}(am)^3+O((am)^4)
\bigg\}
\eeq
shows that such cut-off effects scale as $O(am)$.
For arbitrary $d$ and generic $r$ one starts from the quadratic equation
$(r^2-1)\cosh^2(aE)-2r(r+am)\cosh(aE)+1+(r+am)^2=0$ whereupon
\beq
\cosh(aE)=\frac{r(r+am)\pm\sqrt{1+2ram+(am)^2}}{r^2-1}
\eeq
out of which only the second solution (with negative sign) is physical, since it is the one which agrees,
in the limit $r\to1$, with the solution found in this special case.
This yields the expansion
\beq
aE=am
\bigg\{
1-\frac{r}{2}am+\frac{3r^2-1}{6}(am)^2-\frac{[5r^2-3]r}{8}(am)^3+O((am)^4)
\bigg\}
\eeq
which, again, in the special case $r=1$ is found to agree with the previous expansion.
The lesson is that cut-off effects of Wilson fermions are linear in $am$.
It is impossible to get rid of this undesirable term through a clever choice of $r$,
since for $r=0$ we are back to $2^d$ species.


\subsection{Karsten-Wilczek fermions}

At $a\vec{p}=\vec{0}$ the KW dispersion relation simplifies to $0=-\sinh^2(aE)+(am)^2$
and thus to the form (\ref{static_naive_full}) of the naive action.
Accordingly, the expansion of the rest energy of a static KW fermion in powers of $am$ agrees with 
(\ref{static_naive_series}).
Hence, the KW action yields a $2$ species formulation which maintains the desirable heavy-quark features of the naive discretization.


\subsection{Borici-Creutz fermions}

At $a\vec{p}=\vec{0}$ the BC dispersion relation in $d$ space-time dimensions takes the form
\beq
0=\Big[\ri\sinh(aE)-r\{1-\cosh(aE)\}\Big]^2
+\frac{4\ri r}{d}\sinh(aE)\{1-\cosh(aE)\}
+(am)^2
\label{static_BC}
\eeq
which for $d=2$ simplifies to $0=-\sinh^2(aE)+r^2\{1-\cosh(aE)\}^2+(am)^2$,
while for $d=4$ it takes the form $0=-\sinh^2(aE)+r^2\{1-\cosh(aE)\}^2-\ri r\sinh(aE)\{1-\cosh(aE)\}+(am)^2$.

It seems instructive to first consider the case $r=1$.
In this case the $d=2$ version assumes the compact form $0=2-2\cosh(aE)+(am)^2$,
while the $d=4$ version can be rewritten as $0=[2-\ri\sinh(aE)][1-\cosh(aE)]+(am)^2$.
In $d=2$ dimensions the solution at $r=1$ is
\beq
\cosh(aE)=1+\frac{1}{2}(am)^2
\qquad [d=2, r=1]
\label{static_BC_d2_r1}
\eeq
which expands as
\bdm
aE=am
\bigg\{
1-\frac{1}{24}(am)^2+\frac{3}{640}(am)^4+O((am)^6)
\bigg\}
\qquad [d=2, r=1]
\;.
\edm
In $d=4$ dimensions even at $r=1$ the solution can only be given as the logarithm of the roots of the
polynomial $\ri z^4-(4+2\ri)z^3+(8+4(am)^2)z^2-(4-2\ri)z-\ri$, and a power expansion yields
\beq
aE=am
\bigg\{
1
+\frac{\ri}{4}am
-\frac{19}{96}(am)^2
-\frac{\ri}{8}(am)^3
+\frac{923}{10240}(am)^4
+O((am)^5)
\bigg\}
\qquad [d=4, r=1]
\;.
\label{static_BC_d4_r1}
\eeq

For $r\neq1$ and $d=2$ we notice that eqn.~(\ref{static_BC}) 
is quadratic in $\cosh(aE)$, whereupon
\beq
\cosh(aE)=
\frac{-r^2\pm\sqrt{1+(1-r^2)(am)^2}}{1-r^2}
\qquad [d=2]
\eeq
but only the first solution (with positive sign) is physical, since it is the one which agrees,
in the limit $r\to1$ with the solution (\ref{static_BC_d2_r1}) found previously.
It expands as
\beq
aE=am
\bigg\{
1+\frac{3r^2-4}{24}(am)^2+\frac{35r^4-80r^2+48}{640}(am)^4+O((am)^6)
\bigg\}
\qquad [d=2]
\eeq
and a quick check reveals that each coefficient in the $r=1$ expansion is recovered in that limit.
For $r\neq1$ and $d=4$ the solution of eqn.~(\ref{static_BC}) 
can only be given as the logarithm of the roots of the
polynomial $(\ri r+r^2-1)z^4+(-2\ri r-4r^2)z^3+(4m^2+6r^2+2)z^2+(2\ri r-4r^2)z-1-\ri r+r^2=0$, and a power expansion yields
\bea
aE&=&am
\bigg\{
1
+\frac{\ri r}{4}am
-\frac{3r^2+16}{96}(am)^2
+\frac{\ri[r^3-3r]}{16}(am)^3
-\frac{805r^4-960r^2-768}{10240}(am)^4
\nonumber
\\
&&\qquad
+O((am)^5)
\bigg\}
\qquad [d=4]
\eea
which, for $r\to1$, would indeed simplify to (\ref{static_BC_d4_r1}).

In short, we find that in $d=2$ dimensions the rest-mass of a BC fermion has discretization effects $O((am)^2)$ for generic $r$.
For $r^2=4/3$ they are even pushed to $O((am)^4)$.
By contrast, in $d=4$ dimensions the rest mass of a BC fermion has $O(am)$ cut-off effects, but this order affects only the imaginary part.
Quite generally, it seems that in $d=4$ dimensions the real part of $E/m$ is even in $r$ and $am$,
while the imaginary part is odd in $r$ and $am$.



Another way to see the difference between the cases $d=2$ and $d=4$
is to apply the hyperbolic semi-angle substitution to eqn.~(\ref{static_BC}).
Multiplying it with $(1-t^2)^2$ yields
\bdm
0=-4t^2+8\ri rt^3+4r^2t^4
-\frac{16\ri r}{d} t^3
+(am)^2(1-t^2)^2
\edm
where $t=\tanh(aE/2)$.
Specifically for $d=2$ the troublesome cubic term is gone
\beq
0=-4t^2+4r^2t^4+(am)^2(1-t^2)^2
\qquad [d=2]
\eeq
and the equation is bi-quadratic, while for $d=4$ one ends up with
\beq
0=-4t^2+4\ri rt^3+4r^2t^4+(am)^2(1-t^2)^2
\qquad [d=4]
\eeq
which is a genuine fourth-order equation in $t$.


\section{Check of zero location in Green functions\label{app:C}}


\subsection{Naive fermions}

The denominator of $G_\mr{nai}$ at $am=0$ is $a^2\bar{p}^2=\sum_\mu\sin^2(ap_\mu)$.
It has $16$ zeros in the Brillouin zone, one at $ap_\mu\in\{0,\pi\}$ for each $\mu$,
if the range in each direction is taken to be $]-\frac{\pi}{2},\frac{3\pi}{2}]$.


\subsection{Wilson fermions}

The denominator of $G_\mr{W}$ at $am=0$ is $a^2\bar{p}^2+(\frac{a^2r}{2}\hat{p}^2)^2$;
evidently it is only zero if $a^2\bar{p}^2=0$ and $a^2\hat{p}^2=0$ hold simultaneously.
The first term has $16$ zeros in the Brillouin zone, the second one only one, at $(0,0,0,0)$.
The Wilson term thus lifts $15$ of the $16$ species of the naive action to a level $2r/a$, $4r/a$, $6r/a$ and $8r/a$,
with degeneracies $4,6,4$, and $1$, respectively.


\subsection{Karsten-Wilczek fermions}

The denominator of $G_\mr{KW}$ at $am=0$ is zero if $\sum_{i=1}^{d-1}\bar{p}_i^2+(\bar{p}_d+\frac{ar}{2}\sum_{i=1}^{d-1}\hat{p}_i^2)^2=0$.
This holds if
\beq
0=\sum_i\sin^2(ap_i)
\qquad\land\qquad
0=\sin(ap_d)+2r\sum_i\sin^2(\frac{ap_i}{2})
\label{conditions_KW}
\eeq
hold simultaneously.
The first requirement implies $ap_i\in\{0,\pi\}$ for each $i$, if the range is taken to be $]-\frac{\pi}{2},\frac{3\pi}{2}]$.
Hence we need to evaluate the second requirement for the $2^{d-1}$ spatial momentum configurations, e.g.\
$(0,0,0)$, $(0,0,\pi)$, $(0,\pi,0)$, $(\pi,0,0)$, $(0,\pi,\pi)$, $(\pi,0,\pi)$, $(\pi,\pi,0)$, $(\pi,\pi,\pi)$ for $d=4$.
For $(0,0,0)$ the second requirement reads $0=\sin(ap_d)+0$, and this implies $ap_d\in\{0,\pi\}$.
For each of $(0,0,\pi)$, $(0,\pi,0)$, $(\pi,0,0)$ the second requirement reads $0=\sin(ap_d)+2r$,
which has two solutions (in $ap_d$) for $|r|<\frac{1}{2}$ that merge into one at $|r|=\frac{1}{2}$,
hence the number of flavors changes here by $6$.
For each of $(0,\pi,\pi)$, $(\pi,0,\pi)$, $(\pi,\pi,0)$ the second requirement reads $0=\sin(ap_d)+4r$,
which has two solutions for $|r|<\frac{1}{4}$ that merge into one at $|r|=\frac{1}{4}$,
hence the number of flavors changes here by $6$.
For $(\pi,\pi,\pi)$ the second requirement reads $0=\sin(ap_d)+6r$,
which has two solutions for $|r|<\frac{1}{6}$ that merge into one at $|r|=\frac{1}{6}$.
In summary, $|r|=\frac{1}{6}$ marks the watershed (for $d=4$) between a deformed naive fermion and a $14$ species formulation,
$|r|=\frac{1}{4}$ marks the transition to $8$ species, and
$|r|=\frac{1}{2}$ marks the transition to a minimally doubled lattice fermion
with poles at $(0,0,0,0)$ and $(0,0,0,\pi)$.

\begin{figure}[!tb]
\includegraphics[width=0.99\textwidth]{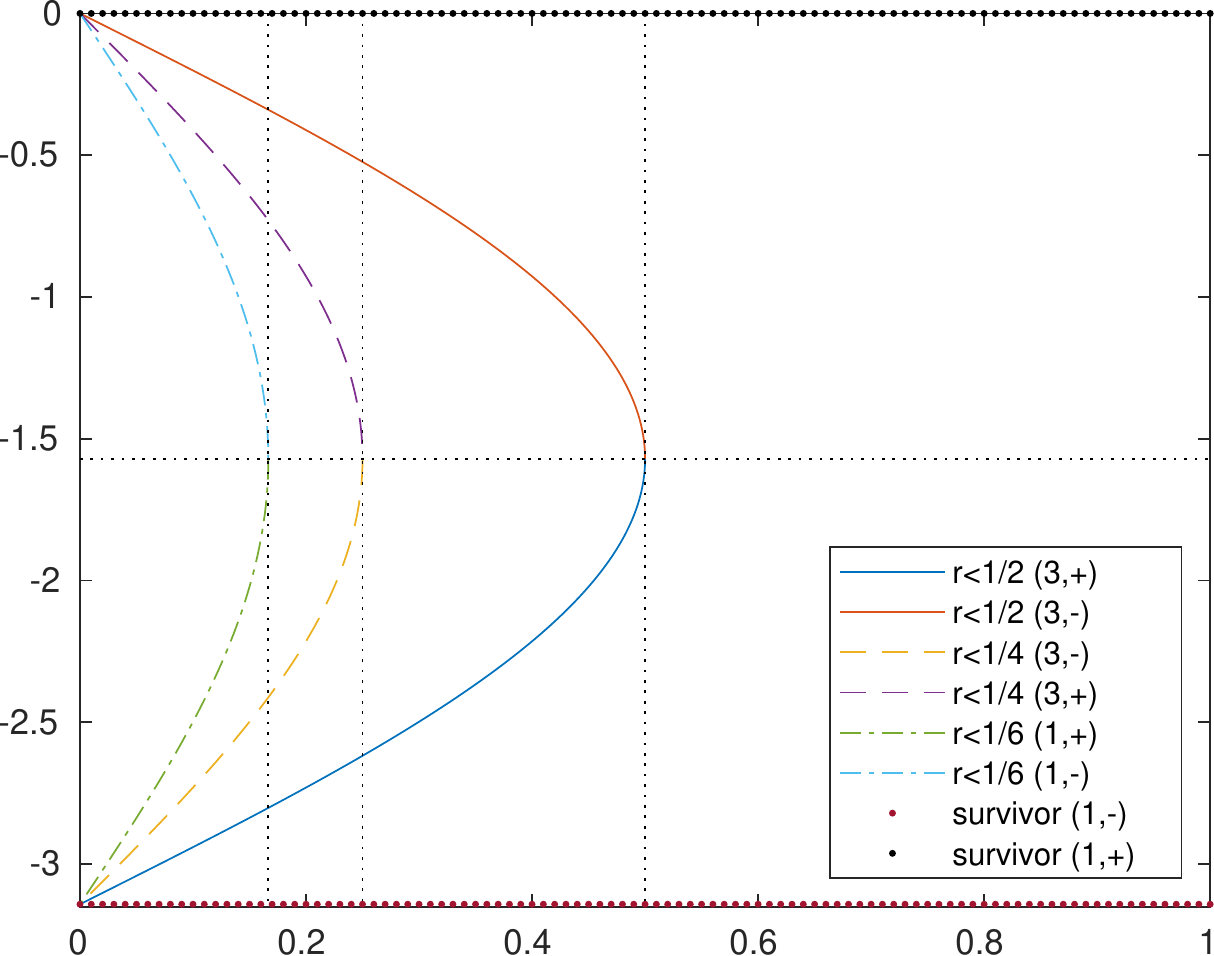}%
\caption{\label{fig:illustration_KW}
Illustration of the free-field pole structure of the Karsten-Wilczek operator in $d=4$ dimensions.
The momentum $ap_4$ is always plotted as a function of the parameter $r$.
The $3$-fold degenerate solution that emerges from $(0,0,\pi)$, $(0,\pi,0)$, or $(\pi,0,0)$, together with $ap_4=\pm\pi$, has correct chirality.
It annihilates, at $r=1/2$, with the $3$-fold degenerate counterpart that emerges from $ap_4=0$ with opposite chirality.
The $3$-fold degenerate solution that emerges from $(0,\pi,\pi)$, $(\pi,0,\pi)$, $(\pi,\pi,0)$, together with $ap_4=\pm\pi$, has opposite chirality.
It annihilates, at $r=1/4$, with the $3$-fold degenerate counterpart that emerges from $ap_4=0$ with correct chirality.
The non-degenerate solution that emerges from $(\pi,\pi,\pi)$, together with $ap_4=\pm\pi$, has correct chirality.
It annihilates, at $r=1/6$, with the non-degenerate counterpart that emerges from $ap_4=0$ with opposite chirality.
The non-degenerate solution that emerges from $(0,0,0)$ and $ap_4=\pm\pi$ has opposite chirality and lives for any $r$.
The non-degenerate solution stemming from the same spatial momentum, but with $ap_4=0$, has correct chirality and lives for any $r$.}
\end{figure}

In view of a similar discussion below for BC fermions, it is perhaps useful to illustrate the solutions to the system (\ref{conditions_KW})
in the $(r,ap_4)$ plane, see Fig.~\ref{fig:illustration_KW}.
The degeneracies and multiplicities of the modes are given in the legend and the caption.
The main mode $(0,0,0,0)$ is labeled ``survivor $(1,+)$'', since it is non-degenerate with correct chirality.
The doubler mode $(0,0,0,\pm\pi)$ is labeled ``survivor $(1,-)$'', since it is non-degenerate with opposite chirality.

\begin{figure}[p]
\includegraphics[width=0.48\textwidth]{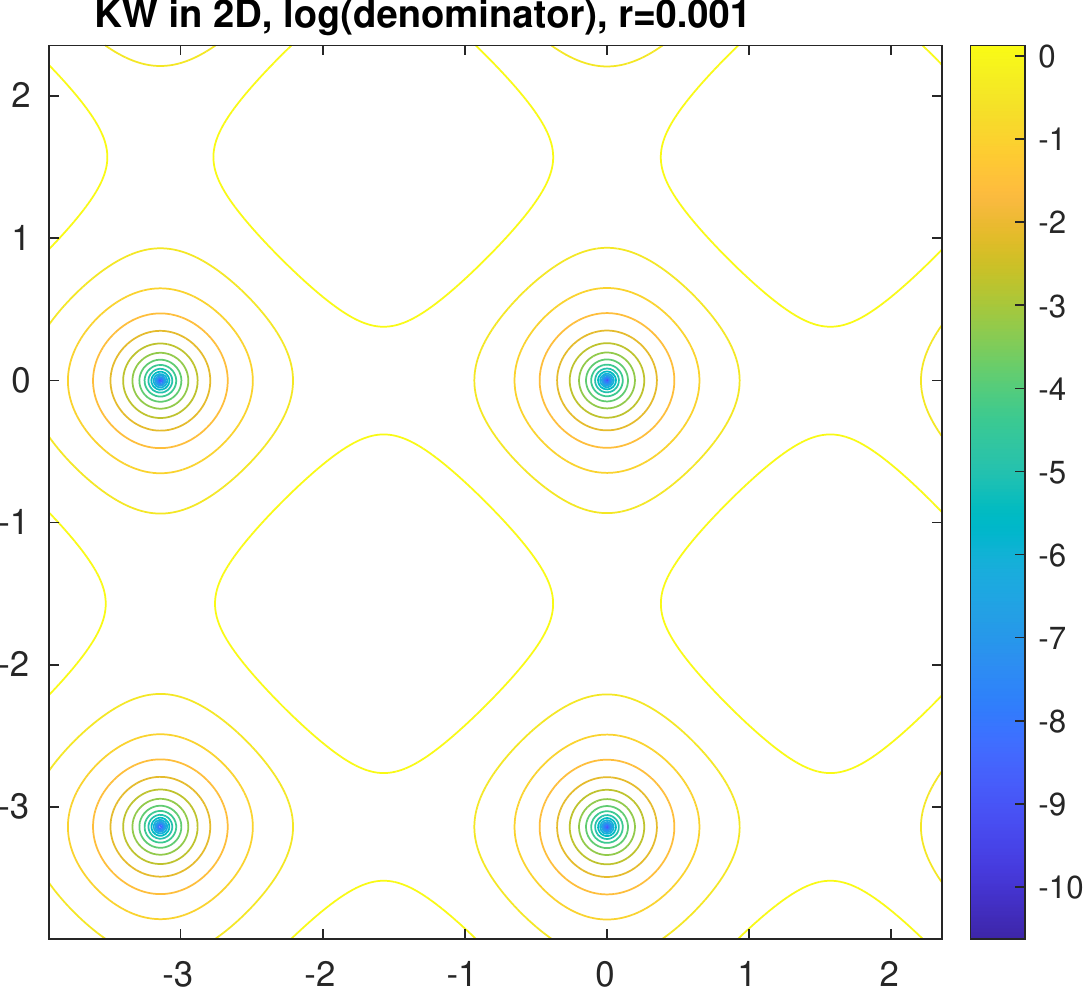}\hfill
\includegraphics[width=0.48\textwidth]{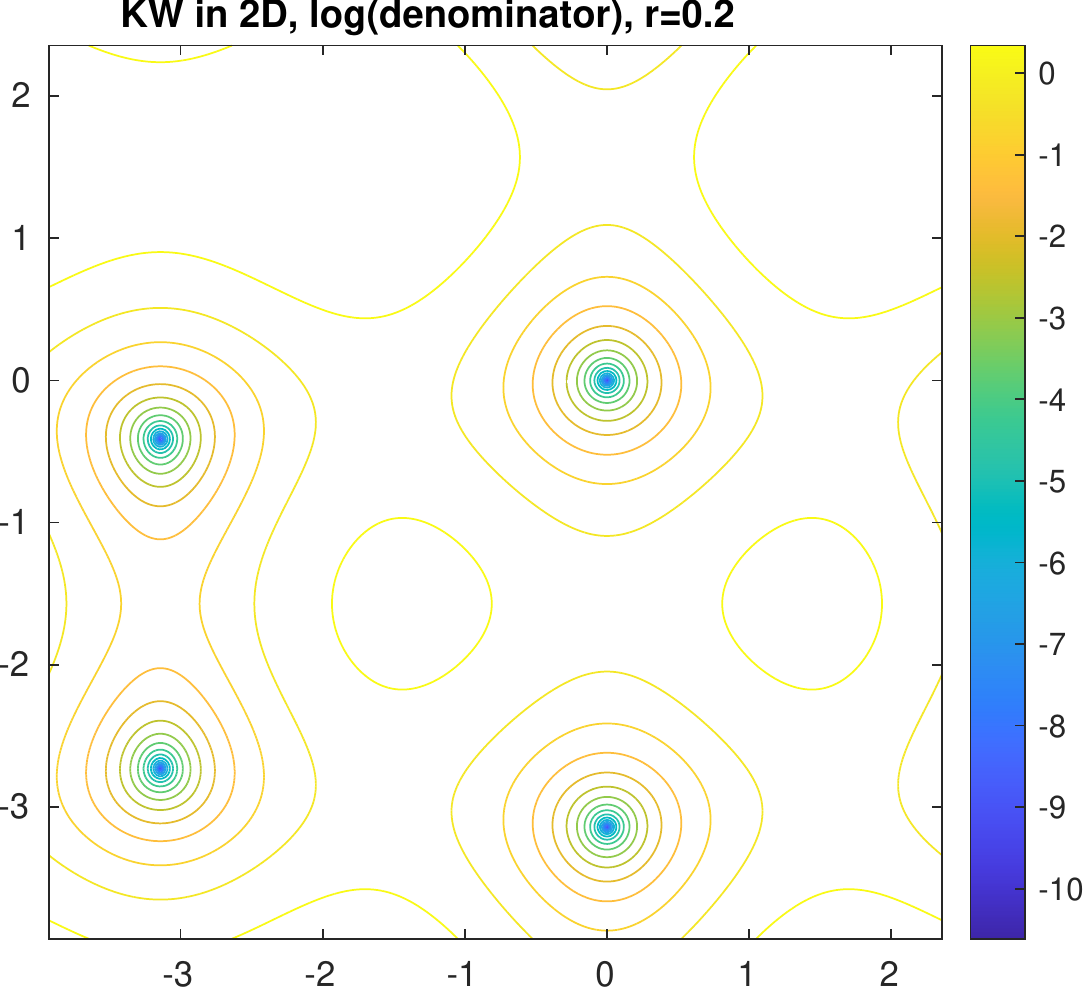}\\
\includegraphics[width=0.48\textwidth]{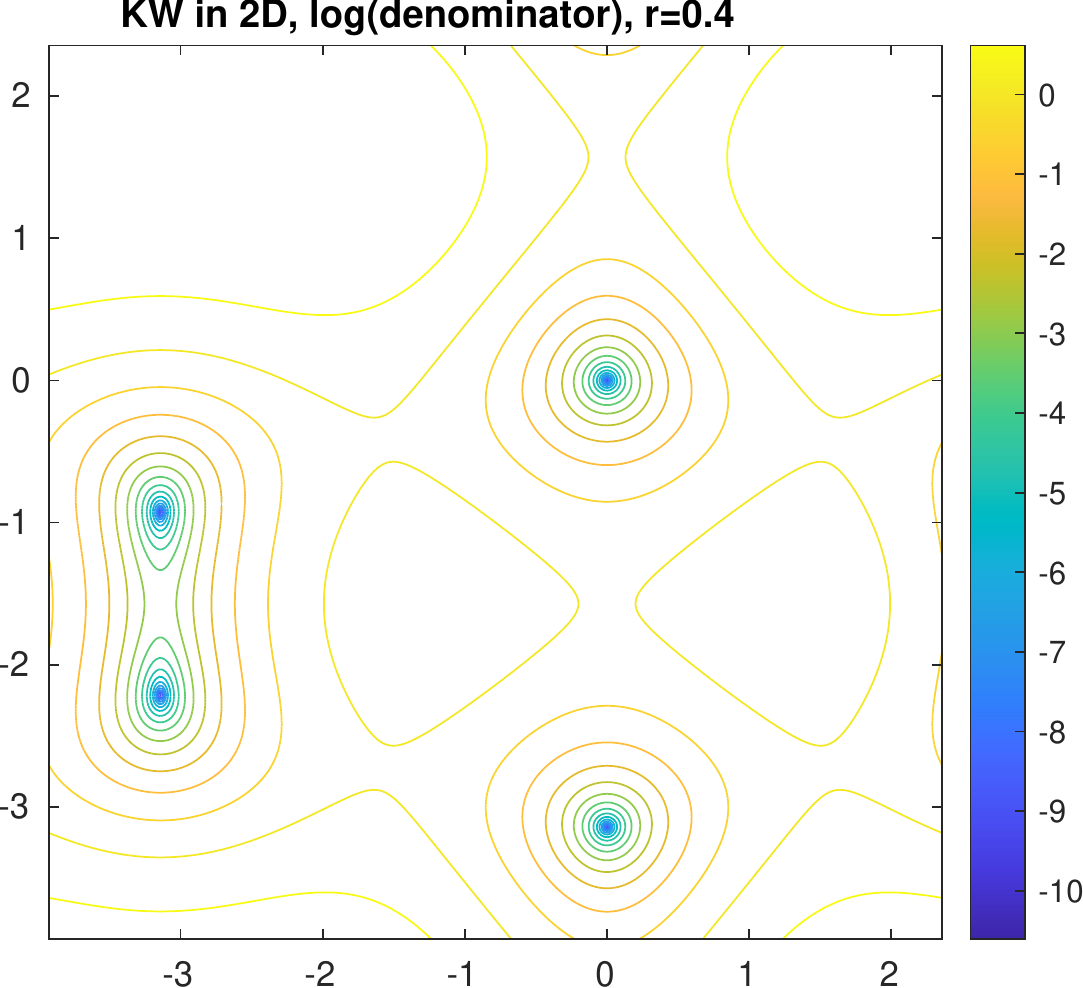}\hfill
\includegraphics[width=0.48\textwidth]{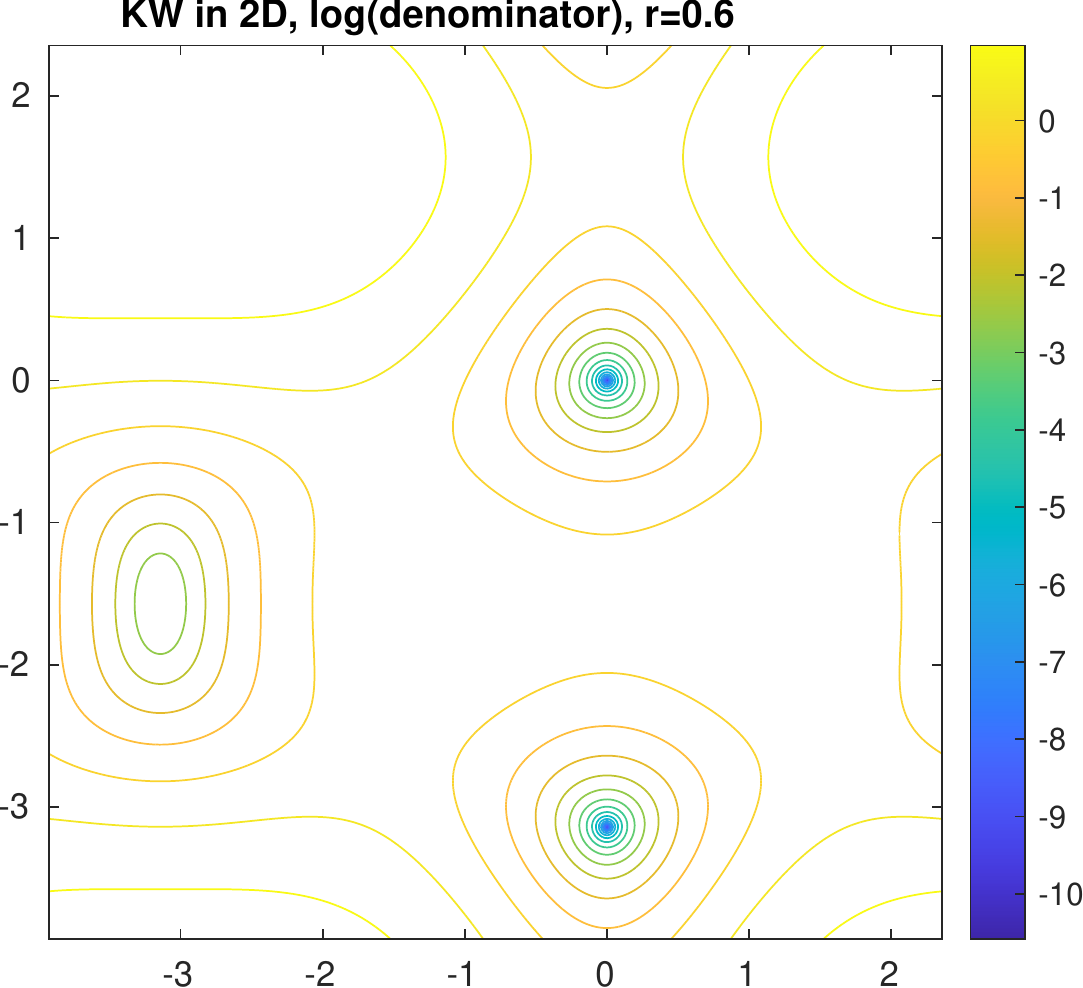}\\
\includegraphics[width=0.48\textwidth]{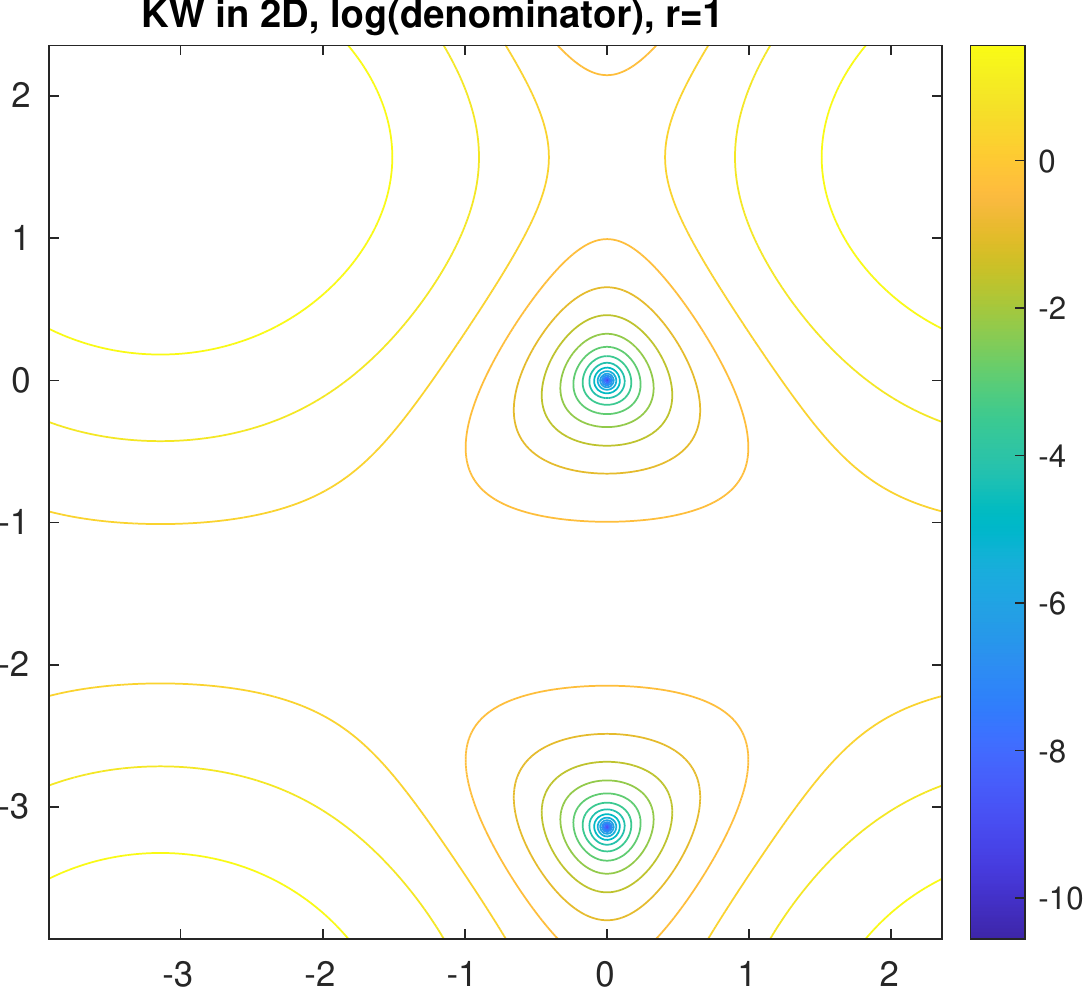}\hfill
\includegraphics[width=0.48\textwidth]{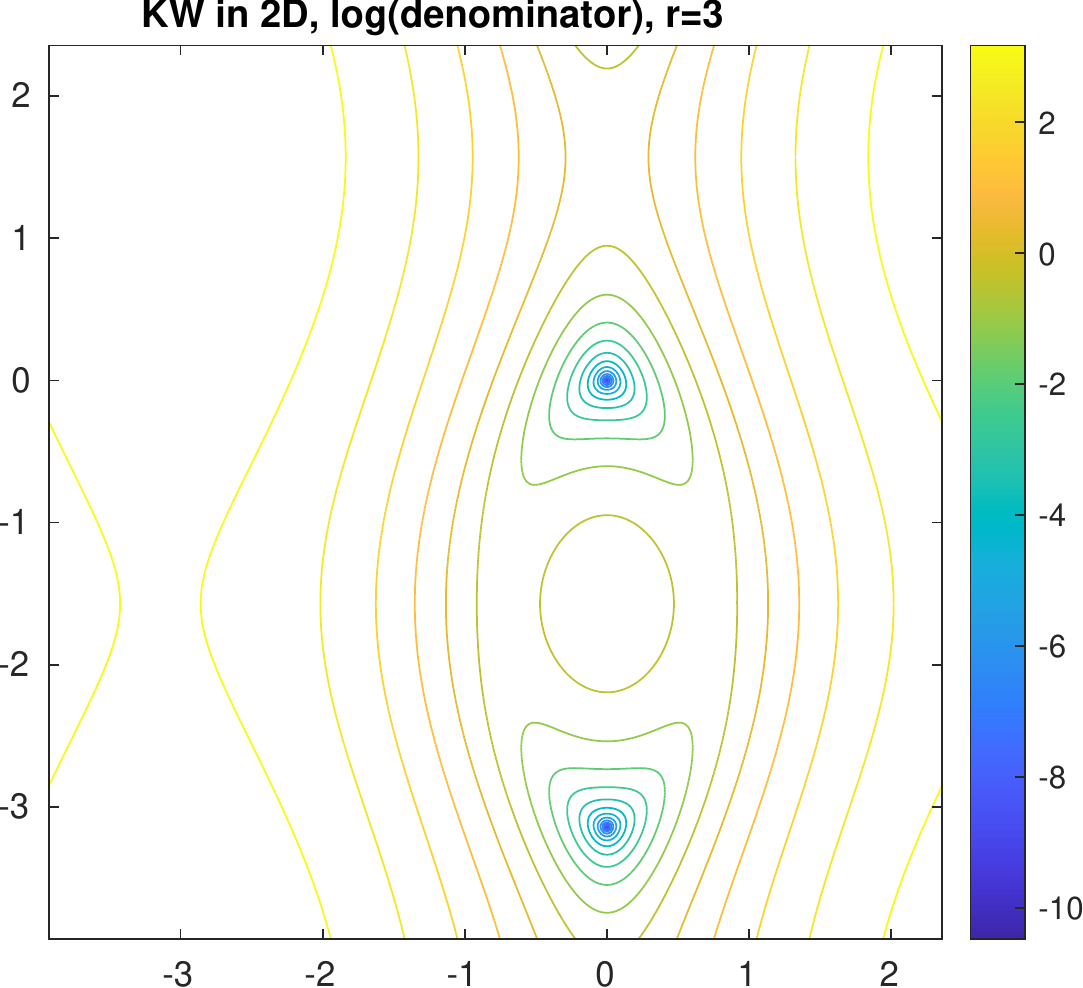}%
\vspace*{-2pt}\caption{\label{fig:contours_kw}
Contour plots of the denominator of the KW propagator
in $d=2$ space-time dimensions for $r\in\{0.001,0.2,0.4,0.6,1,3\}$.
Two poles annihilate at $r=1/2$.}
\end{figure}

A contour plot for KW fermions in $d=2$ dimensions is shown in Fig.~\ref{fig:contours_kw}.
The momentum range is $]-\frac{5}{4}\pi,\frac{3}{4}\pi[$ for both $ap_1$ and $ap_2$.
At $r=0$ one starts with the naive action.
At infinitesimally small $r$ the poles at $(-\pi,0)$ and $(-\pi,-\pi)$ [which have opposite chiralities] start moving towards each other.
At $r=1/2$ they meet at $(-\pi,-\pi/2)$ and annihilate.
The two remaining poles are located at $(0,0)$, with correct chirality on topologically charged backgrounds, and at $(0,-\pi)$, with opposite chirality.
Their position is independent of $r$.


\begin{figure}[p]
\includegraphics[width=0.48\textwidth]{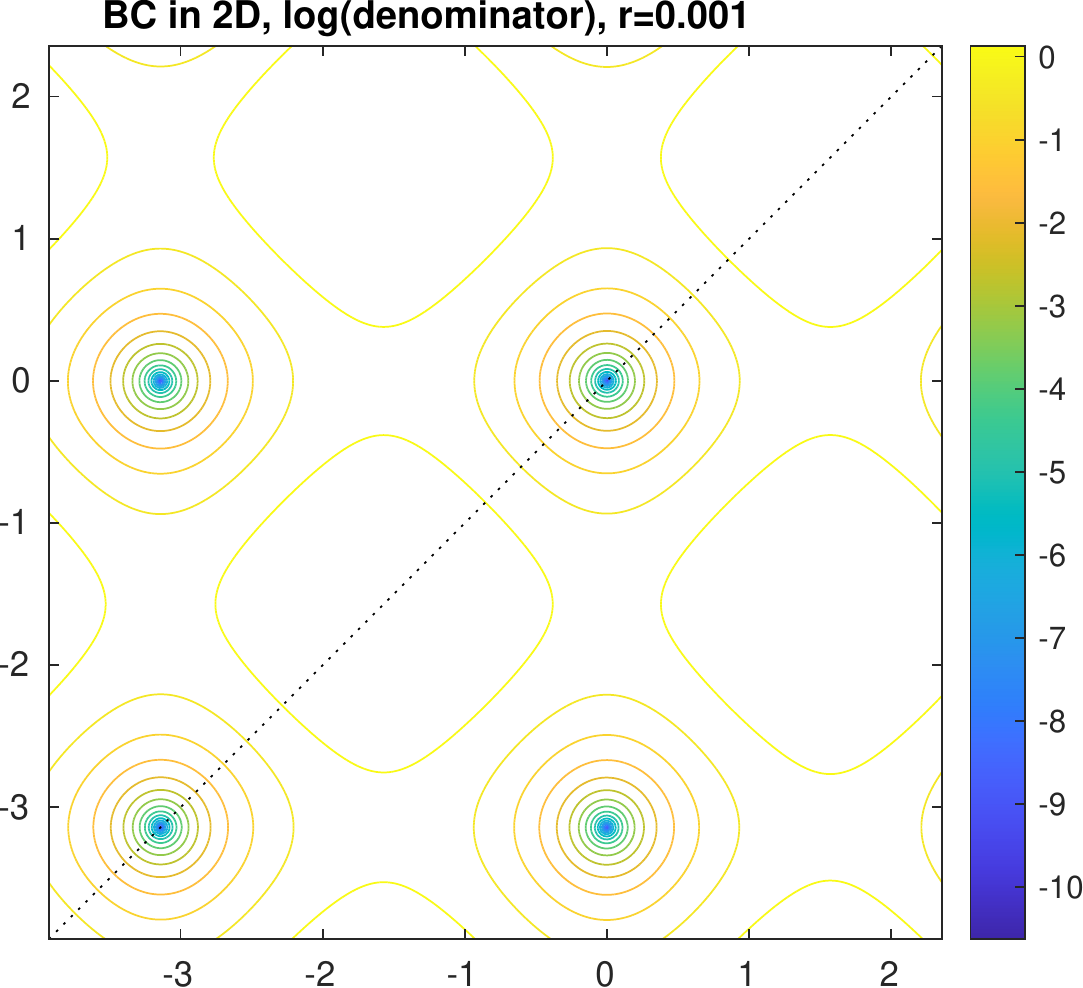}\hfill
\includegraphics[width=0.48\textwidth]{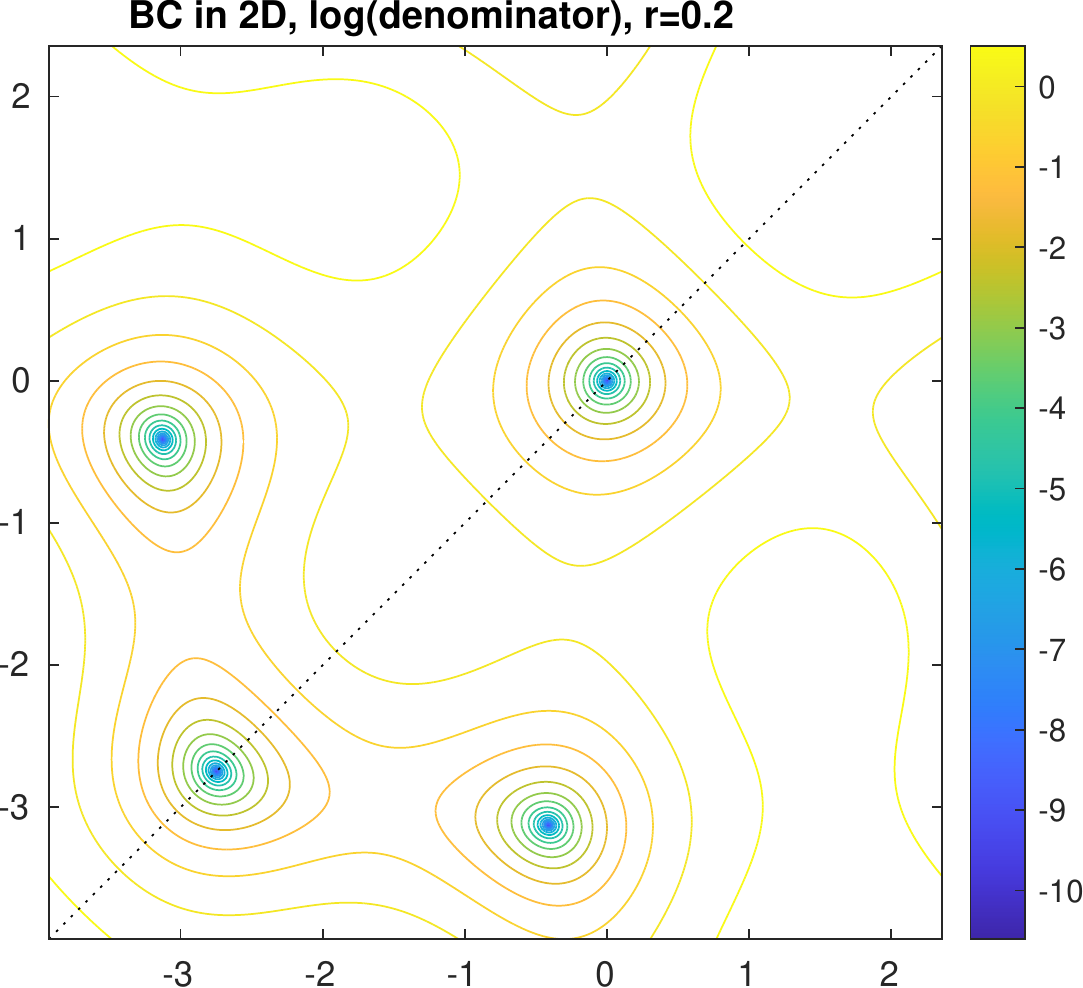}\\
\includegraphics[width=0.48\textwidth]{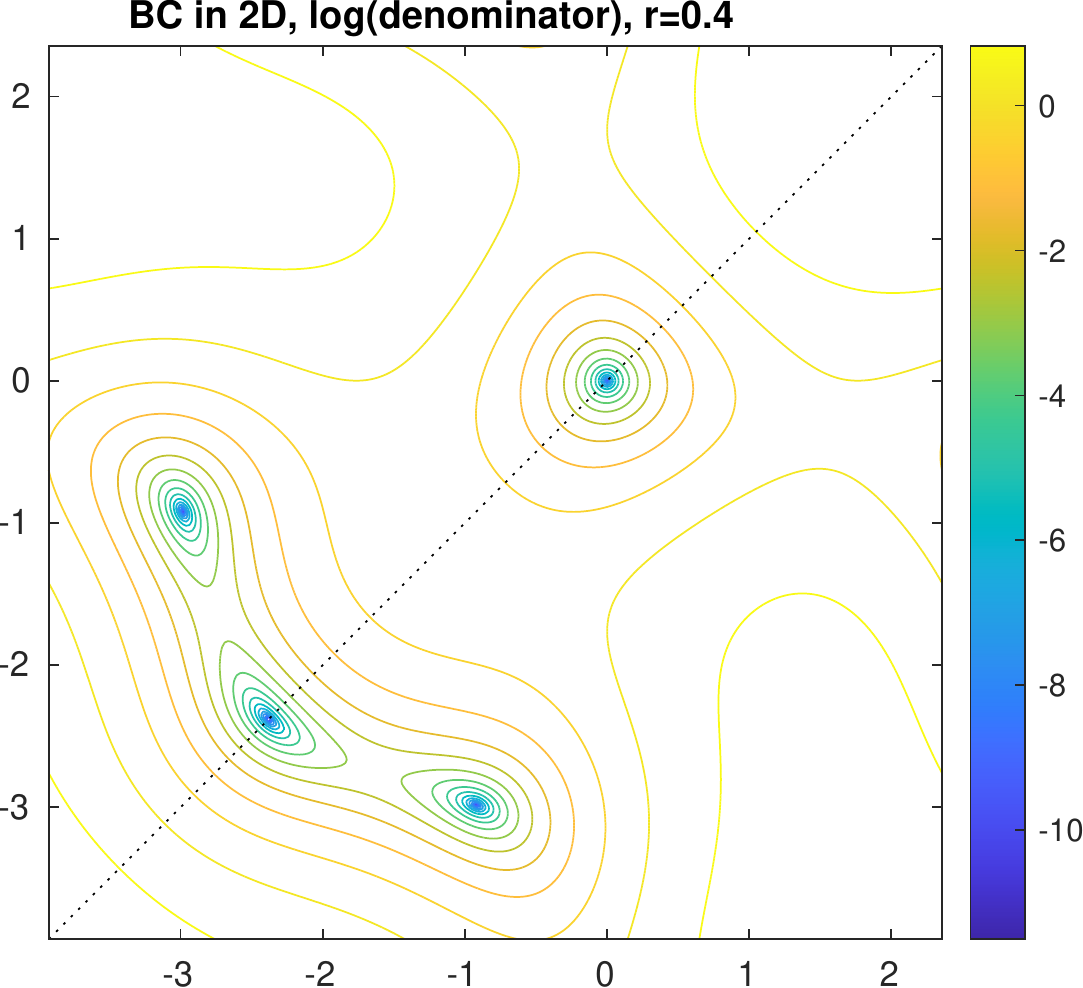}\hfill
\includegraphics[width=0.48\textwidth]{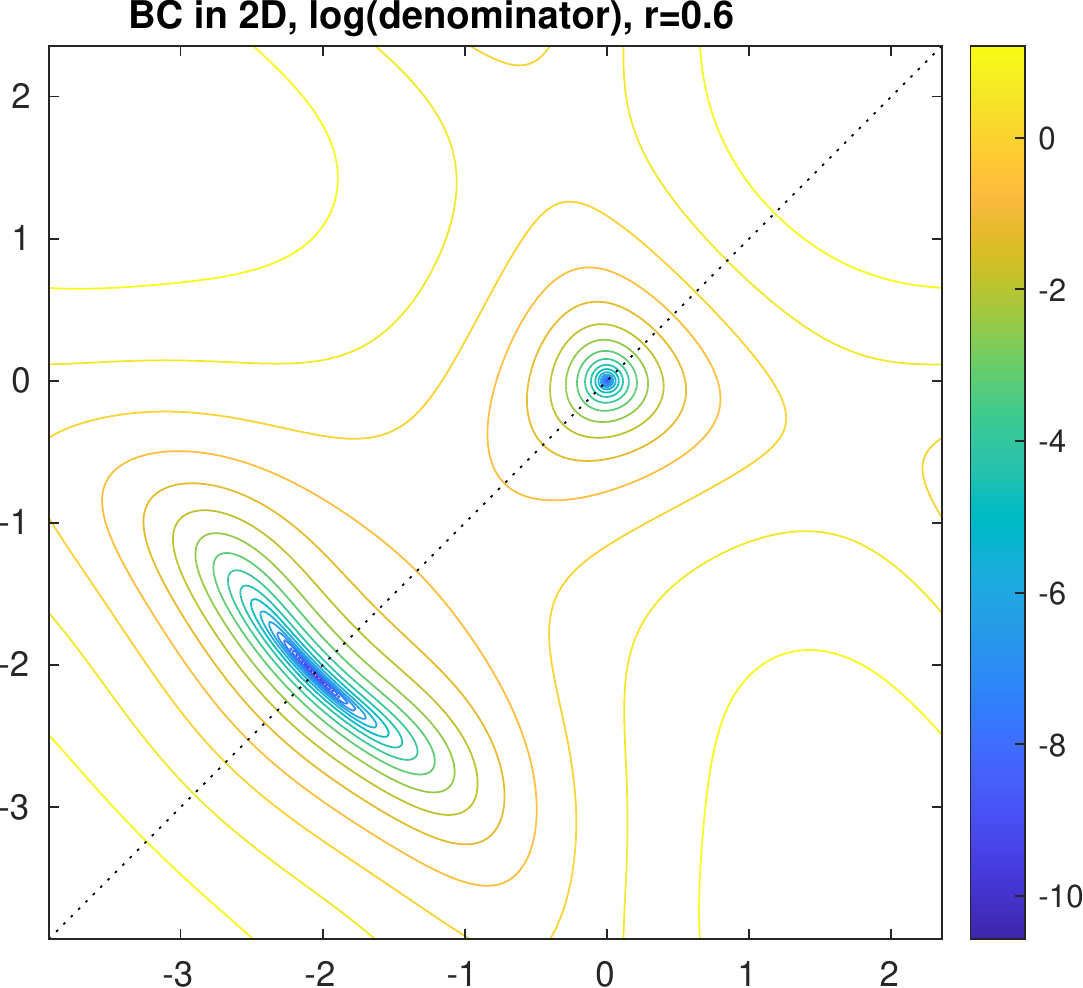}\\
\includegraphics[width=0.48\textwidth]{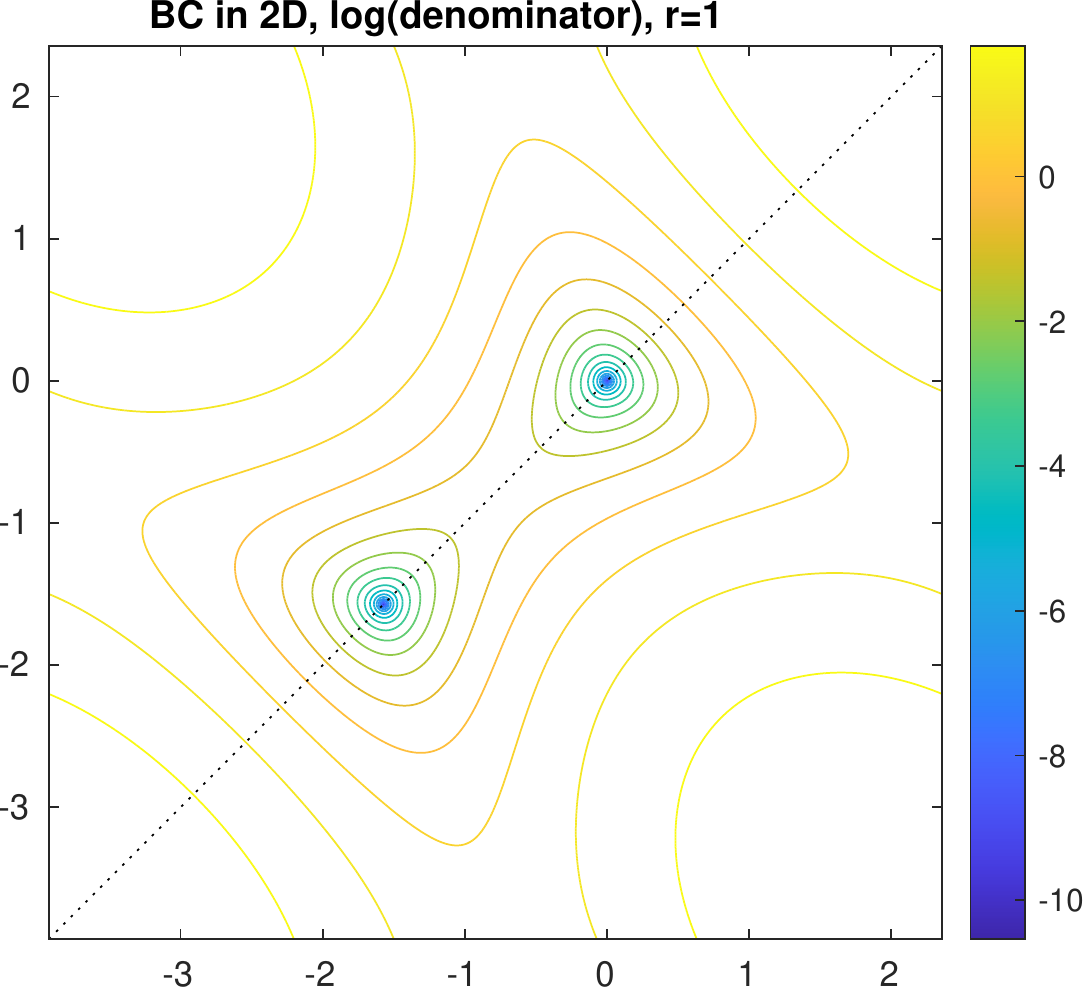}\hfill
\includegraphics[width=0.48\textwidth]{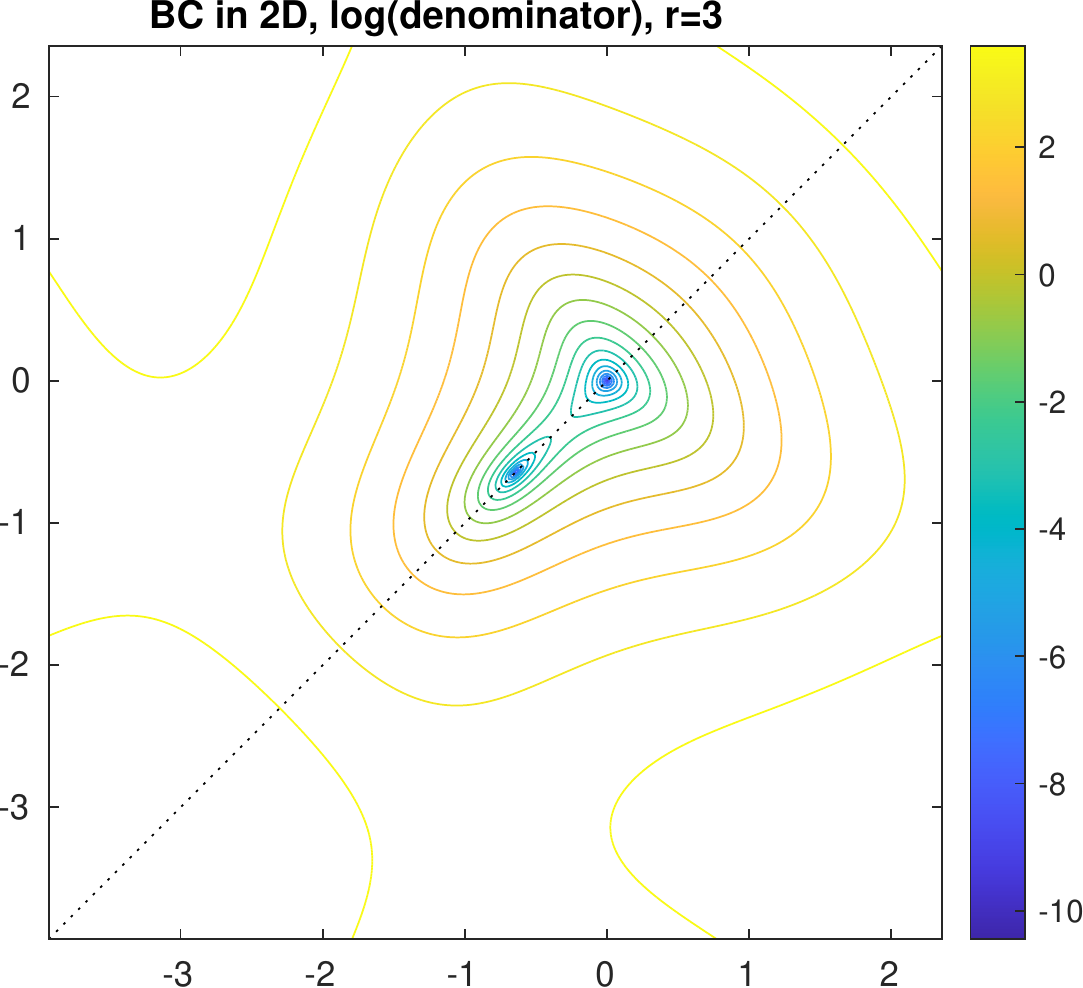}%
\vspace*{-2pt}\caption{\label{fig:contours_bc}
Contour plots of the denominator of the BC propagator
in $d=2$ space-time dimensions for $r\in\{0.001,0.2,0.4,0.6,1,3\}$.
Three poles merge into one at $r=1/\sqrt{3}\simeq0.57735$.}
\end{figure}


\subsection{Borici-Creutz fermions}

For BC fermions in $d=2$ dimensions, eqn.~(\ref{BC_zero_original}) at $am=0$ simplifies to
\bea
0&=&\Big[\sin(ap_1)-r\{1-\cos(ap_1)\}\Big]^2+\Big[\sin(ap_2)-r\{1-\cos(ap_2)\}\Big]^2
\nonumber
\\
&+&2r\Big[\sin(ap_1)+\sin(ap_2)\Big]\Big[2-\cos(ap_1)-\cos(ap_2)\Big]
\label{zeros_for_bc_2d}
\;.
\eea
Let us first search for a symmetric mode, i.e.\ one with $p_1=p_2\equiv p$.
In this case we have
\beq
0=\Big[\cos(ap)-1\Big]\Big[(r^2-1)\cos(ap)-2r\sin(ap)-(r^2+1)\Big]
\eeq
and thus one solution, $ap=0$, is independent of $r$.
To the second square bracket we apply the trigonometric semi-angle substitution $t=\tan(ap/2)$ with $\sin(ap)=2t/(1+t^2)$ and $\cos(t)=(1-t^2)/(1+t^2)$.
Upon multiplying the result with $1+t^2$, the second factor becomes
\beq
0
=(r^2-1)(1-t^2)-2r2t-(r^2+1)(1+t^2)
=-2(rt+1)^2
\eeq
and this yields the $2$-fold zero $t=-1/r$, hence $ap=-2\arctan(1/r)$.

For the non-symmetric modes it is useful to notice that (\ref{zeros_for_bc_2d}) is the sum of two squares
\beq
0=\Big[\sin(ap_1)+r\{1-\cos(ap_2)\}\Big]^2+\Big[\sin(ap_2)+r\{1-\cos(ap_1)\}\Big]^2
\eeq
and one can thus reformulate the condition as a system of two coupled equations
\beq
0=\sin(ap_1)+r\{1-\cos(ap_2)\}
\qquad\land\qquad
0=\sin(ap_2)+r\{1-\cos(ap_1)\}
\;.
\label{zeros_bc_2d}
\eeq
The aforementioned trigonometric semi-angle substitution turns this into
\beq
0=\frac{t_1}{1+t_1^2}+\frac{rt_2^2}{1+t_2^2}
\qquad\land\qquad
0=\frac{t_2}{1+t_2^2}+\frac{rt_1^2}{1+t_1^2}
\eeq
which, after multiplication by $(1+t_1^2)(1+t_2^2)$, leads to the conditions
\beq
0=t_1(1+t_2^2)+rt_2^2(1+t_1^2)
\qquad\land\qquad
0=t_2(1+t_1^2)+rt_1^2(1+t_2^2)
\;.
\eeq
There are four real solutions,
$\{t_1=0,t_2=0\}$,
$\{t_1=-1/r,t_2=-1/r\}$,
\bea
\{\; t_1=\frac{r^2-1+\sqrt{1-2r^2-3r^4}}{2r^3} &,&\quad t_2=\frac{r^2-1-\sqrt{1-2r^2-3r^4}}{2r^3} \;\}
\nonumber
\\
\{\; t_1=\frac{r^2-1-\sqrt{1-2r^2-3r^4}}{2r^3} &,&\quad t_2=\frac{r^2-1+\sqrt{1-2r^2-3r^4}}{2r^3} \;\}
\label{solution_bc_2d}
\eea
where the first two are again symmetric in $p_1 \leftrightarrow p_2$, and the last two interchange under $t_1 \leftrightarrow t_2$.
For the square-root in (\ref{solution_bc_2d}) to be real, one needs $1-2r^2-3r^4\geq0$, and this means $r^2\leq1/3$.
At $r=1/\sqrt{3}$ the solutions become $t_1=t_2=-\sqrt{3}$,
and thus coincide with the symmetric solution, $t=-1/r=-\sqrt{3}$ at this point.
In short, for $0<r<1/\sqrt{3}$ we have a $4$ species action (with two symmetric and two non-symmetric modes),
while for $1/\sqrt{3}<r$ the BC action in $d=2$ dimensions encodes for $2$ species (which live on the diagonal of the Brillouin zone).

A contour plot for BC fermions in $d=2$ dimensions is shown in Fig.~\ref{fig:contours_bc}.
The momentum range is $]-\frac{5}{4}\pi,\frac{3}{4}\pi[$ for both $ap_1$ and $ap_2$.
At $r=0$ one starts with the naive action.
For infinitesimally small $r$ the poles in the $(ap_1,ap_2)$ plane at $(-\pi,0)$ and $(0,-\pi)$ start moving towards the diagonal,
and the pole at $(-\pi,-\pi)$ moves along the diagonal, while the pole at $(0,0)$ stays invariant.
At $r=1/\sqrt{3}$ the three moving poles merge into a single pole.
For $r>1/\sqrt{3}$ one stays with one pole at $(0,0)$, with correct chirality on topologically charged backgrounds, and the merged pole, with opposite chirality.
For $r\to\infty$ the two surviving pole positions are arbitrarily close to each other.

\bigskip

For BC fermions in $d=4$ dimensions, eqn.~(\ref{BC_zero_original}) at $am=0$ simplifies to
\bea
0&=&\Big[\sin(ap_1)-r\{1-\cos(ap_1)\}\Big]^2+...+\Big[\sin(ap_4)-r\{1-\cos(ap_4)\}\Big]^2
\nonumber
\\
&+&r\Big[\sin(ap_1)+...+\sin(ap_4)\Big] \Big[4-\cos(ap_1)-...-\cos(ap_4)\Big]
\;.
\label{zeros_for_bc_4d}
\eea
Let us first focus on a symmetric mode, i.e.\ put $p_1=p_2=p_3=p_4\equiv p$.
In this case we have
\beq
0=\Big[\cos(ap)-1\Big]\Big[(r^2-1)\cos(ap)-2r\sin(ap)-(r^2+1)\Big]
\eeq
exactly as in $d=2$ dimensions, and the solution is again given by $ap=0$ or $ap=-2\arctan(1/r)$.

For the asymmetric modes it is useful to notice that (\ref{zeros_for_bc_4d}) is the sum of four squares
\bea
0&=&\Big[\sin(ap_1)+\frac{r}{2}\{2+\cos(ap_1)-\cos(ap_2)-\cos(ap_3)-\cos(ap_4)\}\Big]^2
\nonumber
\\
 &+&\Big[\sin(ap_2)+\frac{r}{2}\{2-\cos(ap_1)+\cos(ap_2)-\cos(ap_3)-\cos(ap_4)\}\Big]^2
\nonumber
\\
 &+&\Big[\sin(ap_3)+\frac{r}{2}\{2-\cos(ap_1)-\cos(ap_2)+\cos(ap_3)-\cos(ap_4)\}\Big]^2
\nonumber
\\
 &+&\Big[\sin(ap_4)+\frac{r}{2}\{2-\cos(ap_1)-\cos(ap_2)-\cos(ap_3)+\cos(ap_4)\}\Big]^2
\eea
and one can thus reformulate the condition as a set of four coupled equations
\bea
0&=&\sin(ap_1)+\frac{r}{2}\{2+\cos(ap_1)-\cos(ap_2)-\cos(ap_3)-\cos(ap_4)\}
\nonumber
\\
0&=&\sin(ap_2)+\frac{r}{2}\{2-\cos(ap_1)+\cos(ap_2)-\cos(ap_3)-\cos(ap_4)\}
\nonumber
\\
0&=&\sin(ap_3)+\frac{r}{2}\{2-\cos(ap_1)-\cos(ap_2)+\cos(ap_3)-\cos(ap_4)\}
\nonumber
\\
0&=&\sin(ap_4)+\frac{r}{2}\{2-\cos(ap_1)-\cos(ap_2)-\cos(ap_3)+\cos(ap_4)\}
\;.
\label{zeros_for_bc1}
\eea
By adding two successive equations, this system may be reformulated as
\bea
0&=&\sin(ap_1)+\sin(ap_2)+r\{2-\cos(ap_3)-\cos(ap_4)\}
\nonumber
\\
0&=&\sin(ap_2)+\sin(ap_3)+r\{2-\cos(ap_4)-\cos(ap_1)\}
\nonumber
\\
0&=&\sin(ap_3)+\sin(ap_4)+r\{2-\cos(ap_1)-\cos(ap_2)\}
\nonumber
\\
0&=&\sin(ap_4)+\sin(ap_1)+r\{2-\cos(ap_2)-\cos(ap_3)\}
\label{zeros_for_bc2}
\eea
or one might add three and subtract one out of the four equations to obtain
\bea
0&=&-\sin(ap_1)+\sin(ap_2)+\sin(ap_3)+\sin(ap_4)+2r\{1-\cos(ap_1)\}
\nonumber
\\
0&=&+\sin(ap_1)-\sin(ap_2)+\sin(ap_3)+\sin(ap_4)+2r\{1-\cos(ap_2)\}
\nonumber
\\
0&=&+\sin(ap_1)+\sin(ap_2)-\sin(ap_3)+\sin(ap_4)+2r\{1-\cos(ap_3)\}
\nonumber
\\
0&=&+\sin(ap_1)+\sin(ap_2)+\sin(ap_3)-\sin(ap_4)+2r\{1-\cos(ap_4)\}
\;.
\label{zeros_for_bc3}
\eea
Finally, one might add all four equations to obtain
\beq
0=\sin(ap_1)+...+\sin(ap_4)+r\{4-\cos(ap_1)-...-\cos(ap_4)\}
\label{zeros_for_bc4}
\eeq
and an obvious question is which one of the four equivalent systems
(\ref{zeros_for_bc1}), (\ref{zeros_for_bc2}), (\ref{zeros_for_bc3}), or (\ref{zeros_for_bc4})
would be most useful for finding actual solutions.

The last version is useful for finding the symmetric mode.
With $p_1=...=p_4\equiv p$ eqn.~(\ref{zeros_for_bc4}) simplifies to $0=\sin(ap)+r\{1-\cos(ap)\}$,
or $0=\sin(\frac{ap}{2})[\cos(\frac{ap}{2})+r\sin(\frac{ap}{2})]$.
This means $\sin(\frac{ap}{2})=0$ or $\cos(\frac{ap}{2})=-r\sin(\frac{ap}{2})$.
Hence $ap\in\{0,-2\arctan(1/r)\}$, as was found previously.

The last but one version is useful for solutions with 3-to-1 momentum pairing.
Without loss of generality we assume $p_1=p_2=p_3\equiv p$, $p_4\equiv q$, so eqn.~(\ref{zeros_for_bc3}) takes the form
\bea
0&=&1\sin(ap)+\sin(aq)+2r\{1-\cos(ap)\}
\nonumber
\\
0&=&3\sin(ap)-\sin(aq)+2r\{1-\cos(aq)\}
\label{bc_reduced_3to1}
\eea
and the trigonometric semi-angle substitution $t=\tan(ap/2),u=\tan(aq/2)$ turns this into
\bea
0&=& \frac{ t}{1+t^2} + \frac{u}{1+u^2} +r\{1-\frac{1-t^2}{1+t^2}\}
\nonumber
\\
0&=& \frac{3t}{1+t^2} - \frac{u}{1+u^2} +r\{1-\frac{1-u^2}{1+u^2}\}
\;.
\eea
After multiplication by $(1+t^2)(1+u^2)$ one ends up with
\bea
0&=&1t(1+u^2) + u(1+t^2) +2r t^2 \{1+u^2\}
\nonumber
\\
0&=&3t(1+u^2) - u(1+t^2) +2r u^2 \{1+t^2\}
\eea
and the real-valued solutions include $t=u=0$ and $t=u=-r$ (which are the previously found symmetric solutions)
as well as two non-trivial solutions for $|r|\leq1/\sqrt{2}$, namely
\bea
\{\; t=\frac{r(1-s)}{2r^4+r^2+2s-2} &,&\quad u=\frac{2r^2+s-1}{r(2r^2-1)} \;\}
\nonumber
\\
\{\; t=\frac{r(1+s)}{2r^4+r^2-2s-2} &,&\quad u=\frac{2r^2-s-1}{r(2r^2-1)} \;\}
\label{solu_bc_3to1}
\eea
with $s\equiv\sqrt{-2r^4-r^2+1}$.
For $r\to1/\sqrt{2}$ the last two solutions become
\bea
\{\; t\to-\frac{1}{\sqrt{2}} &,&\quad u\to-\infty \;\}
\nonumber
\\
\{\; t\to-\frac{1}{\sqrt{2}} &,&\quad u\to+\infty \;\}
\eea
meaning $ap\to-2\arctan(1/\sqrt{2})$ and $aq\to\mp\pi$.
We also determine the values which the solutions (\ref{solu_bc_3to1}) assume at $r=1/\sqrt{3}$; we find
\bea
\{\; t\to-\sqrt{3}           &,&\quad u\to-\sqrt{3} \;\}
\nonumber
\\
\{\; t\to-\frac{\sqrt{3}}{5} &,&\quad u\to\,3\sqrt{3} \;\}
\eea
which means that only the first one of these two solutions matches onto the symmetric solution at $r=1/\sqrt{3}$.
With respect to the general solution (\ref{solu_bc_3to1}) let us recall that the choice $p_4\equiv q$ was one out of four possibilities,
hence we have eight rather than two non-trivial solutions.

The second version is useful for solutions with 2-to-2 momentum pairing.
Without loss of generality we assume $p_1=p_2\equiv p$, $p_3=p_4\equiv q$, so eqn.~(\ref{zeros_for_bc2}) takes the form
\bea
0&=&\sin(ap)+r\{1-\cos(aq)\}
\nonumber
\\
0&=&\sin(aq)+r\{1-\cos(ap)\}
\label{bc_reduced_2to2}
\eea
but this is identical to the system (\ref{zeros_bc_2d}) for BC fermions in $d=2$ dimensions.
It follows that $\{t=0,u=0\}$ and $\{t=-1/r,u=-1/r\}$ are the symmetric solutions, and
\bea
\{\; t=\frac{r^2-1+\sqrt{1-2r^2-3r^4}}{2r^3} &,&\quad u=\frac{r^2-1-\sqrt{1-2r^2-3r^4}}{2r^3} \;\}
\nonumber
\\
\{\; t=\frac{r^2-1-\sqrt{1-2r^2-3r^4}}{2r^3} &,&\quad u=\frac{r^2-1+\sqrt{1-2r^2-3r^4}}{2r^3} \;\}
\label{solu_bc_2to2}
\eea
are the non-symmetric ones.
Evidently, the second solution emerges from the first one by interchanging $t \leftrightarrow u$.
The square-root is real for $|r|\leq1/\sqrt{3}$, and at this point the non-symmetric solutions
take the form $t=-1/r=-\sqrt{3}, u=-1/r=-\sqrt{3}$, which means that they merge into the symmetric solution.
We recall that the choice $p_1=p_2\equiv p$ was one out of three possibilities, hence we have six rather than two non-symmetric solutions.

The first version of the system was not used at all.
It seems (\ref{zeros_for_bc1}) would be most useful for finding a totally unsymmetric mode, i.e.\ one with pairwise unequal $p_1,p_2,p_3,p_4$.
Apart from having already found $2+8+6=16$ solutions (for $r$ small enough), permutations demanded by invariance under exchange of any axes
would beef up a totally unsymmetric solution to $4!=24$ solutions, and that is too many of them.

Overall, we thus arrive at the following picture for BC fermions in $d=4$ dimensions.
For infinitesimally small $r$ there is an invariant solution, $ap=(0,0,0,0)$,
a symmetric solution, $ap=-\arctan(1/r)(2,2,2,2)$,
eight solutions with 3-to-1 momentum pairing of the type (\ref{solu_bc_3to1}),
and six solutions with 2-to-2 momentum pairing of the type (\ref{solu_bc_2to2}).
The first two solutions have the correct chirality on topologically charged backgrounds,
the 3-to-1 paired solutions all have opposite chirality,
and the 2-to-2 paired solutions have correct chiralities again.
At $r=1/\sqrt{3}$ a dramatic merger and exchange of chiralities takes place, since
the 2-to-2 paired solutions cease to exist, but their chiralities are transferred to the remaining modes
in the sense that the 3-to-1 paired solutions fall into two sub-categories (the four of them who passed through the central point now have correct chirality),
and the symmetric solution gets flipped to opposite chirality at this point.
In short, for $r$ slightly above $1/\sqrt{3}$ the BC action has $10$ species (five of each chirality).
At $r=1/\sqrt{2}$ the second change takes place, since the eight solutions with 3-to-1 pairing annihilate each other
(in two different points of the Brillouin zone, and they can do so, since four of them have correct chirality, and four of them have opposite chirality).
Slightly above this value of $r$ the BC action is minimally doubled, i.e.\ has one species of each chirality.

\begin{figure}[!tb]
\includegraphics[width=0.99\textwidth]{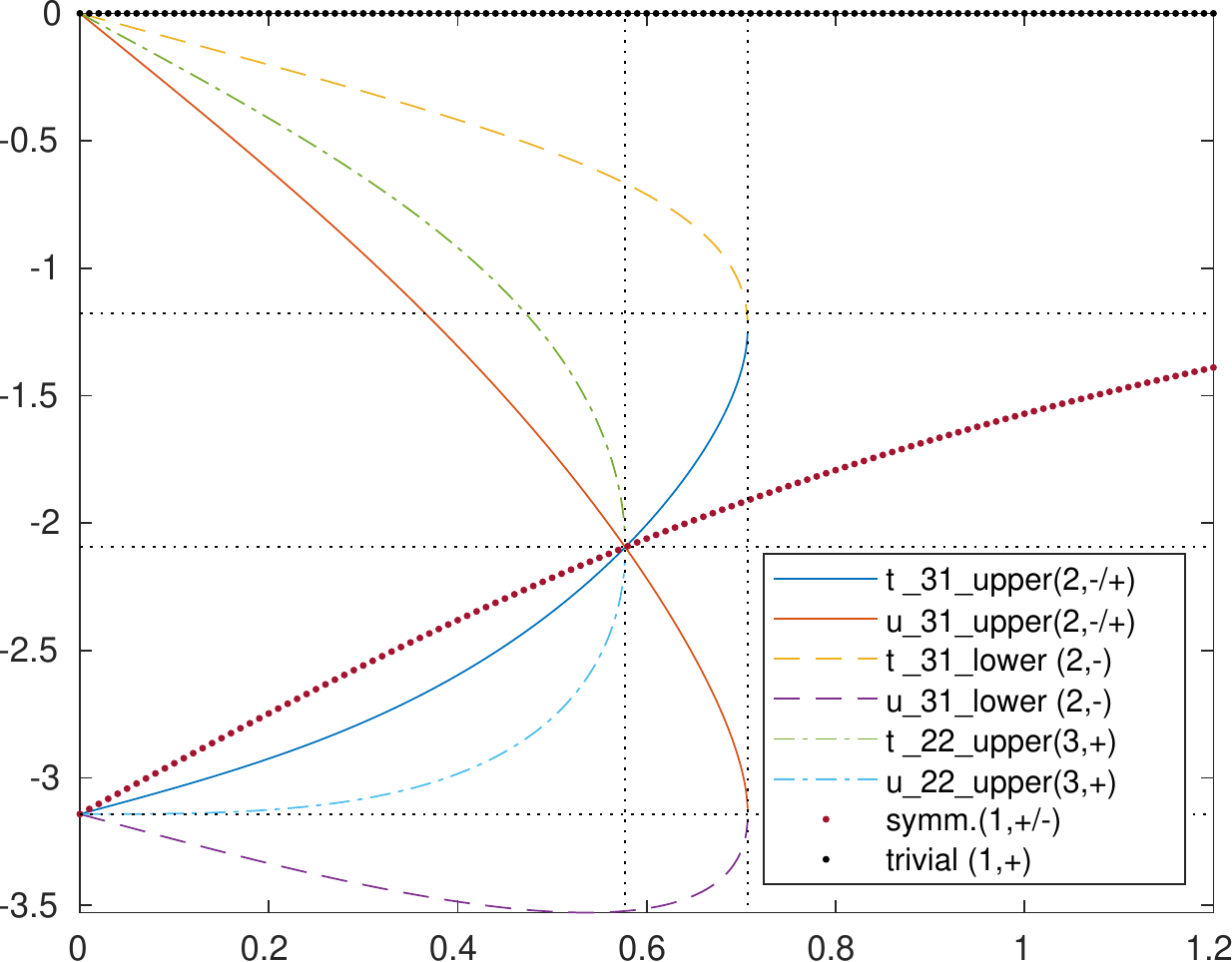}%
\caption{\label{fig:illustration_BC}
Illustration of the free-field pole structure of the Borici-Creutz operator in $d=4$ dimensions.
Throughout, the momentum $ap$ or $aq$ is plotted as a function of the lifting parameter $r$.
The full/dashed lines give $2\arctan(t),2\arctan(u)$ with the 3-to-1 solutions $t,u$ defined in the upper/lower line of (\ref{solu_bc_3to1}).
The dash-dotted lines give $2\arctan(t),2\arctan(u)$ with the 2-to-2 solutions $t,u$ defined in the upper line of (\ref{solu_bc_2to2}).
The lower line of that system interchanges $t \leftrightarrow u$, and would give the same graph.
The fat-dotted lines indicate the symmetric solution $-2\arctan(1/r)$ and the trivial solution.
The horizontal dotted lines are at lattice momentum $-3\pi/8$, $-2\pi/3$, and $-\pi$, respectively.
The vertical dotted lines are at $r=1/\sqrt{2}$, and $r=1/\sqrt{3}$, respectively.
In $d=2$ dimensions all 3-to-1 paired solutions would be absent,
the dash-dotted curve would refer to (\ref{solution_bc_2d}),
and the fat-dotted curves would be unchanged.}
\end{figure}

Following a similar attempt in the KW case, we try to illustrate the various modes in the $(r,ap)$ or $(r,aq)$ plane in Fig.~\ref{fig:illustration_BC}.
At $r=1/\sqrt{3}$ the dramatic merger and exchange of chiralities takes place, as discussed above.
At $r=1/\sqrt{2}$ the second reduction in the number of species takes place, since at this point all 3-to-1 paired solutions annihilate each other.
The trivial solution $(0,0,0,0)$ has correct chirality, symmetric solution has correct chirality for $r<1/\sqrt{3}$ and opposite chirality for $r>1/\sqrt{3}$.

\begin{figure}[p]
\includegraphics[width=0.48\textwidth]{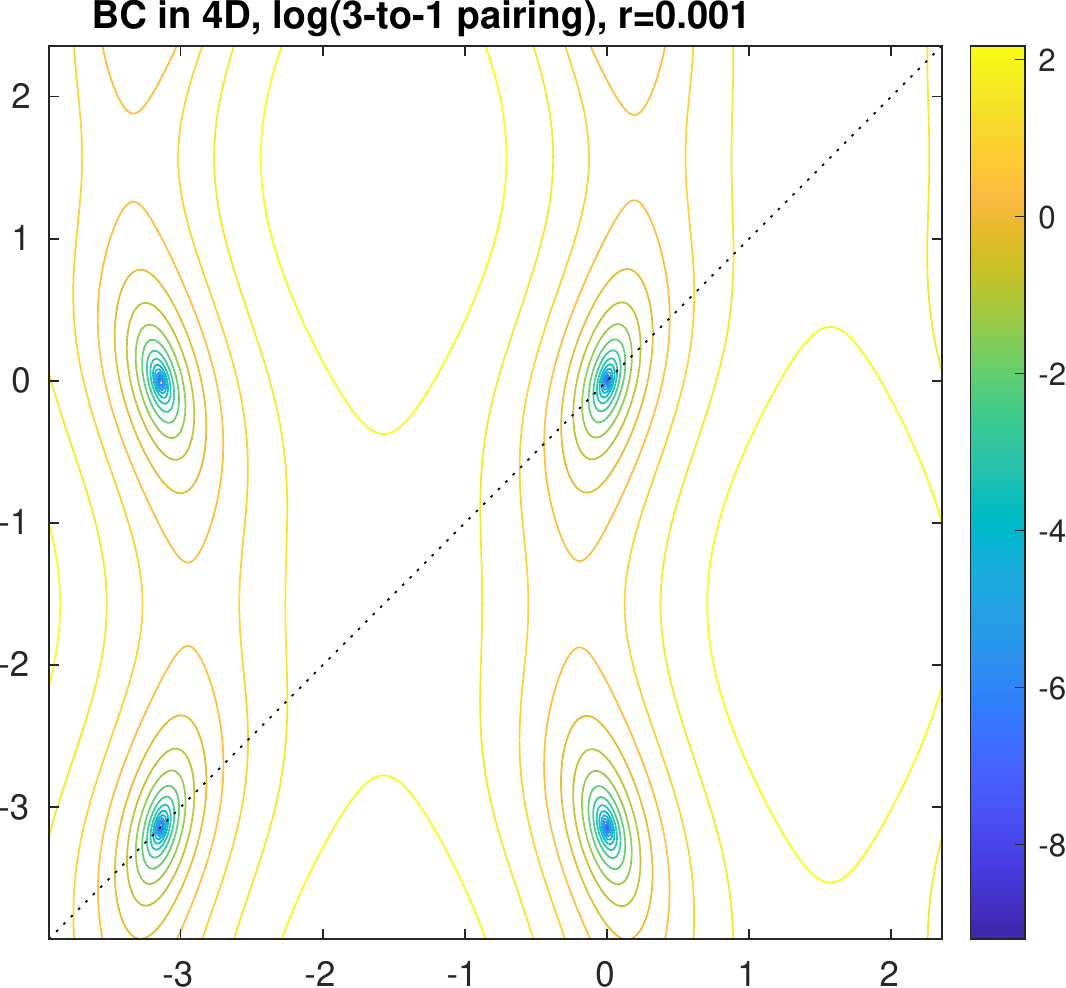}\hfill
\includegraphics[width=0.48\textwidth]{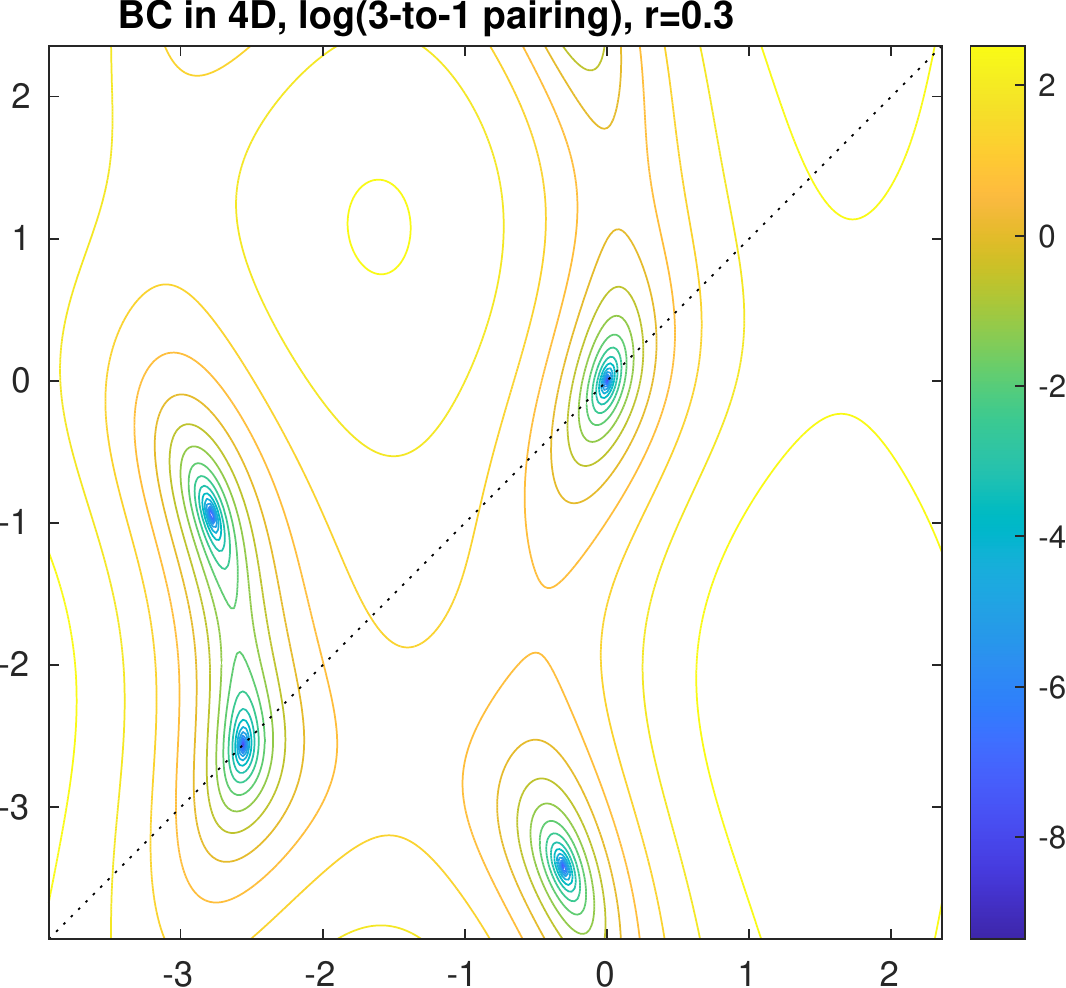}\\
\includegraphics[width=0.48\textwidth]{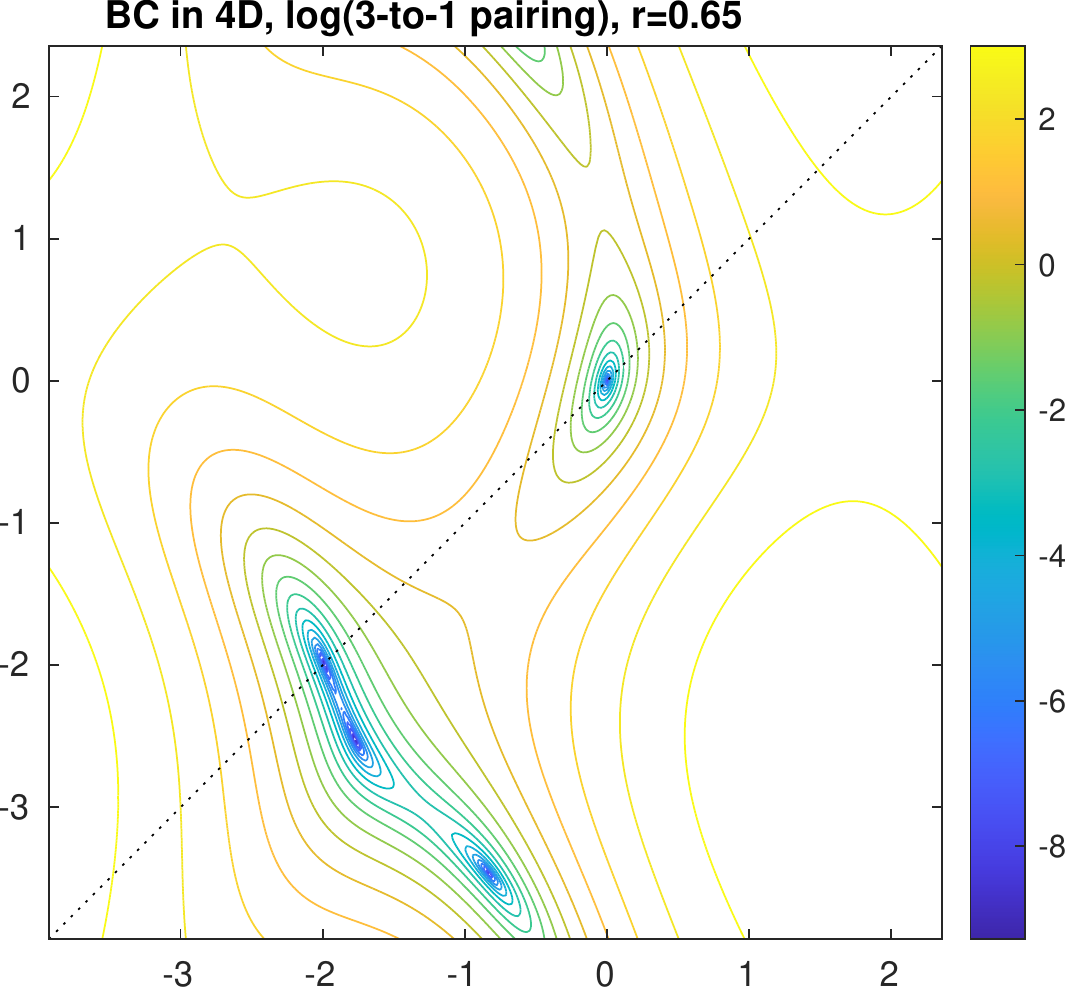}\hfill
\includegraphics[width=0.48\textwidth]{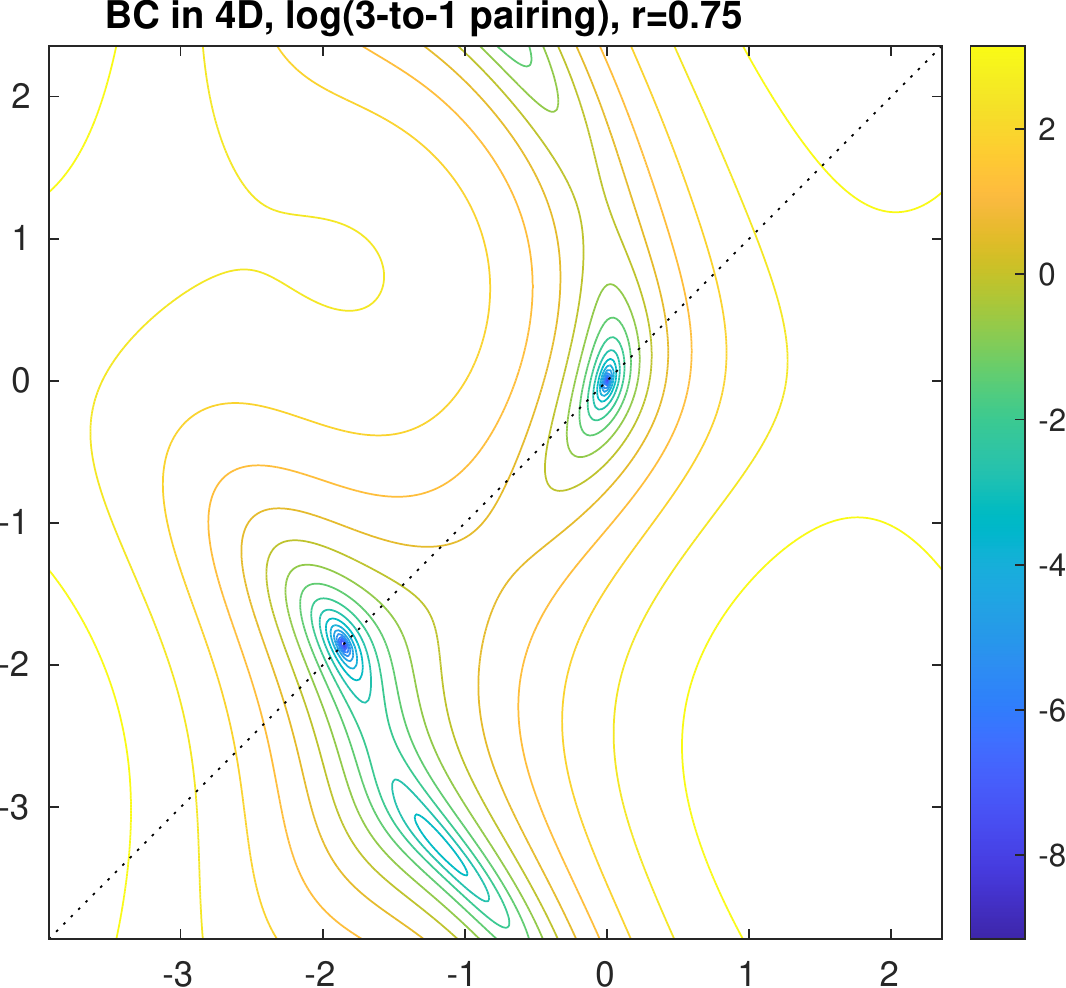}\\
\includegraphics[width=0.48\textwidth]{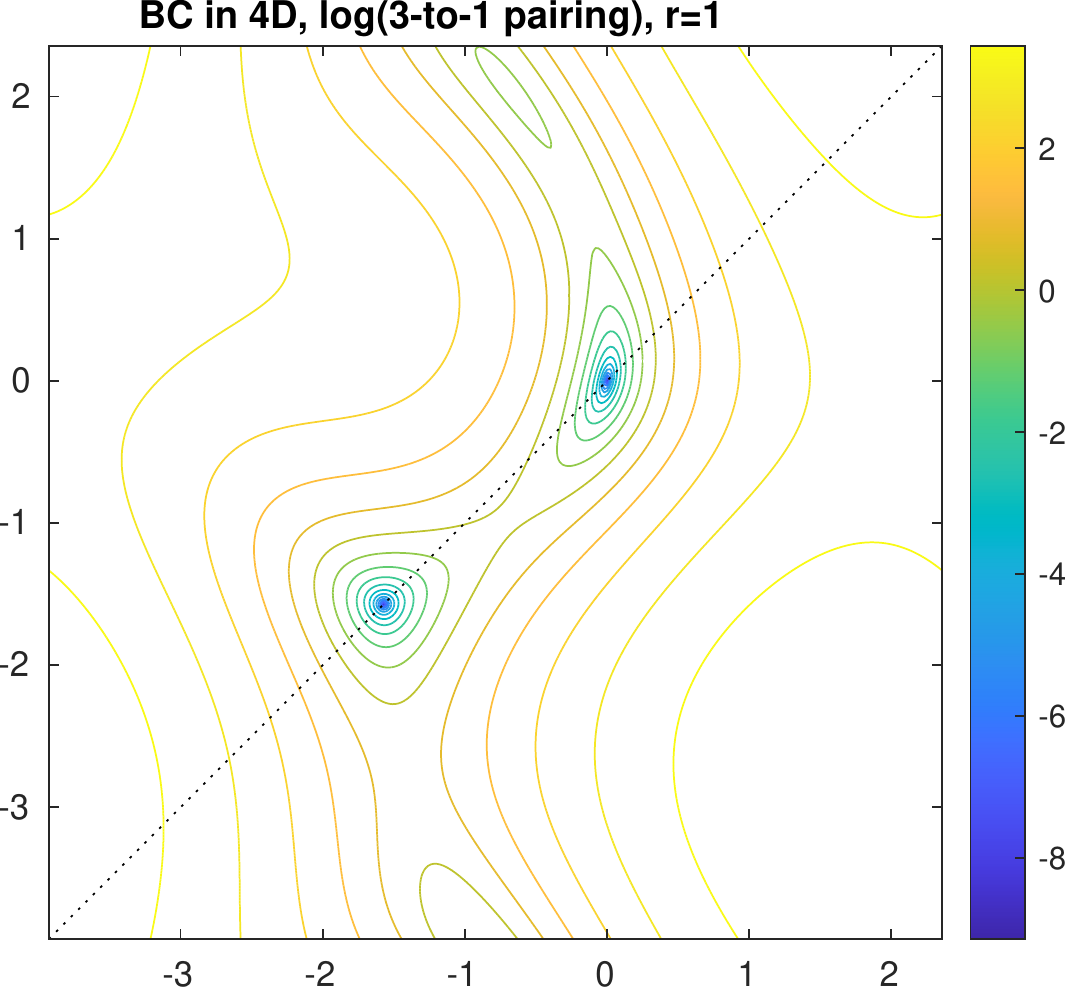}\hfill
\includegraphics[width=0.48\textwidth]{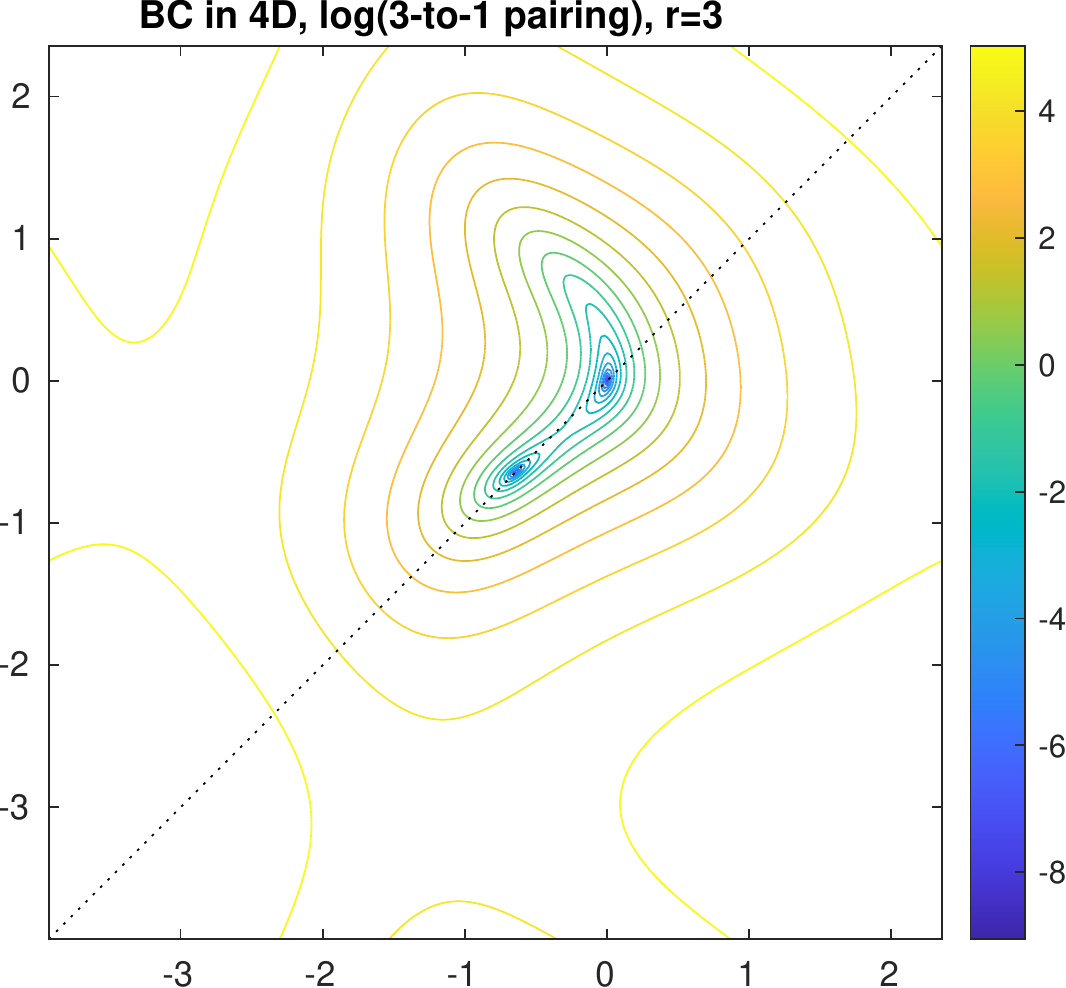}%
\vspace*{-7pt}\caption{\label{fig:contours_3to1}
Contour plots for the solutions to (\ref{bc_reduced_3to1}) for $r\in\{0.001,0.3,0.65,0.75,1,3\}$.
There are touch- and endpoints at $r=1/\sqrt{3}\simeq0.57735$, and $r=1/\sqrt{2}=0.70711$.
See text for details.}
\end{figure}

In $d=4$ dimensions it is more difficult to visualize the moving of the various poles as a function of $r$ than in $d=2$ dimensions.
While it seems impossible to visualize the original system (\ref{zeros_for_bc_4d}), we can visualize each one of
the successor relations (\ref{zeros_for_bc1}), (\ref{zeros_for_bc2}), (\ref{zeros_for_bc3}), and (\ref{zeros_for_bc4})
under the assumption of the associate momentum pairing.
Eqn.~(\ref{zeros_for_bc1}) would be most useful without any pairing, but we just learned that this cannot yield a solution.
Eqn.~(\ref{zeros_for_bc2}) is most useful with 2-to-2 pairing, and the reduced form, eqn.~(\ref{bc_reduced_2to2}),
can be visualized as a contour plot of $[\sin(ap)+r\{1-\cos(aq)\}]^2+[\sin(aq)+r\{1-\cos(ap)\}]^2$.
But this is identical to the functional that was visualized in the $d=2$ case, so the figure would look like Fig.~\ref{fig:contours_bc},
with the axes indicating the joint momenta $p$ and $q$, respectively.
Eqn.~(\ref{zeros_for_bc3}) is most useful with 3-to-1 pairing, and the reduced form, eqn.~(\ref{bc_reduced_3to1}),
can be visualized as a contour plot of $[\sin(ap)+\sin(aq)+2r\{1-\cos(ap)\}]^2+[3\sin(ap)-\sin(aq)+2r\{1-\cos(aq)\}]^2$.
Here $p$ is the $3$-fold momentum, and $q$ is the single momentum.
The pertinent contours, with momentum range $]-\frac{5}{4}\pi,\frac{3}{4}\pi[$ for both $ap$ and $aq$, are shown in Fig.~\ref{fig:contours_3to1}.
For small $r$ one sees the trivial solution $t=u=0$, the symmetric solution $t=u=-r$, as well as the upper line of (\ref{solu_bc_3to1}).
At $r=1/\sqrt{3}$ the pole above the diagonal hits the symmetric solution, and the dramatic exchange of chiralities
(which involves solutions which are not visualized in Fig.~\ref{fig:contours_3to1}) takes place.
And at $r=1/\sqrt{2}$ the two poles below the diagonal annihilate each other, and after this point the BC action is minimally doubled.


\section{Hyperdiagonal propagation of Borici-Creutz fermions\label{app:D}}


Recall that the substitution $p_4\to\ri E$ would bring eqn.~(\ref{BC_zero_original}) to the form (\ref{BC_zero_standard}).
If we let the fermion propagate along the hyperdiagonal direction, we should use the
substitution $\ri E=(p_1+p_2)/\sqrt{2}$ in $d=2$ dimensions, and $\ri E=(p_1+p_2+p_3+p_4)/2$ in $d=4$ dimensions.
The ``spatial'' momenta should be orthogonal to this direction, hence $q\equiv(p_1-p_2)/\sqrt{2}$ in $d=2$ dimensions,
and $q_1\equiv(-p_1+p_2+p_3-p_4)/2,q_2\equiv(p_1-p_2+p_3-p_4)/2,q_3\equiv(p_1+p_2-p_3-p_4)/2$ in $d=4$ dimensions,
since this definition establishes the orthogonality relation $q_1\perp q_2\perp q_3 \perp q_1$.

In $d=2$ dimensions, eqn.~(\ref{BC_zero_original}) simplifies to
\bea
0&=&\Big[\sin(ap_1)-r\{1-\cos(ap_1)\}\Big]^2+\Big[\sin(ap_2)-r\{1-\cos(ap_2)\}\Big]^2
\nonumber
\\
&+&2r\Big[\sin(ap_1)+\sin(ap_2)\Big]\Big[2-\cos(ap_1)-\cos(ap_2)\Big]
+(am)^2
\eea
and with $p_1=(q+\ri E)/\sqrt{2}$ and $p_2=(-q+\ri E)/\sqrt{2}$ it takes the form
\bea
0&=&
-2\ri r\sinh(\sqrt{2}aE)+(r^2-1)\cos(\sqrt{2}aq)\cosh(\sqrt{2}aE)
\nonumber
\\
&+&4\cos(\frac{aq}{\sqrt{2}})(-r^2\cosh(\frac{aE}{\sqrt{2}})+\ri r\sinh(\frac{aE}{\sqrt{2}}))
+1+3r^2+(am)^2
\eea
which is a quartic equation in $e^{aE/\sqrt{2}}$.
Specifically at $q=0$ it simplifies to
\bea
0&=&
-2\ri r\sinh(\sqrt{2}aE)+(r^2-1)\cosh(\sqrt{2}aE)
\nonumber
\\
&+&4(-r^2\cosh(\frac{aE}{\sqrt{2}})+\ri r\sinh(\frac{aE}{\sqrt{2}}))
+1+3r^2+(am)^2
\eea
and upon setting $r=1$ it further simplifies to
\beq
0=4 \Big[1-\cosh\Big(\frac{aE}{\sqrt{2}}\Big)\Big] \Big[1+\ri\sinh\Big(\frac{aE}{\sqrt{2}}\Big)\Big]+(am)^2
\;.
\eeq
This equation has formally four solutions
\beq
aE=\sqrt{2}\ln
\Big(
\mbox{RootOf}(-\ri+\ri\,\_z^4+(2-2\ri)\,\_z^3-(4+m^2)\,\_z^2+(2+2\ri)\,\_z)
\Big)
\eeq
out of which the physical one expands as
\beq
aE=am
\bigg\{
1-\frac{\sqrt{2}\ri}{4}am-\frac{1}{3}(am)^2+\frac{\sqrt{2}\ri}{4}(am)^3+\frac{17}{40}(am)^4+O((am)^5)
\bigg\}
\qquad[r=1]
\;.
\eeq
For generic $r$ the expanded solution reads
\bea
aE&=&am
\bigg\{
1-\frac{\sqrt{2}\ri r}{4}am-(\frac{r^2}{4}+\frac{1}{12})(am)^2+\frac{\sqrt{2}\ri r(5r^2+3)}{32}(am)^3
\nonumber
\\
&&\qquad
+(\frac{7r^4}{32}+\frac{3r^2}{16}+\frac{3}{160})(am)^4+O((am)^5)
\bigg\}
\eea
which, in the limit $r\to1$, is seen to coincide with the previous expansion.

As an aside we mention that choosing the propagation direction orthogonal to the hyperdiagonal axis,
i.e.\ $p_1=(q+\ri E)/\sqrt{2}$ and $p_2=(q-\ri E)/\sqrt{2}$, yields
\bea
0&=&(4(r^2-1)\cos^{2}(q/\sqrt{2})-2(r^2-1))\cosh^2(E/\sqrt{2})
\nonumber
\\
&+&(-4r^2\cos(q/\sqrt{2})+4r\sin(q/\sqrt{2}))\cosh(E/\sqrt{2})
\nonumber
\\
&+&(-2r^2+2)\cos^2(q/\sqrt{2})-4r\sin(q/\sqrt{2})\cos(q/\sqrt{2})+4r^2+m^2
\eea
which is a quadratic equation in $\cosh(aE/\sqrt{2})$.
Specifically at $q=0$ it simplifies to
\beq
0=2(r^2-1)\cosh^2\Big(\frac{aE}{\sqrt{2}}\Big)-4r^2\cosh\Big(\frac{aE}{\sqrt{2}}\Big)+2r^2+2+(am)^2
\label{diagonal_d2}
\eeq
and upon setting $r=1$ it further simplifies to $0=-4\cosh(aE/\sqrt{2})+4+(am)^2$.
This equation is linear in $\cosh(aE/\sqrt{2})$ and yields $\cosh(aE/\sqrt{2})=1+(am)^2/4$
which, in turn, expands as
\beq
aE=am\bigg\{1-\frac{1}{48}(am)^2+\frac{3}{2560}(am)^4+O((am)^6)\bigg\}
\qquad[r=1]
\;.
\eeq
The quadratic equation~(\ref{diagonal_d2}) has the unique physical solution
\beq
\cosh\Big(\frac{aE}{\sqrt{2}}\Big)=\frac{2r^2-\sqrt{4-2(r^2-1)(am)^2}}{2(r^2-1)}
\eeq
since the other mathematical solution does not match onto the $r=1$ case,
and expands as
\beq
aE=am\bigg\{
1
+\frac{3r^2-4}{48}(am)^2
+\frac{35r^4-80r^2+48}{2560}(am)^4
+O((am)^6)
\bigg\}
\;.
\eeq
In this peculiar case choosing $r^2=4/3$ shifts the leading cut-off effects in $aE$ to $O((am)^4)$.

In $d=4$ dimensions, eqn.~(\ref{BC_zero_original}) simplifies to
\bea
0&=&\Big[\sin(ap_1)-r\{1-\cos(ap_1)\}\Big]^2+...+\Big[\sin(ap_4)-r\{1-\cos(ap_4)\}\Big]^2
\nonumber
\\
&+&r\Big[\sin(ap_1)+...+\sin(ap_4)\Big] \Big[4-\cos(ap_1)-...-\cos(ap_4)\Big]
+(am)^2
\eea
and with $p_1=(-q_1+q_2+q_3+\ri E)/2$, $p_2=(q_1-q_2+q_3+\ri E)/2$, $p_3=(q_1+q_2-q_3+\ri E)/2$
and $p_4=(-q_1-q_2-q_3+\ri E)/2$ it takes the form
\bea
0&=&8\sin(\frac{aq_1}{2})\sin(\frac{aq_2}{2})\sin(\frac{aq_3}{2})
\Big[
\ri r^2\sinh(\frac{aE}{2})
+r\cosh(\frac{aE}{2})
\Big]
\nonumber
\\
&+&8\cos(\frac{aq_1}{2})\cos(\frac{aq_2}{2})\cos(\frac{aq_3}{2})
\Big[-r^2\cosh(\frac{aE}{2})+\ri r\sinh(\frac{aE}{2})\Big]
\nonumber
\\
&-&2\ri(r^2-1)\sin(aq_1)\sin(aq_2)\sin(aq_3)\sinh(aE)
\nonumber
\\
&+&2(r^2-1)\cos(aq_1)\cos(aq_2)\cos(aq_3)\cosh(aE)
\nonumber
\\
&+&2\ri r\Big[\cos(aq_1)\cos(aq_2)\cos(aq_3)-\cos(aq_1)-\cos(aq_2)-\cos(aq_3)\Big]\sinh(aE)
\nonumber
\\
&+&2r\sin(aq_1)\sin(aq_2)\sin(aq_3)\cosh(aE)
+6r^2+2+(am)^2
\eea
which is a bi-quadratic equation in $\cosh(aE/2)$ and $\sinh(aE/2)$.
At zero momentum orthogonal to the (++++) propagation direction, i.e.\ at $q_1=q_2=q_3=0$, it simplifies to
\beq
0=4(r^2-1)\cosh^2(\frac{aE}{2})+8(r^2+\ri r\sinh(\frac{aE}{2}))(1-\cosh(\frac{aE}{2}))-4(r^2-1)+(am)^2
\label{diagonal_d4}
\eeq
and upon setting $r=1$ it further simplifies to
\beq
0=8 \Big[1-\cosh(\frac{aE}{2})\Big] \Big[1+\ri\sinh(\frac{aE}{2})\Big] + (am)^2
\;.
\eeq
This equation has formally four solutions, but only
\beq
aE=2\log\Big(\frac{1-\ri}{4}\Big[2\ri+am+\sqrt{4+4\ri am+(am)^2}\Big]\Big)
\eeq
is physical, since it expands as
\beq
aE=am
\bigg\{
1-\frac{\ri}{4}am-\frac{1}{6}(am)^2+\frac{\ri}{8}(am)^3+\frac{17}{160}(am)^4+O((am)^5)
\bigg\}
\eeq
while the remaining ones have a constant imaginary part and/or start with a negative slope in $am$.
The quartic equation~(\ref{diagonal_d4}) has the unique physical solution
\beq
aE=2\log\Big(\frac{1-\ri r}{2(1+r^2)}\Big[2\ri r+am+\sqrt{4+4\ri ram+(am)^2}\Big]\Big)
\eeq
since the remaining ones do not match onto the $r=1$ case, and it expands as
\beq
aE=am
\bigg\{
1
-\frac{\ri r}{4}am
-\frac{3r^2+1}{24}(am)^2
+\frac{\ri r(5r^2+3)}{64}(am)^3
+\frac{35r^4+30r^2+3}{640}(am)^4
+O((am)^5)
\bigg\}
\eeq
which, in the limit $r\to1$, is found to reproduce the previous result.


In short, for BC fermions with propagation in the hyperdiagonal direction we find similar properties than with the standard propagation along the $d$-th axis.
In $d=2$ and $d=4$ dimensions the rest mass of a BC fermion with diagonal propagation direction has $O(am)$ cut-off effects, but this order affects only the imaginary part.
The coefficient of the $O((am)^2)$ cut-off effects is $-[3r^2+1]/12$ in $d=2$ dimensions and $-[3r^2+1]/24$ in $d=4$ dimensions.
Again, it seems that the real part of $E/m$ is even in $r$ and $am$, while the imaginary part is odd in $r$ and $am$.
Overall, we do not see any compelling advantage of the (++) or (++++) propagation direction over the standard propagation in the $d$-th direction.
A peculiarity of $d=2$ space-time dimensions is that the propagation direction can be chosen orthogonal
to the hyperdiagonal direction, and in this case the $O(am)$ cut-off effects disappear,
and for $r^2=4/3$ the leading cut-off effect in the heavy-quark mass is pushed to $O((am)^4)$.


\section{Spectral bounds\label{app:E}}


\subsection{Karsten-Wilczek operator}

Plugging in the momenta in eqn.~(\ref{momrep_KW}) at $m=0$ yields
\beq
aD_\mr{KW}/\ri=\sum_{i=1}^{d-1}\ga_i\sin(ap_i)+\ga_d\Big\{\sin(ap_d)+2r\sum_i\sin^2(\frac{ap_i}{2})\Big\}^2
\eeq
and with $\om_\mr{KW}\equiv\max(\la_\mr{KW}/\ri)$ it follows that the symmetry among the spatial axes implies
\beq
\om_\mr{KW}^2=(d-1)\sin^2(ap)+\Big\{\sin(ap_d)+2(d-1)r\sin^2(\frac{ap}{2})\Big\}^2
\eeq
for some appropriately chosen momentum configuration.
Obviously, setting $ap_d=\pi/2$ helps to reach the maximum.
Using $2\sin^2(ap/2)=1-\cos(ap)$ we thus need to maximize
\beq
\om_\mr{KW}^2=(d-1)\sin^2(ap)+\Big\{1+(d-1)r[1-\cos(ap)]\Big\}^2
\eeq
over $p\in[-\pi,\pi]$.
We need to keep in mind that at either endpoint we have the value
\beq
\om_\mr{KW}^2=\Big\{1+2(d-1)r\Big\}^2=
\left\{
\begin{array}{ll}
(1+6r)^2 & [d=4]\\
(1+2r)^2 & [d=2]
\end{array}
\right.
\;.
\eeq
Taking the derivative with respect to $ap$ and setting it to zero yields
\beq
0=\cos(ap)+\Big\{1+(d-1)r[1-\cos(ap)]\Big\}r
\eeq
and this leads to the solution
\beq
\cos(ap)=\frac{(d-1)r^2+r}{(d-1)r^2-1}=
\left\{
\begin{array}{ll}
(3r^2+r)/(3r^2-1) & [d=4]\\
r/(r-1)           & [d=2]
\end{array}
\right.
\;.
\eeq
Plugging the $d=4$ result into the general expression yields
\beq
\om_\mr{KW}^2=3[1-\frac{(3r^2+r)^2}{(3r^2-1)^2}]+\Big\{1+3r[1-\frac{3r^2+r}{3r^2-1}]\Big\}^2=\frac{4+6r}{1-3r^2}
\eeq
and equating this with the endpoint value shows that the switching beween the two solutions happens at $r=1/3$.
Plugging the $d=2$ result into the general expression yields
\beq
\om_\mr{KW}^2=1-\frac{r^2}{(r-1)^2}+\Big\{1+r[1-\frac{r}{r-1}]\Big\}^2=\frac{2}{1-r}
\eeq
and equality with the endpoint value is reached at $r=1/2$.

To summarize, in $d=4$ dimensions we find the spectral bound~(\ref{specbound_KW}).
In $d=2$ dimensions
\beq
|\mr{Im}(\la_\mr{KW})|\leq
\left\{
\begin{array}{ll}
\sqrt{2/(1-r)} & r\leq1/2\\
1+2r           & r\geq1/2
\end{array}
\right.
\eeq
and the bound for large $r$ generalizes to $1+2(d-1)r$ in $d$ dimensions.
For $r\to0$ the general result tends to $\sqrt{d}$, which is the upper bound of the staggered free-field eigenvalue spectrum.

The upper envelope of the numerical data in Fig.~\ref{fig:linlog_KW} is well consistent with the bound~(\ref{specbound_KW}).
For $r<1/3$ the value is (depending on the volume) very close to the bound; this is unsurprising, since the bound comes from an ``internal value''.
For $r>1/3$ the bound is saturated by the numerical value; again this is unsurprising, since the bound stems from an ``endpoint value''.


\subsection{Borici-Creutz operator}

Plugging in the momenta in eqn.~(\ref{momrep_BC}) at $m=0$, and using eqn.~(\ref{def_dualgamma})  yields
\beq
aD_\mr{BC}/\ri=\sum_\mu\ga_\mu\sin(ap_\mu)+2r\sum_\mu\Big[\frac{2}{\sqrt{d}}\Gamma-\ga_\mu\Big]\sin^2(ap_\mu/2)
\eeq
and with the definition~(\ref{def_big}) we obtain the expression
\beq
aD_\mr{BC}/\ri=\sum_\mu\ga_\mu\sin(ap_\mu)+\frac{4r}{d}\sum_\nu\ga_\nu\cdot\sum_\mu\sin^2(ap_\mu/2)-2r\sum_\mu\ga_\mu\sin^2(ap_\mu/2)
\eeq
where we may interchange the indices $\mu\leftrightarrow\nu$ in the middle term.
This yields
\beq
aD_\mr{BC}/\ri=\sum_\mu\ga_\mu
\Big\{
\sin(ap_\mu)+\frac{4r}{d}\sum_\nu\sin^2(ap_\nu/2)-2r\sin^2(ap_\mu/2)
\Big\}
\eeq
and with $\om_\mr{BC}\equiv\max(\la_\mr{BC}/\ri)$ it follows that the momentum symmetry implies
\beq
\om_\mr{BC}^2=d\Big\{
\sin(ap)+2r\sin^2(ap/2)
\Big\}^2
\;.
\eeq
Using $2\sin^2(ap/2)=1-\cos(ap)$ we thus need to maximize
\beq
\om_\mr{BC}^2=d\Big\{\sin(ap)+r[1-\cos(ap)]\Big\}^2
\eeq
over $p\in[-\pi,\pi]$, and we take the liberty to maximize \emph{or} minimize, instead,
\beq
\sin(ap)+r[1-\cos(ap)]
\eeq
over the same interval.
At either endpoint the original expression takes the value $d4r^2$.
Taking the derivative of the alternative expression with respect to $ap$, and setting it to zero yields
\beq
0=\cos(ap)+r\sin(ap)
\eeq
which finds the solutions
\beq
ap=\arctan(\frac{-1}{r})+\pi\,\mb{Z}
\;.
\eeq
Plugging in the version which realizes the global maximum of the original expression,
and using $\sin(\arctan(x))=x/\sqrt{1+x^2}$, $\cos(\arctan(x))=1/\sqrt{1+x^2}$, we find
\beq
\om_\mr{BC}^2
=d\Big\{\frac{1/r}{\sqrt{1+1/r^2}}+r\Big[1+\frac{1}{\sqrt{1+1/r^2}}\Big]\Big\}^2
=d\Big\{r+\sqrt{1+r^2}\Big\}^2
\;.
\eeq

In summary, since this value is always larger than the endpoint value, we have
\beq
|\mr{Im}(\la_\mr{BC})|\leq\sqrt{d}(r+\sqrt{1+r^2})
\eeq
which was quoted as eqn.~(\ref{specbound_BC}) for $d=4$.
In the limit $r\to0$, it takes the value $\sqrt{d}$, which is known to be the upper bound of the staggered free-field eigenvalue spectrum.

The upper envelope of the numerical data in Fig.~\ref{fig:linlog_BC} is well consistent with the bound~(\ref{specbound_BC}).
The bound is usually not saturated (except for $r=0$ and $r=1$), but the numerical value is (depending on the volume) very close to the bound.



\end{document}